\documentclass[aps,prd,amsmath,amssymb,superscriptaddress,showkeys,showpacs,twocolumn,floatfix]{revtex4-1}

\usepackage[utf8]{inputenc}

\usepackage{diagbox}

\usepackage{mathtools}
\usepackage{amsfonts}
\usepackage{mathrsfs}
\usepackage{bbm}
\usepackage{slashed}

\usepackage{graphicx}
\usepackage{color}
\usepackage{array}
\usepackage{esint}
\usepackage{placeins}
\usepackage{booktabs}
\usepackage{makecell}
\usepackage{epstopdf}
\usepackage[caption=false]{subfig}

\usepackage{xspace}
\usepackage{siunitx}
\usepackage{hyperref}
\usepackage[nameinlink]{cleveref}
\usepackage{appendix}

\usepackage{xifthen}
\usepackage{xcolor}
\hypersetup{
	colorlinks,
	linkcolor={red!75!black},
	citecolor={blue!75!black},
	urlcolor={blue!75!black}
}
\usepackage{comment}

\setkeys{Gin}{width=0.48\textwidth}
\graphicspath{
	{./figures/}
	{../figures/}
}

\def\s0#1#2{\mbox{\small{$ \frac{#1}{#2} $}}}
\def\0#1#2{\frac{#1}{#2}}

\usepackage{xcolor}

\usepackage{ulem}

\begin{document}
	
	\title{A quantitative analysis of Gravitational Wave spectrum sourced from First-Order  Chiral Phase Transition of QCD}
	
	\author{Hui-wen Zheng}
	 \email{zhenghuiwen@stu.pku.edu.cn}
	\affiliation{Department of Physics and State Key Laboratory of Nuclear\\ Physics and Technology, Peking University, Beijing 100871, China.}	

	\author{Fei Gao}
	\email{fei.gao@bit.edu.cn}
	\affiliation{School of Physics, Beijing Institute of Technology, 100081 Beijing, China}

	\author{Ligong Bian}
	\email{lgbycl@cqu.edu.cn}
	\affiliation{Department of Physics and Chongqing Key Laboratory for Strongly Coupled Physics, Chongqing University, Chongqing 401331, China.}
	\affiliation{Center for High Energy Physics, Peking University, Beijing 100871, China.}


\author{Si-xue Qin}
	\email{sqin@cqu.edu.cn}
	\affiliation{Department of Physics and Chongqing Key Laboratory for Strongly Coupled Physics, Chongqing University, Chongqing 401331, China.}

	\author{Yu-xin Liu}
	\email{yxliu@pku.edu.cn}
	\affiliation{Department of Physics and State Key Laboratory of Nuclear\\ Physics and Technology, Peking University, Beijing 100871, China.}
	\affiliation{Center for High Energy Physics, Peking University, Beijing 100871, China.}
	\affiliation{Collaborative Innovation Center of Quantum Matter, Beijing 100871, China.}
	
	\date{\today}
	
	\begin{abstract}
	We investigate the cosmological first-order chiral phase transition of  QCD, and for the first time calculate its parameters which can determine the gravitational wave spectrum. With the state-of-the-art calculation from the functional QCD method,
    we found that   the large chemical potential of QCD phase transition results in  very weak and fast first-order phase transitions at the  temperature  lower than $\mathcal{O}(10^2)$  MeV.
    These results further suggest that the GW signals of NANOGrav are very unlikely sourced from the chiral phase transition of   QCD.   

        %

	\end{abstract}
	\maketitle
	
	{ \bf Introduction.}
     {The space-based interferometers, such as $\mu$Ares~\cite{Sesana:2019vho}, Taiji~\cite{Hu:2017mde,Ruan:2018tsw}, Tianqin~\cite{TianQin:2015yph,Zhou:2023rop}, the Laser Interferometer Space
    Antenna (LISA)~\cite{LISA:2017pwj,Baker:2019nia,Boileau:2021gbr,Schmitz:2020rag}, the Big Bang Observer (BBO)~\cite{Crowder:2005nr,Harry:2006fi} and the Deci-Hertz Interferometer Gravitational Wave Observatory (DECIGO)~\cite{Seto:2001qf,Yagi:2011wg,Isoyama:2018rjb}, offer a frequency window from $10^{-7}\, {\rm Hz}$ to $10\, {\rm Hz}$ for detecting the stochastic gravitational waves background (SGWB).}
    %
    %
    Moreover, the evidences of the SGWB at nanohz are detected by the pulsar timing arrays (PTA) collaborations, including 
    the North American Nanohertz Observatory for Gravitational Waves (NANOGrav)~\cite{NANOGrav:2020bcs,NANOGrav:2023gor}, the European Pulsar Timing Array (EPTA)~\cite{EPTA:2021crs,EPTA:2023fyk},  the Parkes Pulsar Timing Array (PPTA)~\cite{Goncharov:2021oub,Reardon:2023gzh}, the International Pulsar Timing Array (IPTA)~\cite{Antoniadis:2022pcn}, and the Chinese Pulsar Timing Array (CPTA)~\cite{Xu:2023wog}.  { The generation mechanism of SGWB is  a highly debated topic with  enormous models including supermassive black hole binaries (SMBHBs)~\cite{Rajagopal:1994zj,Phinney:2001di,Jaffe:2002rt,Wyithe:2002ep}, curvature perturbations~\cite{Ananda:2006af,Baumann:2007zm}, and new-physics models such as first-order phase transition (FOPT)~\cite{Losada:1996ju,Cline:1996mga,Neronov:2020qrl,Li:2021qer,Nakai:2020oit,Ratzinger:2020koh,Kosowsky:1992rz,Caprini:2010xv,Xue:2021gyq,Ellis:2023tsl,Addazi:2023jvg,Ghosh:2023aum,Rezapour:2022iqq,Ahmadvand:2017xrw,Athron:2023mer,Han:2023olf,Salvio:2023blb,Qin:2024idc,Gouttenoire:2023bqy}, cosmic strings~\cite{Siemens:2006yp,Blasi:2020mfx,Buchmuller:2020lbh,Lazarides:2023ksx,Wang:2023len,Samanta:2020cdk,Bian:2022tju}, and domain walls~\cite{Hiramatsu:2013qaa,Ferreira:2022zzo,Kitajima:2023cek,Blasi:2023sej,Li:2023yzq,Bian:2022qbh,Babichev:2023pbf,Sakharov:2021dim,Zhang:2023nrs,Gouttenoire:2023ftk}, etc.}
    %
    %
    Amongst the  studies, it has been estimated that the SGWB can be induced by  a strong  FOPT at about $T_*\sim 1-10$ MeV~\cite{NANOGrav:2023hvm,Bian:2023dnv,Bai:2021ibt}, which  makes the cosmic QCD phase transition (PT) a competitive candidate. Moreover, the recent studies suggest that the large lepton asymmetries can induce the  QCD FOPT in early Universe, and give the observed value of  baryon asymmetry via the  so called ``sphaleron freeze-in" mechanism~\cite{Gao:2021nwz, Gao:2023djs}. Thus, a strong  QCD phase transition may be a simple solution for multiple problems in cosmology. 
   
    The nonperturbative nature of QCD makes it difficult to study its PT dynamics, especially when it comes to the quantitative information of thermodynamics quantities and the effective potential of QCD. Recently, the quantitative truncation scheme together with the computation of thermodynamics quantities are reached~\cite{Lu:2023mkn}, and a novel method to determine the effective potential  was proposed in Ref.~\cite{Zheng:2023tbv}. These progresses make the quantitative analysis of the QCD PT viable through 
    { the Dyson-Schwinger equation (DSE)~\cite{Roberts:1994dr,Alkofer:2000wg,Fischer:2006ub,Gao:2021wun,Lu:2023mkn,Zheng:2023tbv}.}
    For the first time, we quantitatively determine the strength, the duration, and the percolation temperature $T_*$ of the QCD PT. 
    We then compute the gravitational waves (GWs) spectrum from QCD PT and found that the signal from QCD PT is unlikely to  match with the NANOGrav measurements and the being constructed GW detectors in the frequency range from nanohertz to hertz.

    
    {\bf Cosmic trajectories}
    In the QCD epoch, the baryon number $Y_B$, the individual lepton flavour number $Y_{L_f}$, and the electric charge $Q$ asymmetry can be considered as the conserved quantities during the QCD PT~\cite{Schwarz:2009ii}. Thus, without any further assumptions, one can consider the conservation equations:
    \begin{equation}
\label{eq:five_conse_eq}
          {{Y}_{L_f }}= \frac{{{n}_{f }}+{{n}_{{{\nu }_{f }}}}}{s}, \quad  
         {{Y}_{B}}= \sum\limits_{i}{\frac{{{B}_{i}}{{n}_{i}}}{s}}, \quad  
         Q= \sum\limits_{i}{\frac{{{Q}_{i}}{{n}_{i}}}{s}},
    \end{equation}
    with $s$ the total entropy density of the Universe, $n_i$ the net number density (i.e., the particle
    minus the anti-particle number density of the species $i$), $f=(e,\mu,\tau)$, $i=(e,\mu,\tau, u, d, s, c)$, $B_i$ the baryon number of species $i$, $Q_i$ the electric charge of species $i$, $Y_B=8.7\times 10^{-11}$ the observed baryon asymmetry~\cite{Planck:2018vyg}, $Q=0$ the charge neutrality for universe. $Y_{L_f}$ is less constrained, allowing for a larger lepton asymmetry, provided it satisfies the big bang nucleosynthesis (BBN) and the cosmic microwave background (CMB) constraint $\left| {{Y}_{{{L}_{e}}}}+{{Y}_{{{L}_{\mu }}}}+{{Y}_{{{L}_{\tau }}}} \right|<1.2\times{{10}^{-2}}$~\cite{Pitrou:2018cgg,Oldengott:2017tzj,Popa:2008tb,Simha:2008mt,Serpico:2005bc,Gelmini:2020ekg}. 
    Under this constraint, the individual $Y_{L_f}$ will decrease  once the neutrino oscillations start at $T_*\sim 10\,\rm{MeV}$~\cite{Pastor:2008ti,Mangano:2010ei,Mangano:2011ip}. Therefore, the large lepton asymmetries exist only for $T_*>10\,\rm{MeV}$.
    For $T_*<10\,\rm{MeV}$, the diminished $Y_{L_f}$ will lower the lepton and the quark chemical potentials to satisfy the BBN constraint.
    

    Under the conditions of the kinetic and the chemical equilibrium~\cite{Schwarz:2009ii}, the net leptons and the photons number densities are expressed by Fermi-Dirac distributions and Bose-Einstein distributions, respectively. In the QCD sector, the nonperturbative property  prevents the QCD matter from being simply represented by the free particle distributions. Therefore, we apply the DSE to accomplish the complete calculations of the QCD phase diagram, the thermodynamic  quantities and especially here the effective potential, which is for the first time obtained and applied to compute the gravitational wave signals. 
    
    

    \begin{figure}[t]
        \centering
        \includegraphics[width=1\linewidth]{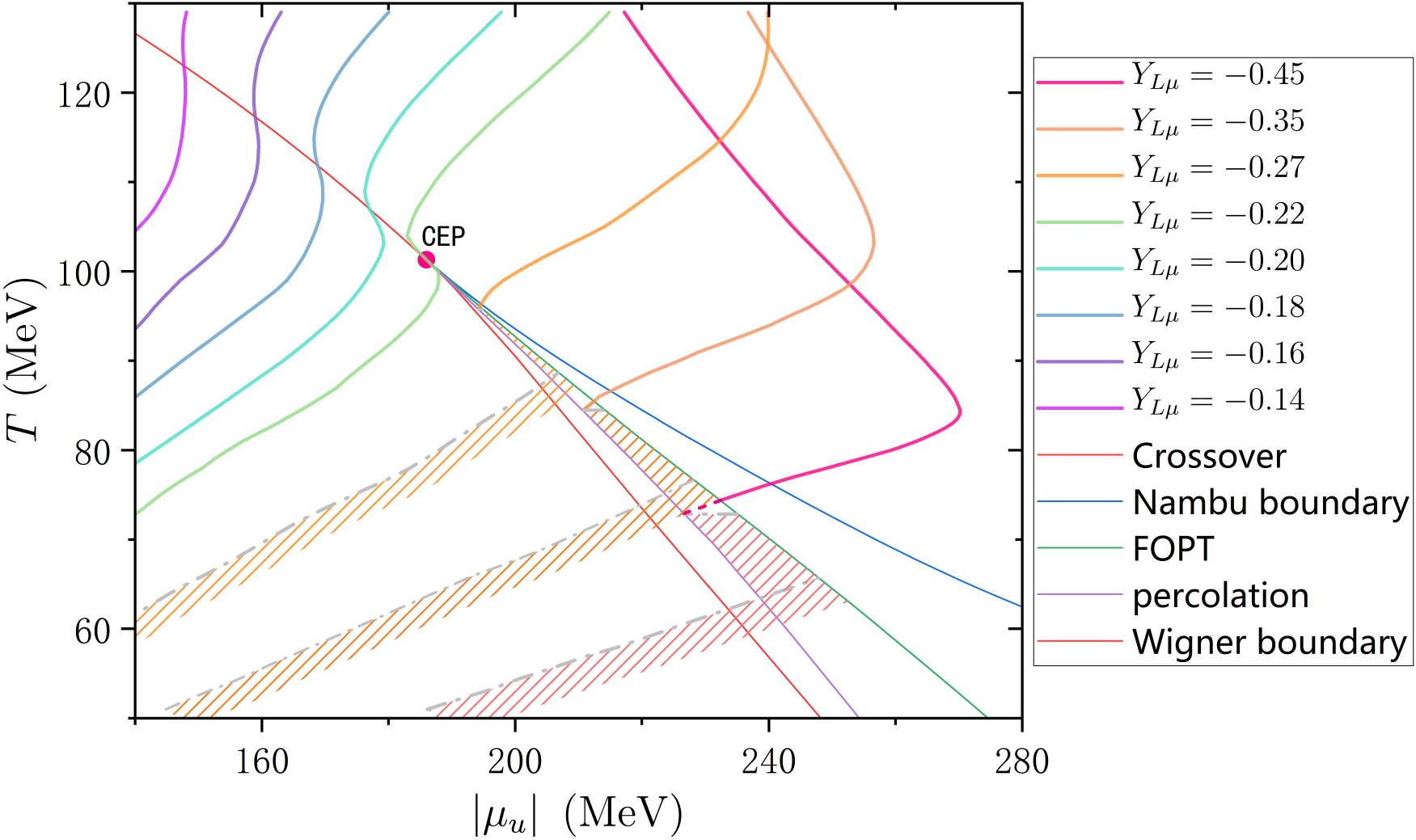}
        \caption{Cosmic trajectories in the scenario $Y_{L_e}=0$ and $Y_{L_\mu}=-Y_{L_\tau}$ in QCD phase diagram. 
        %
        The cosmic trajectories intersect the FOPT line at $(T_c, \mu_c)$ and the percolation line at $(T_*,\mu_*)$.
        Solid line: the determined cosmic trajectory; dashed line: the supercooling stage for a cosmic trajectory before the percolation and after the FOPT; gray dot-dashed line: the upper bound for the possible cosmic trajectory; shadow region: an area contains possible cosmic trajectory.
        }
        \label{fig:costraj_all}
    \end{figure}


    {This work focuses on the scenario \((Y_{L_e}=0, Y_{L_\mu}=-Y_{L_\tau})\), which satisfies BBN constraints and can induce the QCD FOPT~\cite{Middeldorf-Wygas:2020glx,Gao:2021nwz}. Fig.~\ref{fig:costraj_all} shows how the cosmic trajectory depends on \(Y_{L_\mu}\). The critical value \(Y_{L_\mu}=-0.22\) triggers the FOPT, which is consistent with the previous study~\cite{Gao:2023djs}.  
    As \(Y_{L_\mu}\) decreases from \(-0.14\) to \(-0.22\), the cosmic trajectory quickly approaches the CEP. When \(Y_{L_\mu}<-0.22\), the trajectory enters the FOPT region.}
    
    \begin{figure}[t]
        \centering
        \includegraphics[width=1\linewidth]{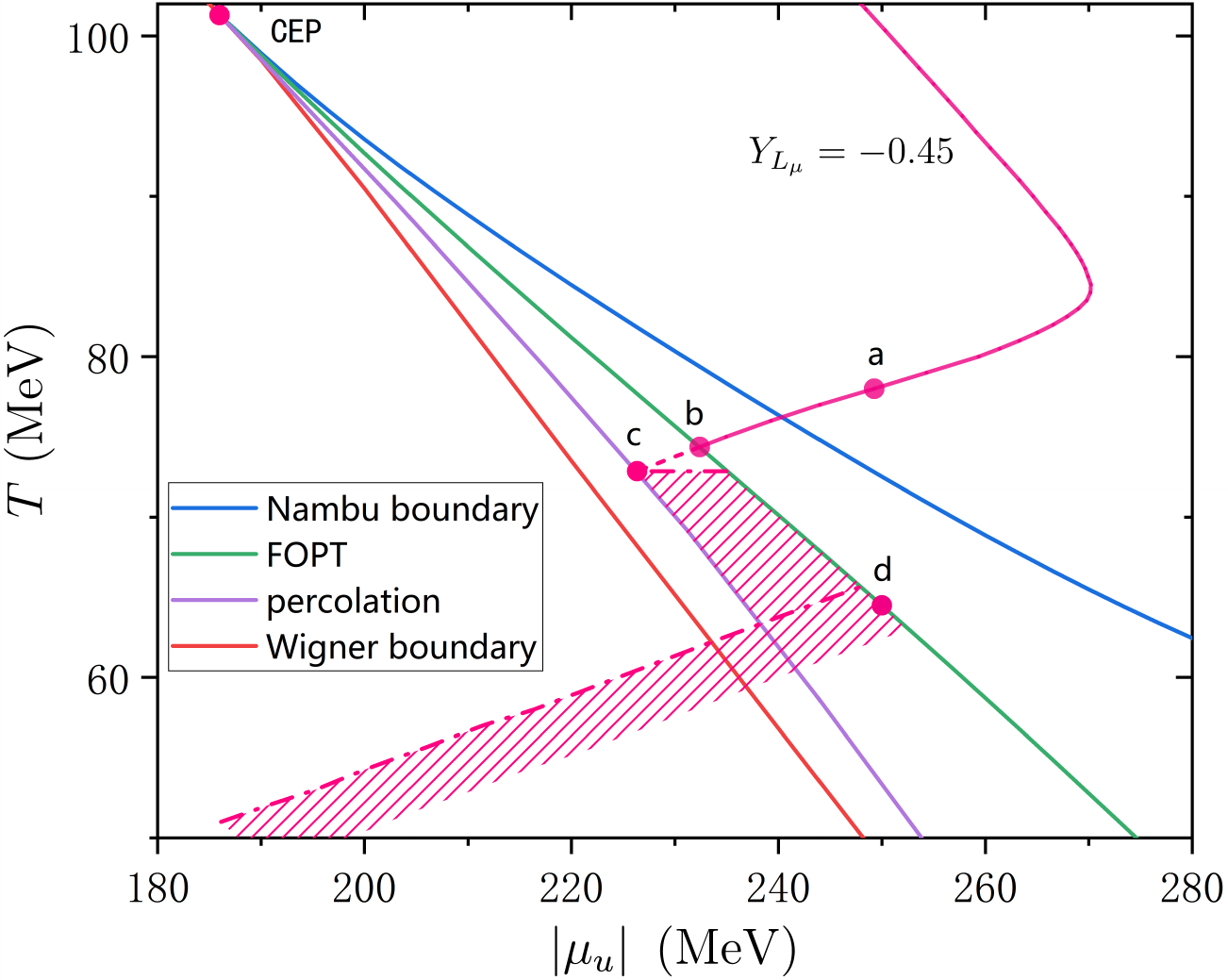}
        \caption{{The cosmic trajectory across the FOPT region in the scenario $Y_{L_e}=0,\, Y_{L_\mu}=-Y_{L_\tau}$, where $Y_{L_\mu}=-0.45$. 
        %
        %
        Solid line: the determined cosmic trajectory, such as $a$-$b$; dashed line: the supercooling stage for a cosmic trajectory before the percolation and after the FOPT, such as $b$-$c$; dot-dashed line: the upper bound for the possible cosmic trajectory; shadow region: an area contains possible cosmic trajectory.}}
        \label{fig:costraj_peda}
    \end{figure}


   {When the cosmic trajectory crosses the   FOPT, there can be GW emission during the bubble nucleation.  
   Before nucleation, the QCD effective potential, explained in the next section, can be used to study the process.  After nucleation,  when {part} of the Universe is converted to the true vacuum, the entropy is no longer conserved due to the non-equilibrium effect. The upper bound of the cosmic trajectory is determined under the condition that $s/n_q$ is conserved after the PT as shown in Fig.~\ref{fig:costraj_all}. }
   {The   evolution   after the percolation stage is influenced by the dynamics of the bubble expansion, the increase in total entropy in the system due to non-equilibrium processes, and their interplay. The  bubble expansion modifies the volume fraction of the true vacuum within the Hubble volume after percolation, which in turn affects extensive thermodynamic quantities. Meanwhile, the increase in total entropy, driven by non-equilibrium processes, causes the  evolution to deviate from equilibrium. Both   have  significant impacts on conservation laws, making the determination of the cosmic trajectory challenging, with only the  upper limit of the  stage after percolation  being determined. Nevertheless, our main focus is the GW signals generated during  the percolation stage,  which can be fully determined within the information before the percolation stage. 
    }

 {
    We illustrate the  detailed evolution in Fig.~\ref{fig:costraj_peda}  with $Y_{L_\mu} = -0.45$ along the path.
    %
    During the $a$-$b$ period, the symmetric vacuum energy  is lower. At this stage, the universe is filled with the symmetric vacuum, i.e., the Wigner vacuum in QCD.
    In the subsequent $b$-$c$ period, the energy of the Wigner vacuum exceeds that of  Nambu vacuum, yet both are in local minima. This is then a supercooling period with  the universe remaining in    Wigner vacuum.
    At the point $c$, the Wigner vacuum initiates the decay into the true vacuum, leading to the nucleation and the percolation of bubbles.
    %
    %
    {The \(c\)-\(d\) process remains unclear due to the absence of calculations for non-equilibrium effects. However, cosmic trajectories involving these processes are constrained within the shaded region. The dot-dashed line above point \(d\) marks the upper bound of the cosmic trajectory determined by the true vacuum under the conservation of $s/n$, while \(s/n\) generally increases due to the non-equilibrium effects. Near point \(d\), the Gibbs phase equilibrium condition requires the cosmic trajectory aligns with the FOPT line. 
    At point \(d\), the Universe is filled with the Nambu vacuum, marking the end of the phase transition.}
    }
    
    {\bf PT parameters.}
    %
    {The QCD FOPT mechanism is expected to generate gravitational waves (GWs). To account for the causality tail in GW signals~\cite{Franciolini:2023wjm}, we apply the GW spectrum modeled by NANOGrav.}
    The GW spectrum is written as~\cite{NANOGrav:2023hvm,Jinno:2016vai,Hindmarsh:2015qta,Espinosa:2010hh,Ellis:2020awk,Ellis:2019oqb,Guo:2020grp,Weir:2017wfa}:
    \begin{align}
        {{\Omega }_{b}}\left( f \right)=\mathcal{D}\,{{\tilde{\Omega }}_{b}}{{\left( \frac{{{\kappa }_{b}}{{\alpha }_{*}}}{1+{{\alpha }_{*}}} \right)}^{2}}{{\left( {{H}_{*}}{{R}_{*}} \right)}^{2}}S\left( f/{{f}_{b}}\  \right),\\
        {{\Omega }_{s}}\left( f \right)=\mathcal{D}\,{{\widetilde{\Omega }}_{s}}\Upsilon \left( {{\tau }_{sw}} \right){{\left( \frac{{{\kappa }_{s}}\alpha_* }{1+\alpha_* } \right)}^{2}}\left( {{H}_{*}}{{R}_{*}} \right)\,\mathcal{S}\left( {f}/{{{f}_{s}}}\; \right), 
    \end{align}
    with $\alpha_*$ the PT strength, $\beta/H_*=(8\pi)^{1/3}v_w/(H_*R_*)$ being the inverse PT duration~\cite{Espinosa:2010hh} and $v_w$ the bubble wall velocity. 
    %
    %
    Both  parameters requires the knowledge of the QCD effective potential, which we obtain  by applying the homotopy method to the QCD theory~\cite{Zheng:2023tbv}. 
    The PT parameters are calculated on the percolation line, which corresponds to the percolation temperature $T_*$, {i.e.}, the temperature when $\sim 34\%$ of the volume of the universe is converted into the true vacuum determined  with the effective potential.
    %


   The QCD effective potential comes from a generalized Legendre transformation to the Cornwall, Jackiw and Tomboulis (CJT)  generating functional~\cite{Zheng:2023tbv}. This transformation converts the variable of the CJT effective potential, {i.e.}, the quark propagator, to the self-energy, which is the key point to keep the effective potential to be bounded from below~\cite{Haymaker:1986tt}. The new effective potential is written as~\cite{Zheng:2023tbv}:
    \begin{align}
	{{\Gamma }_{\Sigma}}\left[ \Sigma  \right]=&~\text{TrLn}{{\left[ S_{0}^{-1}+\Sigma  \right]}^{-1}}-\text{Tr}\left[ 1 \right]\notag\\&-             {{\Gamma }_{2}}\left[ {{\left( S+L \right)}_{ }} \right]+\text{Tr}\left[ {{\left( S+L \right)}_{ }}\Sigma  \right],\label{eq:AF_sigma}
    \end{align}
    with $S_0$ the free quark propagator, $S$ the dressed quark propagator, $\Sigma=\Sigma \left[(S+L) \right]\equiv\frac{\delta {{\Gamma }_{2}}\left[ \left( S+L \right) \right]}{\delta \left( S+L \right)}$ the self-energy, $L$ the external source, and $\Gamma_2$ the sum of all two-particle-irreducible diagrams as a functional of $S+L$.
    


    Moreover, we apply the homotopy method, {i.e.}, a simplified version of the functional integration, 
    to determine the effective potential from the quantitative truncation scheme of the DSE~\cite{Zheng:2023tbv}:
    \begin{align}    \label{AF_first_deri_1}
            \Gamma_\Sigma[M_q]& :=\int_{0}^{\chi} \,d\xi \, \text{Tr}\left\{ \frac{d\Sigma_{\xi}}{d\xi } \left[ -{{\left( S_{0}^{-1} + \Sigma_{\xi}\right)}^{-1}} + {{S}_{\xi }} \right] \right\}\, ,
    \end{align}
    where  $M_q$ the dynamic quark mass that depends on  the homotopy parameter and
    
    \begin{align}\label{eq:homoS}
        \Sigma_{\xi} = \Sigma[S_{\xi}], \quad S_{\xi}=\left[(1-\xi) S_1^{-1} + \xi S_2^{-1}\right]^{-1},
    \end{align}
  {with $\chi$ and $\xi$ being the homotopy parameter, $S_1$ and $S_2$ being the Wigner and Nambu$+$ solutions, and \(\Sigma\) the self energy  whose explicit expression is placed in the supplement.
    In the functional space, a one-dimensional line parameterized by \(\chi\) and \(\xi\), on which the effective potential lies, is chosen.}
 
    \begin{figure}[t]
        \centering
        \includegraphics[width=1\linewidth]{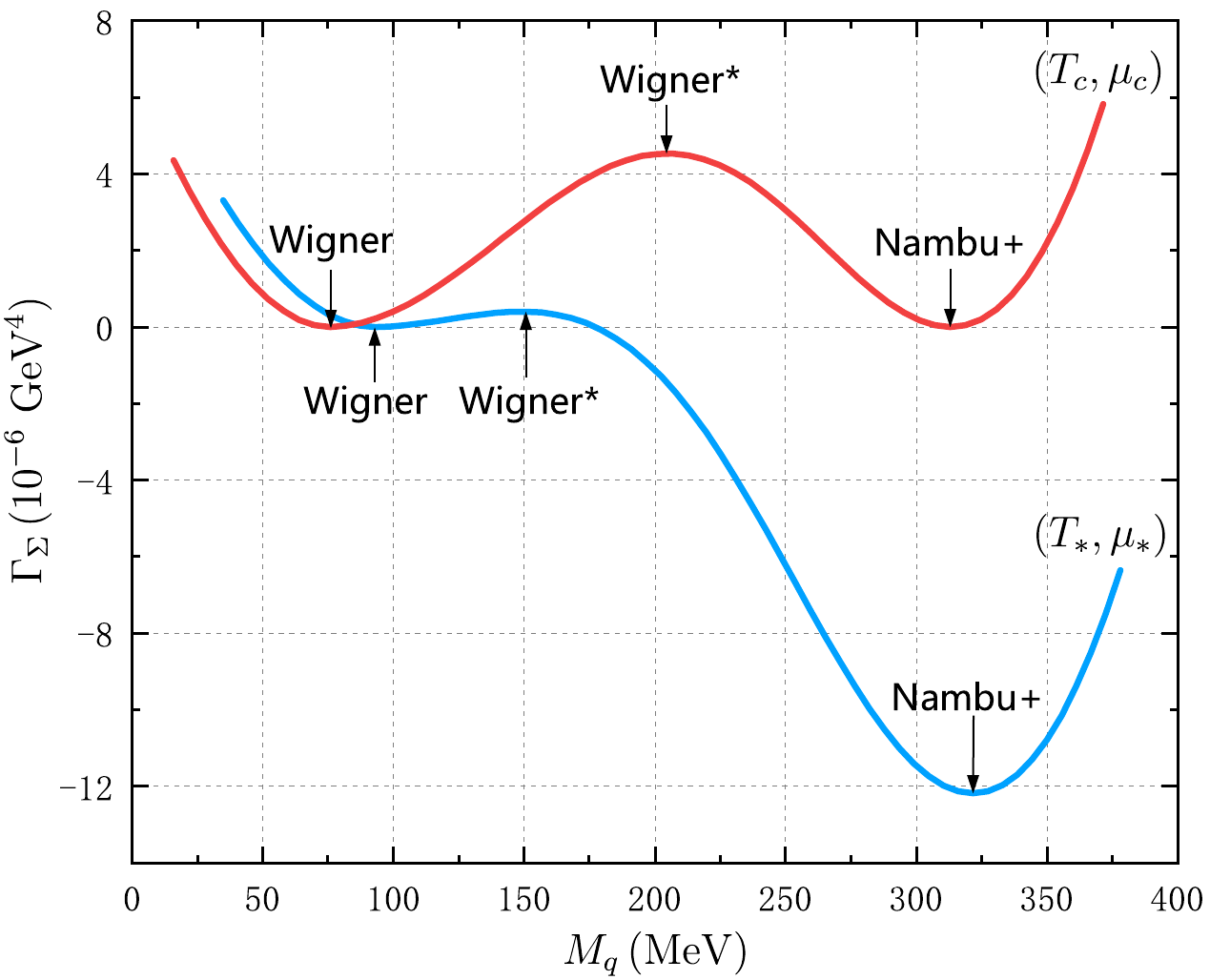}
        \caption{{The $N_f=2$ effective potential evolving from $(T_c, \mu_c)$ to $(T_*, \mu_*)$ for the $Y_{L_\mu}=-0.45$ cosmic trajectory.
        The dynamic quark mass $M_q$ is the order parameter of the effective potential and the effective potential is defined to be zero at the Wigner solution.}
        }
        \label{fig:eff_temp}
    \end{figure}
       We  illustrate the  effective potential $\Gamma_\Sigma$ in Fig.~\ref{fig:eff_temp}, where the Wigner vacuum is the false vacuum, the Nambu$+$ is the true vacuum and the Wigner$*$ is the unstable vacuum indicating the occurrence of FOPT. At the PT temperature $T_c$, the two vacuums are degenerate. Vacuum bubbles merge around the percolation temperature $T_*$. At that time, the false vacuum has an energy larger than the other. {Since the homotopy method guarantees the functional relation between the effective potential and the equation of motion of QCD,  the effective potential incorporates all the information  from the non-perturbative DSE, which naturally includes the temperature effects and all the temperature-induced infrared effects.}

    {Using the effective potential, we calculate the decay rate of the false vacuum using the python package~\cite{Ekstedt:2023sqc}:
        \begin{align}
            \Gamma (T,\mu )=&\frac{\sqrt{\left| {{\lambda }_{-}} \right|}}{2\pi }{{\left| \frac{\operatorname{det}'(-{{\nabla }^{2}}+{{{{V}''}}_{\rm eff}}({{\phi }_{b}}))}{\det (-{{\nabla }^{2}}+{{{{V}''}}_{\rm eff}}({{\phi }_{F}}))} \right|}^{-1/2}}\notag\\&\times{{\left( \frac{{{S}_{3}}}{2\pi T} \right)}^{\frac{3}{2}}}\exp \left( -\frac{{{S}_{3}}}{T} \right),
        \end{align}
        with
        \begin{equation}
            {{S}_{3}}=4\pi \int_{0}^{\infty }{dr{{r}^{2}}\left[ \frac{1}{2}{{\left( \frac{d\phi_b }{dr} \right)}^{2}}+{{V}_{\rm eff}}\left( \phi_b  \right) \right]}.
        \end{equation}
   Here, $\phi_b$ is the bounce solution, $\phi_F$ is false vacuum solution, $\lambda_{-}$ is the negative eigenvalue of the functional determinant, $S_3$ is the 3-dimensional Euclidean actions for the $O(3)$-symmetric bounce solutions, $r$ is the bubble radius, $\phi=M_q$ is the dynamical quark mass, ${{V}''_{\rm eff}}={{d}^{2}}{{V}_{\rm eff}}/d{{\phi }^{2}}$, and $V_{\rm eff}=\Gamma_\Sigma$ is the effective potential calculated from the DSE with the homotopy method \cite{Zheng:2023tbv}. 
   It is important to note that QCD lacks a tree-level potential that describes the dynamical quark mass $M_q$. Therefore, we must construct an effective theory with $M_q$ as the degrees of freedom and $V_{eff}$ as the tree-level potential. Furthcrmore, based on this effective theory, we can also incorporate one-loop corrections to $V_{eff}$.
   {The evolution of the decay rate  and the percolation  temperature with the evolution details    in the supplemental material.}
   
   {\color{brown} 
   
   }


    {We calculate the bubble wall velocity and find it to be lower than both the Jouguet velocity and the speed of sound in the background radiation, which indicates the subsonic deflagration mode, with details putting in the Supplemental Material. The GW spectrum parameter \(\kappa_s\) is determined as in Ref.~\cite{Espinosa:2010hh}. 
    Additionally, the local thermal equilibrium (LTE) effects could influence the bubble wall velocity~\cite{Ai:2023see,Krajewski:2024gma}. They are not included here because \(\alpha_*\) is small in our study, which makes the bubble wall velocity highly sensitive to the enthalpy ratio between the broken and symmetric phases (\(\Psi_n\)), and introduces significant uncertainty in determining \(v_w\). Since \(v_w\) does not qualitatively affect our conclusions on the GW spectrum, LTE corrections are omitted.}
      
   
    }

    \begin{figure}[htbp]
        \centering
        \subfloat{\includegraphics[width=0.45\textwidth]{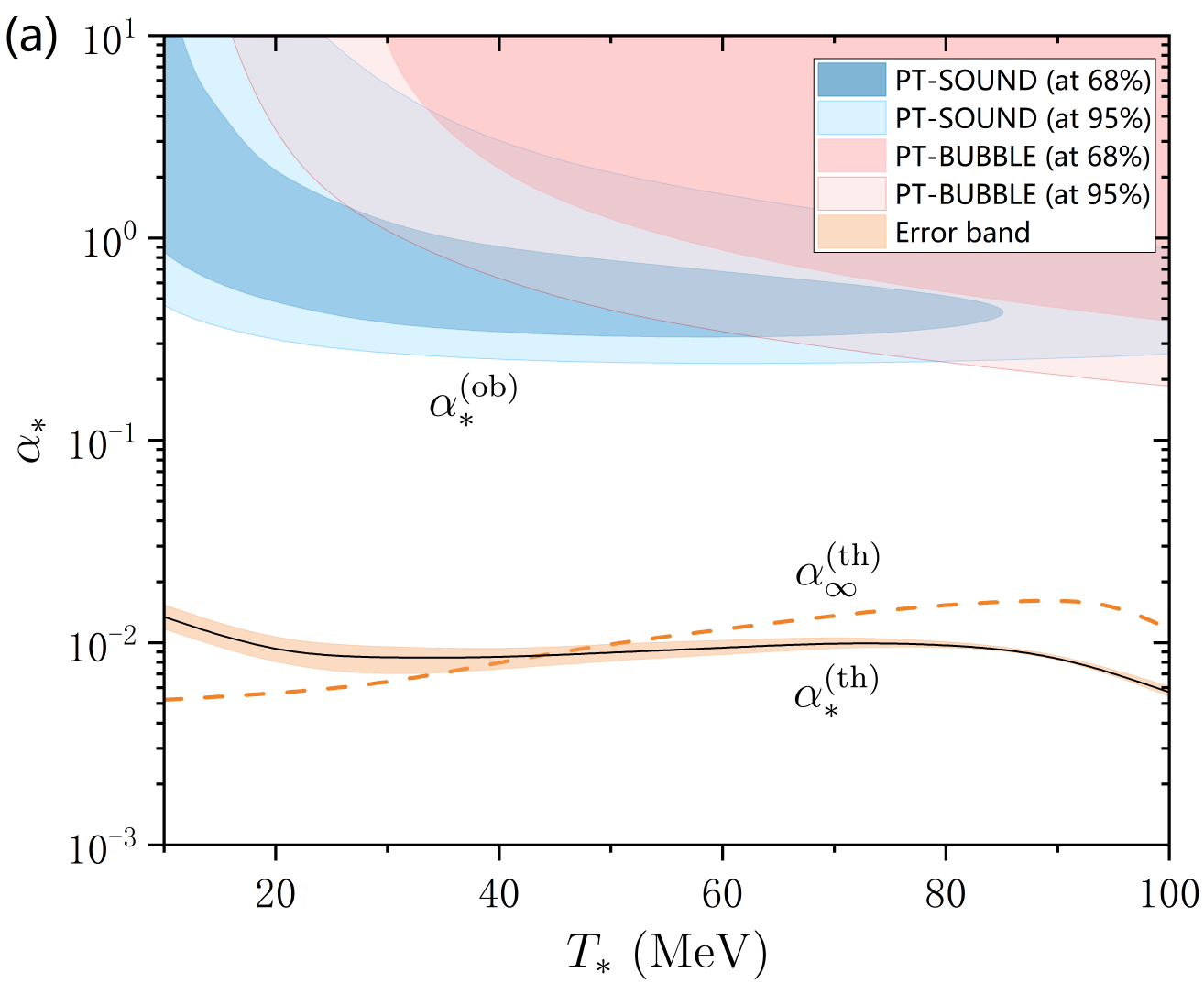}\label{fig:alpha}}
        \hfill
        \subfloat{\includegraphics[width=0.45\textwidth]{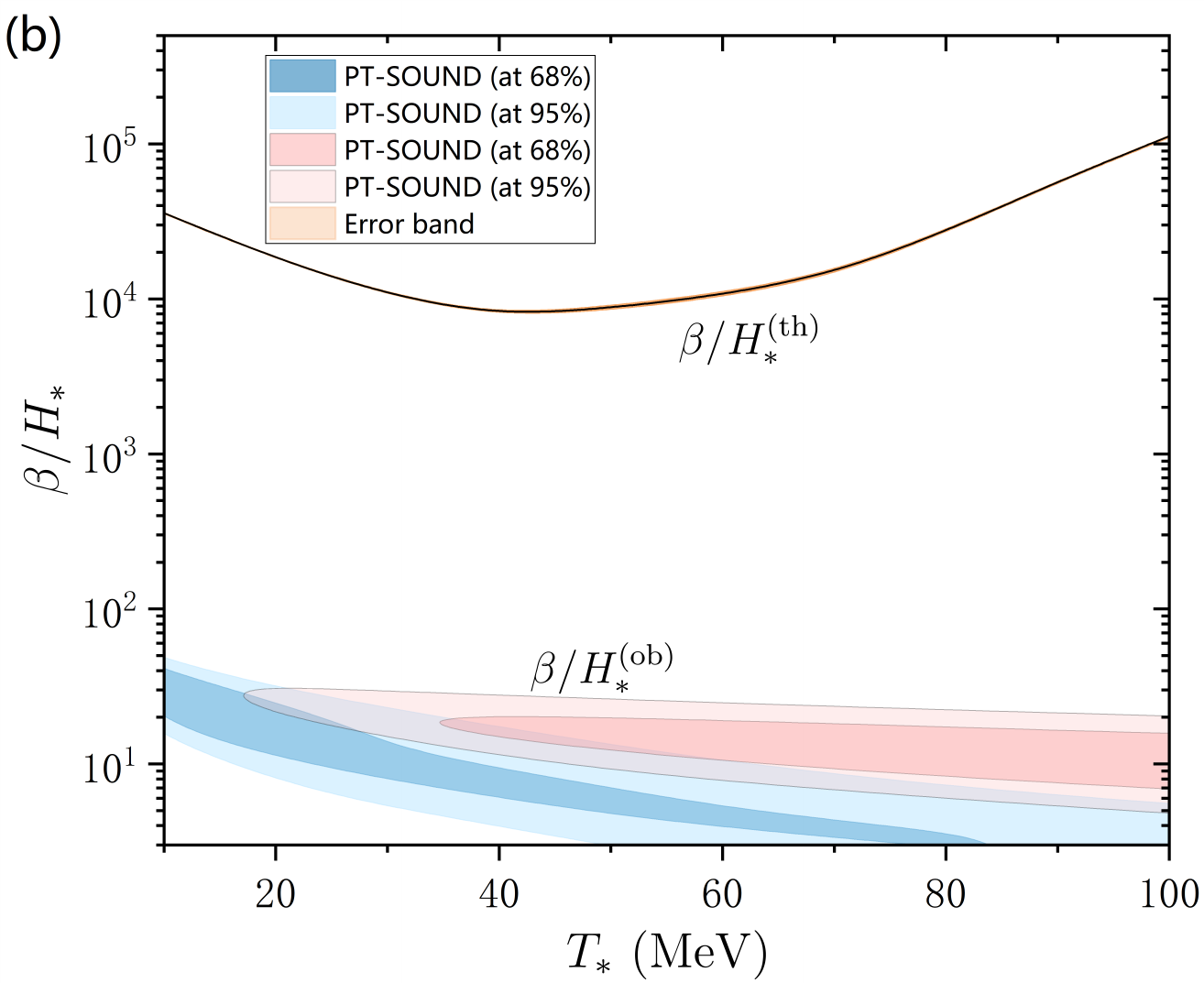}\label{fig:beta}}
        
        \caption{(a) The PT strength $\alpha_*$. (b) The inverse PT duration $\beta/H_*$. The temperature dependence of $\alpha_*$ and $\beta/H_*$ is on the percolation line. The shaded region: the NANOGrav constraints for the PT-SOUND and the PT-BUBBLE model~\cite{NANOGrav:2023hvm}. {  The superscripts of \(\alpha_*\) and \(\beta/H_*\) are as follows: "th" denotes theory, and "ob" denotes observation.}
        }
        \label{fig:alpha_beta}
    \end{figure}

    The PT strength $\alpha_*$ depends on the available energy released from the PT, normalized to the energy densities of the background radiation:
        \begin{equation}
            \alpha_* ={{\left. \frac{\Delta \theta }{{{\rho }_{r}}} \right|}_{(T_*,\mu_*)}},
        \end{equation}
        with
        \begin{align}
            \Delta \theta =\frac{1}{4}T\Delta s&+\frac{1}{4}\mu_q\Delta {{n}_{q}}-\frac{1+1/c_{s,b}^2}{4}\Delta P,\quad
            c_{s,b}^2 = {{\left. \frac{dp}{de} \right|}_{s,b}},\notag\\
            {{\rho }_{r}}(T,\mu)&=\sum\limits_{a}{\int{\frac{{{d}^{3}}p}{{{(2\pi )}^{3}}}E_a{{f}_{F/B}}(E_a,T,\mu )}}\;.
        \end{align}
    Here, $E_a=\sqrt{p^2+m_a^2}$, {$\theta$ is the pseudotrace anomaly~\cite{Giese:2020rtr}}, $s$ is the entropy density, $n_q$ is the quark number density and  $\rho_r$ collects electron, muon, all neutrinos, photon and $u/d$ quarks.
    The $c_{s,b}^2$ is the speed of sound for the Nambu phase, where the derivative is along the {Isentropic trajectory, $p_b$ is the respective pressure and $e_b$ is the energy density.
    A critical value $\alpha_\infty$ that can determine whether the bubble walls can runaway is also considered~\cite{Ellis:2019oqb,Bodeker:2009qy}, which also determines the parameter $\kappa_b$.


     For the FOPT with finite chemical potential, the inverse PT duration $\beta/H_*$ can be obtained as:
    \begin{equation}\label{eq:beta/H}
        \beta /{{H}_{*}}=4\rho_r \frac{d}{d\rho_r }{{\left. \left( \frac{{{S}_{3}}}{T} \right) \right|}_{{(T_*,\mu_*)}}},
    \end{equation}
    where the derivative lies along the cosmic trajectory. The details of above equations can be found in the supplementary material.
 {In summary, the QCD effective potential completely determines the  parameters and the pattern of the GW spectrum as depicted in the next section.}

    \begin{figure}[t]
        \centering
        \includegraphics[width=1\linewidth]{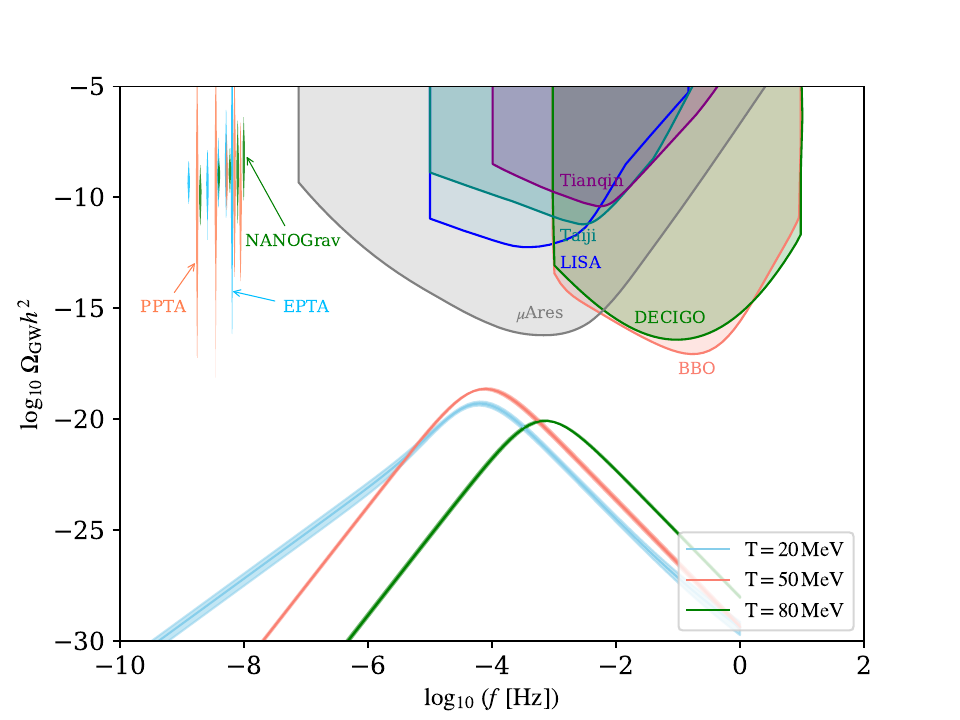}
        \caption{The GW spectra of the QCD FOPT, {where $T=[20,\,50,\,80]\,{\rm MeV}$ correspond to points on the percolation line in the phase diagram.} The violins plot the first five frequency bins of the datasets, including: EPTA~\cite{EPTA:2023fyk}, PPTA~\cite{Reardon:2023gzh,Reardon:2023zen,Zic:2023gta}, and NANOGrav~\cite{NANOGrav:2023gor,NANOGrav:2023hvm,NANOGrav:2023hde}. The sensitivity curves includes: $\mu$Ares~\cite{Sesana:2019vho}, Taiji~\cite{Hu:2017mde,Ruan:2018tsw}, Tianqin~\cite{TianQin:2015yph,Zhou:2023rop}, LISA~\cite{LISA:2017pwj,Baker:2019nia,Boileau:2021gbr,Schmitz:2020rag}, BBO~\cite{Crowder:2005nr,Harry:2006fi} and DECIGO~\cite{Seto:2001qf,Yagi:2011wg,Isoyama:2018rjb}. }
        \label{fig:GW_spectrum}
    \end{figure}

 { \bf Numerical results.} 
    After a complete computation for the GW spectrum, we find that neither  $\alpha_*$ and $\beta/H_* $ over the whole chiral PT can be  matched  with the NANOGrav measurements as shown in  Fig.~\ref{fig:alpha_beta}. {The error bands {in Fig.~\ref{fig:alpha_beta} and Fig.~\ref{fig:GW_spectrum}, coming from the computation of QCD sector with varying the chiral condensate by 10$\%$, are negligible compared to the uncertainties in the gravitational wave models.}
    After a complete computation for the GW spectrum, we find that neither  $\alpha_*$ and $\beta/H_* $ over the whole chiral PT can be  matched  with the NANOGrav measurements as shown in  Fig.~\ref{fig:alpha_beta}. {The error bands {in  Fig.~\ref{fig:alpha_beta} and Fig.~\ref{fig:GW_spectrum}, coming from the computation of QCD sector with varying the chiral condensate by 10$\%$, are negligible compared to the uncertainties in the gravitational wave models.}
    %
    
    In detail,  for $\alpha_*$, its value is at least one order of magnitude smaller comprared to the NANOGrav constraints. Such a small value is  mainly due to the large energy density of the background radiation. The large energy density comes from the large  chemical potentials of the leptons, which is a direct result of the conservation conditions in Eq.~(\ref{eq:five_conse_eq}). In particular, the chemical potentials of the electron and electron neutrino are roughly same as  the quark's, { while those of muon neutrinos and tau neutrinos are more than three times than the quark's.} For $\beta/H_*$, the value we obtained is three orders of magnitude larger than the one to match with the analysis from NANOGrav collaboration. This is mainly because the cosmic trajectory  leans more towards the chemical potential, causing the derivative  of $S_3/T$ along the trajectory to be large. Meanwhile, in the Eq.~(\ref{eq:beta/H}), $\rho_r$ outside of the derivative also maintains a large value due to the contribution of the non-vanishing chemical potential along the percolation line.

    
   %
   Based on the obtained PT parameters $\alpha_*$ and $\beta/H_*$, we show our result of the GWs sourced from the QCD FOPT in Fig.~\ref{fig:GW_spectrum}.
   To be specific, for $T_*>42.9\,{\rm MeV}$, where $\alpha<\alpha_\infty$, we only consider the PT-SOUND model from the NANOGrav. for $T_*<42.9\,{\rm MeV}$, where $\alpha>\alpha_\infty$, both the bubble collision and the sound wave effects are taken into account for the GW spectrum.
   The quantitative computation here disapproves the QCD first-order chiral phase transition as a source of the SGWB to explain the signals detected by NANOGrav.


    {\bf Conclusion.}
    In this \textit{Letter}, we calculate the cosmic trajectories at large quark chemical potential, and for the first time quantitatively determine the PT parameters of the QCD FOPT, through their relation to the thermodynamic quantities and effective potential from the DSE.
    
    %
    We firstly verify that a large lepton asymmetry can induce the  QCD FOPT. We consider the scenario ($Y_{L_e}=0, Y_{L_\mu}=-Y_{L_\tau}$), and find that   the FOPT occurs when $Y_{L_\mu}<-0.22$. This conclusion is essentially from the conservation laws during the Universe evolution without further assumptions. 
    {We further compute the PT parameters via the QCD effective potential, and then  determine the GW spectrum through the sound shell model.}    

 The obtained PT strength $\alpha_*$ is $\mathcal{O}(10^{-2})$, the inverse PT duration $\beta/H_*$ is $\mathcal{O}(10^{4})$.
    The values deviate  significantly from the measurements of NANOGrav mainly due to the large energy density of the background radiation. This large energy density arises from the large chemical potentials of the leptons, which is a direct result of the conservation laws.
    { We compare the GW spectra computed from PT parameters with violin plots and GW detector sensitivity curves. Our analysis shows that the SGWB signal from the QCD chiral FOPT is too weak for detection by NANOGrav or current and upcoming GW detectors.}
    


     { Additionally, it may be worth exploring the scenario in Ref.~\cite{Bodeker:2021mcj,Sagunski:2023ynd,Giese:2020znk,Bodeker:2021mcj,vonHarling:2017yew,Iso:2017uuu},  
     which provides an exit of electroweak (EW) supercooling with QCD phase transition. This involves a beyond-Standard-Model (BSM) scalar field that acts as the coefficient of the Higgs field's quadratic term. The consequences are twofold: first, EW symmetry remains unbroken until the QCD FOPT occurs, resulting in a QCD phase transition with six massless quarks and a critical endpoint (CEP) at lower chemical potential; Second, the  breaking of the Higgs and BSM fields releases significant vacuum energy, potentially exceeding the energy density of the thermal bath, which could produce detectable gravitational waves.}

		\subparagraph{Acknowledgments}
	F.G. and H.Z. thank Isabel M. Oldengott and Yi Lu for helpful discussions. 	
 This work is supported by the National  Science Foundation of China under Grants  No.12175007,  No.12247107 and No.12305134. L.B.
is supported by the National Natural Science Foundation of
China (NSFC) under Grants No. 12075041, No. 12322505,
and No. 12347101. L.B. also acknowledges
Chongqing Talents: Exceptional Young Talents Project No.
cstc2024ycjh-bgzxm0020.

    \bibliography{GW}

\begin{thebibliography}{122}%
\makeatletter
\providecommand \@ifxundefined [1]{%
 \@ifx{#1\undefined}
}%
\providecommand \@ifnum [1]{%
 \ifnum #1\expandafter \@firstoftwo
 \else \expandafter \@secondoftwo
 \fi
}%
\providecommand \@ifx [1]{%
 \ifx #1\expandafter \@firstoftwo
 \else \expandafter \@secondoftwo
 \fi
}%
\providecommand \natexlab [1]{#1}%
\providecommand \enquote  [1]{``#1''}%
\providecommand \bibnamefont  [1]{#1}%
\providecommand \bibfnamefont [1]{#1}%
\providecommand \citenamefont [1]{#1}%
\providecommand \href@noop [0]{\@secondoftwo}%
\providecommand \href [0]{\begingroup \@sanitize@url \@href}%
\providecommand \@href[1]{\@@startlink{#1}\@@href}%
\providecommand \@@href[1]{\endgroup#1\@@endlink}%
\providecommand \@sanitize@url [0]{\catcode `\\12\catcode `\$12\catcode `\&12\catcode `\#12\catcode `\^12\catcode `\_12\catcode `\%12\relax}%
\providecommand \@@startlink[1]{}%
\providecommand \@@endlink[0]{}%
\providecommand \url  [0]{\begingroup\@sanitize@url \@url }%
\providecommand \@url [1]{\endgroup\@href {#1}{\urlprefix }}%
\providecommand \urlprefix  [0]{URL }%
\providecommand \Eprint [0]{\href }%
\providecommand \doibase [0]{http://dx.doi.org/}%
\providecommand \selectlanguage [0]{\@gobble}%
\providecommand \bibinfo  [0]{\@secondoftwo}%
\providecommand \bibfield  [0]{\@secondoftwo}%
\providecommand \translation [1]{[#1]}%
\providecommand \BibitemOpen [0]{}%
\providecommand \bibitemStop [0]{}%
\providecommand \bibitemNoStop [0]{.\EOS\space}%
\providecommand \EOS [0]{\spacefactor3000\relax}%
\providecommand \BibitemShut  [1]{\csname bibitem#1\endcsname}%
\let\auto@bib@innerbib\@empty
\bibitem [{\citenamefont {Sesana}\ \emph {et~al.}(2021)\citenamefont {Sesana} \emph {et~al.}}]{Sesana:2019vho}%
  \BibitemOpen
  \bibfield  {author} {\bibinfo {author} {\bibfnamefont {A.}~\bibnamefont {Sesana}} \emph {et~al.},\ }\href {\doibase 10.1007/s10686-021-09709-9} {\bibfield  {journal} {\bibinfo  {journal} {Exper. Astron.}\ }\textbf {\bibinfo {volume} {51}},\ \bibinfo {pages} {1333} (\bibinfo {year} {2021})},\ \Eprint {http://arxiv.org/abs/1908.11391} {arXiv:1908.11391 [astro-ph.IM]} \BibitemShut {NoStop}%
\bibitem [{\citenamefont {Hu}\ and\ \citenamefont {Wu}(2017)}]{Hu:2017mde}%
  \BibitemOpen
  \bibfield  {author} {\bibinfo {author} {\bibfnamefont {W.-R.}\ \bibnamefont {Hu}}\ and\ \bibinfo {author} {\bibfnamefont {Y.-L.}\ \bibnamefont {Wu}},\ }\href {\doibase 10.1093/nsr/nwx116} {\bibfield  {journal} {\bibinfo  {journal} {Natl. Sci. Rev.}\ }\textbf {\bibinfo {volume} {4}},\ \bibinfo {pages} {685} (\bibinfo {year} {2017})}\BibitemShut {NoStop}%
\bibitem [{\citenamefont {Ruan}\ \emph {et~al.}(2020)\citenamefont {Ruan}, \citenamefont {Guo}, \citenamefont {Cai},\ and\ \citenamefont {Zhang}}]{Ruan:2018tsw}%
  \BibitemOpen
  \bibfield  {author} {\bibinfo {author} {\bibfnamefont {W.-H.}\ \bibnamefont {Ruan}}, \bibinfo {author} {\bibfnamefont {Z.-K.}\ \bibnamefont {Guo}}, \bibinfo {author} {\bibfnamefont {R.-G.}\ \bibnamefont {Cai}}, \ and\ \bibinfo {author} {\bibfnamefont {Y.-Z.}\ \bibnamefont {Zhang}},\ }\href {\doibase 10.1142/S0217751X2050075X} {\bibfield  {journal} {\bibinfo  {journal} {Int. J. Mod. Phys. A}\ }\textbf {\bibinfo {volume} {35}},\ \bibinfo {pages} {2050075} (\bibinfo {year} {2020})},\ \Eprint {http://arxiv.org/abs/1807.09495} {arXiv:1807.09495 [gr-qc]} \BibitemShut {NoStop}%
\bibitem [{\citenamefont {Luo}\ \emph {et~al.}(2016)\citenamefont {Luo} \emph {et~al.}}]{TianQin:2015yph}%
  \BibitemOpen
  \bibfield  {author} {\bibinfo {author} {\bibfnamefont {J.}~\bibnamefont {Luo}} \emph {et~al.} (\bibinfo {collaboration} {TianQin}),\ }\href {\doibase 10.1088/0264-9381/33/3/035010} {\bibfield  {journal} {\bibinfo  {journal} {Class. Quant. Grav.}\ }\textbf {\bibinfo {volume} {33}},\ \bibinfo {pages} {035010} (\bibinfo {year} {2016})},\ \Eprint {http://arxiv.org/abs/1512.02076} {arXiv:1512.02076 [astro-ph.IM]} \BibitemShut {NoStop}%
\bibitem [{\citenamefont {Zhou}\ \emph {et~al.}(2023)\citenamefont {Zhou}, \citenamefont {Cheng},\ and\ \citenamefont {Ren}}]{Zhou:2023rop}%
  \BibitemOpen
  \bibfield  {author} {\bibinfo {author} {\bibfnamefont {K.}~\bibnamefont {Zhou}}, \bibinfo {author} {\bibfnamefont {J.}~\bibnamefont {Cheng}}, \ and\ \bibinfo {author} {\bibfnamefont {L.}~\bibnamefont {Ren}},\ }\href@noop {} {\  (\bibinfo {year} {2023})},\ \Eprint {http://arxiv.org/abs/2306.14439} {arXiv:2306.14439 [gr-qc]} \BibitemShut {NoStop}%
\bibitem [{\citenamefont {Amaro-Seoane}\ \emph {et~al.}(2017)\citenamefont {Amaro-Seoane} \emph {et~al.}}]{LISA:2017pwj}%
  \BibitemOpen
  \bibfield  {author} {\bibinfo {author} {\bibfnamefont {P.}~\bibnamefont {Amaro-Seoane}} \emph {et~al.} (\bibinfo {collaboration} {LISA}),\ }\href@noop {} {\  (\bibinfo {year} {2017})},\ \Eprint {http://arxiv.org/abs/1702.00786} {arXiv:1702.00786 [astro-ph.IM]} \BibitemShut {NoStop}%
\bibitem [{\citenamefont {Baker}\ \emph {et~al.}(2019)\citenamefont {Baker} \emph {et~al.}}]{Baker:2019nia}%
  \BibitemOpen
  \bibfield  {author} {\bibinfo {author} {\bibfnamefont {J.}~\bibnamefont {Baker}} \emph {et~al.},\ }\href@noop {} {\  (\bibinfo {year} {2019})},\ \Eprint {http://arxiv.org/abs/1907.06482} {arXiv:1907.06482 [astro-ph.IM]} \BibitemShut {NoStop}%
\bibitem [{\citenamefont {Boileau}\ \emph {et~al.}(2022)\citenamefont {Boileau}, \citenamefont {Jenkins}, \citenamefont {Sakellariadou}, \citenamefont {Meyer},\ and\ \citenamefont {Christensen}}]{Boileau:2021gbr}%
  \BibitemOpen
  \bibfield  {author} {\bibinfo {author} {\bibfnamefont {G.}~\bibnamefont {Boileau}}, \bibinfo {author} {\bibfnamefont {A.~C.}\ \bibnamefont {Jenkins}}, \bibinfo {author} {\bibfnamefont {M.}~\bibnamefont {Sakellariadou}}, \bibinfo {author} {\bibfnamefont {R.}~\bibnamefont {Meyer}}, \ and\ \bibinfo {author} {\bibfnamefont {N.}~\bibnamefont {Christensen}},\ }\href {\doibase 10.1103/PhysRevD.105.023510} {\bibfield  {journal} {\bibinfo  {journal} {Phys. Rev. D}\ }\textbf {\bibinfo {volume} {105}},\ \bibinfo {pages} {023510} (\bibinfo {year} {2022})},\ \Eprint {http://arxiv.org/abs/2109.06552} {arXiv:2109.06552 [gr-qc]} \BibitemShut {NoStop}%
\bibitem [{\citenamefont {Schmitz}(2020)}]{Schmitz:2020rag}%
  \BibitemOpen
  \bibfield  {author} {\bibinfo {author} {\bibfnamefont {K.}~\bibnamefont {Schmitz}},\ }\href {\doibase 10.3390/sym12091477} {\bibfield  {journal} {\bibinfo  {journal} {Symmetry}\ }\textbf {\bibinfo {volume} {12}},\ \bibinfo {pages} {1477} (\bibinfo {year} {2020})},\ \Eprint {http://arxiv.org/abs/2005.10789} {arXiv:2005.10789 [hep-ph]} \BibitemShut {NoStop}%
\bibitem [{\citenamefont {Crowder}\ and\ \citenamefont {Cornish}(2005)}]{Crowder:2005nr}%
  \BibitemOpen
  \bibfield  {author} {\bibinfo {author} {\bibfnamefont {J.}~\bibnamefont {Crowder}}\ and\ \bibinfo {author} {\bibfnamefont {N.~J.}\ \bibnamefont {Cornish}},\ }\href {\doibase 10.1103/PhysRevD.72.083005} {\bibfield  {journal} {\bibinfo  {journal} {Phys. Rev. D}\ }\textbf {\bibinfo {volume} {72}},\ \bibinfo {pages} {083005} (\bibinfo {year} {2005})},\ \Eprint {http://arxiv.org/abs/gr-qc/0506015} {arXiv:gr-qc/0506015} \BibitemShut {NoStop}%
\bibitem [{\citenamefont {Harry}\ \emph {et~al.}(2006)\citenamefont {Harry}, \citenamefont {Fritschel}, \citenamefont {Shaddock}, \citenamefont {Folkner},\ and\ \citenamefont {Phinney}}]{Harry:2006fi}%
  \BibitemOpen
  \bibfield  {author} {\bibinfo {author} {\bibfnamefont {G.~M.}\ \bibnamefont {Harry}}, \bibinfo {author} {\bibfnamefont {P.}~\bibnamefont {Fritschel}}, \bibinfo {author} {\bibfnamefont {D.~A.}\ \bibnamefont {Shaddock}}, \bibinfo {author} {\bibfnamefont {W.}~\bibnamefont {Folkner}}, \ and\ \bibinfo {author} {\bibfnamefont {E.~S.}\ \bibnamefont {Phinney}},\ }\href {\doibase 10.1088/0264-9381/23/15/008} {\bibfield  {journal} {\bibinfo  {journal} {Class. Quant. Grav.}\ }\textbf {\bibinfo {volume} {23}},\ \bibinfo {pages} {4887} (\bibinfo {year} {2006})},\ \bibinfo {note} {[Erratum: Class.Quant.Grav. 23, 7361 (2006)]}\BibitemShut {NoStop}%
\bibitem [{\citenamefont {Seto}\ \emph {et~al.}(2001)\citenamefont {Seto}, \citenamefont {Kawamura},\ and\ \citenamefont {Nakamura}}]{Seto:2001qf}%
  \BibitemOpen
  \bibfield  {author} {\bibinfo {author} {\bibfnamefont {N.}~\bibnamefont {Seto}}, \bibinfo {author} {\bibfnamefont {S.}~\bibnamefont {Kawamura}}, \ and\ \bibinfo {author} {\bibfnamefont {T.}~\bibnamefont {Nakamura}},\ }\href {\doibase 10.1103/PhysRevLett.87.221103} {\bibfield  {journal} {\bibinfo  {journal} {Phys. Rev. Lett.}\ }\textbf {\bibinfo {volume} {87}},\ \bibinfo {pages} {221103} (\bibinfo {year} {2001})},\ \Eprint {http://arxiv.org/abs/astro-ph/0108011} {arXiv:astro-ph/0108011} \BibitemShut {NoStop}%
\bibitem [{\citenamefont {Yagi}\ and\ \citenamefont {Seto}(2011)}]{Yagi:2011wg}%
  \BibitemOpen
  \bibfield  {author} {\bibinfo {author} {\bibfnamefont {K.}~\bibnamefont {Yagi}}\ and\ \bibinfo {author} {\bibfnamefont {N.}~\bibnamefont {Seto}},\ }\href {\doibase 10.1103/PhysRevD.83.044011} {\bibfield  {journal} {\bibinfo  {journal} {Phys. Rev. D}\ }\textbf {\bibinfo {volume} {83}},\ \bibinfo {pages} {044011} (\bibinfo {year} {2011})},\ \bibinfo {note} {[Erratum: Phys.Rev.D 95, 109901 (2017)]},\ \Eprint {http://arxiv.org/abs/1101.3940} {arXiv:1101.3940 [astro-ph.CO]} \BibitemShut {NoStop}%
\bibitem [{\citenamefont {Isoyama}\ \emph {et~al.}(2018)\citenamefont {Isoyama}, \citenamefont {Nakano},\ and\ \citenamefont {Nakamura}}]{Isoyama:2018rjb}%
  \BibitemOpen
  \bibfield  {author} {\bibinfo {author} {\bibfnamefont {S.}~\bibnamefont {Isoyama}}, \bibinfo {author} {\bibfnamefont {H.}~\bibnamefont {Nakano}}, \ and\ \bibinfo {author} {\bibfnamefont {T.}~\bibnamefont {Nakamura}},\ }\href {\doibase 10.1093/ptep/pty078} {\bibfield  {journal} {\bibinfo  {journal} {PTEP}\ }\textbf {\bibinfo {volume} {2018}},\ \bibinfo {pages} {073E01} (\bibinfo {year} {2018})},\ \Eprint {http://arxiv.org/abs/1802.06977} {arXiv:1802.06977 [gr-qc]} \BibitemShut {NoStop}%
\bibitem [{\citenamefont {Arzoumanian}\ \emph {et~al.}(2020)\citenamefont {Arzoumanian} \emph {et~al.}}]{NANOGrav:2020bcs}%
  \BibitemOpen
  \bibfield  {author} {\bibinfo {author} {\bibfnamefont {Z.}~\bibnamefont {Arzoumanian}} \emph {et~al.} (\bibinfo {collaboration} {NANOGrav}),\ }\href {\doibase 10.3847/2041-8213/abd401} {\bibfield  {journal} {\bibinfo  {journal} {Astrophys. J. Lett.}\ }\textbf {\bibinfo {volume} {905}},\ \bibinfo {pages} {L34} (\bibinfo {year} {2020})},\ \Eprint {http://arxiv.org/abs/2009.04496} {arXiv:2009.04496 [astro-ph.HE]} \BibitemShut {NoStop}%
\bibitem [{\citenamefont {Agazie}\ \emph {et~al.}(2023{\natexlab{a}})\citenamefont {Agazie} \emph {et~al.}}]{NANOGrav:2023gor}%
  \BibitemOpen
  \bibfield  {author} {\bibinfo {author} {\bibfnamefont {G.}~\bibnamefont {Agazie}} \emph {et~al.} (\bibinfo {collaboration} {NANOGrav}),\ }\href {\doibase 10.3847/2041-8213/acdac6} {\bibfield  {journal} {\bibinfo  {journal} {Astrophys. J. Lett.}\ }\textbf {\bibinfo {volume} {951}},\ \bibinfo {pages} {L8} (\bibinfo {year} {2023}{\natexlab{a}})},\ \Eprint {http://arxiv.org/abs/2306.16213} {arXiv:2306.16213 [astro-ph.HE]} \BibitemShut {NoStop}%
\bibitem [{\citenamefont {Chen}\ \emph {et~al.}(2021)\citenamefont {Chen} \emph {et~al.}}]{EPTA:2021crs}%
  \BibitemOpen
  \bibfield  {author} {\bibinfo {author} {\bibfnamefont {S.}~\bibnamefont {Chen}} \emph {et~al.} (\bibinfo {collaboration} {EPTA}),\ }\href {\doibase 10.1093/mnras/stab2833} {\bibfield  {journal} {\bibinfo  {journal} {Mon. Not. Roy. Astron. Soc.}\ }\textbf {\bibinfo {volume} {508}},\ \bibinfo {pages} {4970} (\bibinfo {year} {2021})},\ \Eprint {http://arxiv.org/abs/2110.13184} {arXiv:2110.13184 [astro-ph.HE]} \BibitemShut {NoStop}%
\bibitem [{\citenamefont {Antoniadis}\ \emph {et~al.}(2023)\citenamefont {Antoniadis} \emph {et~al.}}]{EPTA:2023fyk}%
  \BibitemOpen
  \bibfield  {author} {\bibinfo {author} {\bibfnamefont {J.}~\bibnamefont {Antoniadis}} \emph {et~al.} (\bibinfo {collaboration} {EPTA, InPTA:}),\ }\href {\doibase 10.1051/0004-6361/202346844} {\bibfield  {journal} {\bibinfo  {journal} {Astron. Astrophys.}\ }\textbf {\bibinfo {volume} {678}},\ \bibinfo {pages} {A50} (\bibinfo {year} {2023})},\ \Eprint {http://arxiv.org/abs/2306.16214} {arXiv:2306.16214 [astro-ph.HE]} \BibitemShut {NoStop}%
\bibitem [{\citenamefont {Goncharov}\ \emph {et~al.}(2021)\citenamefont {Goncharov} \emph {et~al.}}]{Goncharov:2021oub}%
  \BibitemOpen
  \bibfield  {author} {\bibinfo {author} {\bibfnamefont {B.}~\bibnamefont {Goncharov}} \emph {et~al.},\ }\href {\doibase 10.3847/2041-8213/ac17f4} {\bibfield  {journal} {\bibinfo  {journal} {Astrophys. J. Lett.}\ }\textbf {\bibinfo {volume} {917}},\ \bibinfo {pages} {L19} (\bibinfo {year} {2021})},\ \Eprint {http://arxiv.org/abs/2107.12112} {arXiv:2107.12112 [astro-ph.HE]} \BibitemShut {NoStop}%
\bibitem [{\citenamefont {Reardon}\ \emph {et~al.}(2023{\natexlab{a}})\citenamefont {Reardon} \emph {et~al.}}]{Reardon:2023gzh}%
  \BibitemOpen
  \bibfield  {author} {\bibinfo {author} {\bibfnamefont {D.~J.}\ \bibnamefont {Reardon}} \emph {et~al.},\ }\href {\doibase 10.3847/2041-8213/acdd02} {\bibfield  {journal} {\bibinfo  {journal} {Astrophys. J. Lett.}\ }\textbf {\bibinfo {volume} {951}},\ \bibinfo {pages} {L6} (\bibinfo {year} {2023}{\natexlab{a}})},\ \Eprint {http://arxiv.org/abs/2306.16215} {arXiv:2306.16215 [astro-ph.HE]} \BibitemShut {NoStop}%
\bibitem [{\citenamefont {Antoniadis}\ \emph {et~al.}(2022)\citenamefont {Antoniadis} \emph {et~al.}}]{Antoniadis:2022pcn}%
  \BibitemOpen
  \bibfield  {author} {\bibinfo {author} {\bibfnamefont {J.}~\bibnamefont {Antoniadis}} \emph {et~al.},\ }\href {\doibase 10.1093/mnras/stab3418} {\bibfield  {journal} {\bibinfo  {journal} {Mon. Not. Roy. Astron. Soc.}\ }\textbf {\bibinfo {volume} {510}},\ \bibinfo {pages} {4873} (\bibinfo {year} {2022})},\ \Eprint {http://arxiv.org/abs/2201.03980} {arXiv:2201.03980 [astro-ph.HE]} \BibitemShut {NoStop}%
\bibitem [{\citenamefont {Xu}\ \emph {et~al.}(2023)\citenamefont {Xu} \emph {et~al.}}]{Xu:2023wog}%
  \BibitemOpen
  \bibfield  {author} {\bibinfo {author} {\bibfnamefont {H.}~\bibnamefont {Xu}} \emph {et~al.},\ }\href {\doibase 10.1088/1674-4527/acdfa5} {\bibfield  {journal} {\bibinfo  {journal} {Res. Astron. Astrophys.}\ }\textbf {\bibinfo {volume} {23}},\ \bibinfo {pages} {075024} (\bibinfo {year} {2023})},\ \Eprint {http://arxiv.org/abs/2306.16216} {arXiv:2306.16216 [astro-ph.HE]} \BibitemShut {NoStop}%
\bibitem [{\citenamefont {Rajagopal}\ and\ \citenamefont {Romani}(1995)}]{Rajagopal:1994zj}%
  \BibitemOpen
  \bibfield  {author} {\bibinfo {author} {\bibfnamefont {M.}~\bibnamefont {Rajagopal}}\ and\ \bibinfo {author} {\bibfnamefont {R.~W.}\ \bibnamefont {Romani}},\ }\href {\doibase 10.1086/175813} {\bibfield  {journal} {\bibinfo  {journal} {Astrophys. J.}\ }\textbf {\bibinfo {volume} {446}},\ \bibinfo {pages} {543} (\bibinfo {year} {1995})},\ \Eprint {http://arxiv.org/abs/astro-ph/9412038} {arXiv:astro-ph/9412038} \BibitemShut {NoStop}%
\bibitem [{\citenamefont {Phinney}(2001)}]{Phinney:2001di}%
  \BibitemOpen
  \bibfield  {author} {\bibinfo {author} {\bibfnamefont {E.~S.}\ \bibnamefont {Phinney}},\ }\href@noop {} {\  (\bibinfo {year} {2001})},\ \Eprint {http://arxiv.org/abs/astro-ph/0108028} {arXiv:astro-ph/0108028} \BibitemShut {NoStop}%
\bibitem [{\citenamefont {Jaffe}\ and\ \citenamefont {Backer}(2003)}]{Jaffe:2002rt}%
  \BibitemOpen
  \bibfield  {author} {\bibinfo {author} {\bibfnamefont {A.~H.}\ \bibnamefont {Jaffe}}\ and\ \bibinfo {author} {\bibfnamefont {D.~C.}\ \bibnamefont {Backer}},\ }\href {\doibase 10.1086/345443} {\bibfield  {journal} {\bibinfo  {journal} {Astrophys. J.}\ }\textbf {\bibinfo {volume} {583}},\ \bibinfo {pages} {616} (\bibinfo {year} {2003})},\ \Eprint {http://arxiv.org/abs/astro-ph/0210148} {arXiv:astro-ph/0210148} \BibitemShut {NoStop}%
\bibitem [{\citenamefont {Wyithe}\ and\ \citenamefont {Loeb}(2003)}]{Wyithe:2002ep}%
  \BibitemOpen
  \bibfield  {author} {\bibinfo {author} {\bibfnamefont {J.~S.~B.}\ \bibnamefont {Wyithe}}\ and\ \bibinfo {author} {\bibfnamefont {A.}~\bibnamefont {Loeb}},\ }\href {\doibase 10.1086/375187} {\bibfield  {journal} {\bibinfo  {journal} {Astrophys. J.}\ }\textbf {\bibinfo {volume} {590}},\ \bibinfo {pages} {691} (\bibinfo {year} {2003})},\ \Eprint {http://arxiv.org/abs/astro-ph/0211556} {arXiv:astro-ph/0211556} \BibitemShut {NoStop}%
\bibitem [{\citenamefont {Ananda}\ \emph {et~al.}(2007)\citenamefont {Ananda}, \citenamefont {Clarkson},\ and\ \citenamefont {Wands}}]{Ananda:2006af}%
  \BibitemOpen
  \bibfield  {author} {\bibinfo {author} {\bibfnamefont {K.~N.}\ \bibnamefont {Ananda}}, \bibinfo {author} {\bibfnamefont {C.}~\bibnamefont {Clarkson}}, \ and\ \bibinfo {author} {\bibfnamefont {D.}~\bibnamefont {Wands}},\ }\href {\doibase 10.1103/PhysRevD.75.123518} {\bibfield  {journal} {\bibinfo  {journal} {Phys. Rev. D}\ }\textbf {\bibinfo {volume} {75}},\ \bibinfo {pages} {123518} (\bibinfo {year} {2007})},\ \Eprint {http://arxiv.org/abs/gr-qc/0612013} {arXiv:gr-qc/0612013} \BibitemShut {NoStop}%
\bibitem [{\citenamefont {Baumann}\ \emph {et~al.}(2007)\citenamefont {Baumann}, \citenamefont {Steinhardt}, \citenamefont {Takahashi},\ and\ \citenamefont {Ichiki}}]{Baumann:2007zm}%
  \BibitemOpen
  \bibfield  {author} {\bibinfo {author} {\bibfnamefont {D.}~\bibnamefont {Baumann}}, \bibinfo {author} {\bibfnamefont {P.~J.}\ \bibnamefont {Steinhardt}}, \bibinfo {author} {\bibfnamefont {K.}~\bibnamefont {Takahashi}}, \ and\ \bibinfo {author} {\bibfnamefont {K.}~\bibnamefont {Ichiki}},\ }\href {\doibase 10.1103/PhysRevD.76.084019} {\bibfield  {journal} {\bibinfo  {journal} {Phys. Rev. D}\ }\textbf {\bibinfo {volume} {76}},\ \bibinfo {pages} {084019} (\bibinfo {year} {2007})},\ \Eprint {http://arxiv.org/abs/hep-th/0703290} {arXiv:hep-th/0703290} \BibitemShut {NoStop}%
\bibitem [{\citenamefont {Losada}(1997)}]{Losada:1996ju}%
  \BibitemOpen
  \bibfield  {author} {\bibinfo {author} {\bibfnamefont {M.}~\bibnamefont {Losada}},\ }\href {\doibase 10.1103/PhysRevD.56.2893} {\bibfield  {journal} {\bibinfo  {journal} {Phys. Rev. D}\ }\textbf {\bibinfo {volume} {56}},\ \bibinfo {pages} {2893} (\bibinfo {year} {1997})},\ \Eprint {http://arxiv.org/abs/hep-ph/9605266} {arXiv:hep-ph/9605266} \BibitemShut {NoStop}%
\bibitem [{\citenamefont {Cline}\ and\ \citenamefont {Lemieux}(1997)}]{Cline:1996mga}%
  \BibitemOpen
  \bibfield  {author} {\bibinfo {author} {\bibfnamefont {J.~M.}\ \bibnamefont {Cline}}\ and\ \bibinfo {author} {\bibfnamefont {P.-A.}\ \bibnamefont {Lemieux}},\ }\href {\doibase 10.1103/PhysRevD.55.3873} {\bibfield  {journal} {\bibinfo  {journal} {Phys. Rev. D}\ }\textbf {\bibinfo {volume} {55}},\ \bibinfo {pages} {3873} (\bibinfo {year} {1997})},\ \Eprint {http://arxiv.org/abs/hep-ph/9609240} {arXiv:hep-ph/9609240} \BibitemShut {NoStop}%
\bibitem [{\citenamefont {Neronov}\ \emph {et~al.}(2021)\citenamefont {Neronov}, \citenamefont {Roper~Pol}, \citenamefont {Caprini},\ and\ \citenamefont {Semikoz}}]{Neronov:2020qrl}%
  \BibitemOpen
  \bibfield  {author} {\bibinfo {author} {\bibfnamefont {A.}~\bibnamefont {Neronov}}, \bibinfo {author} {\bibfnamefont {A.}~\bibnamefont {Roper~Pol}}, \bibinfo {author} {\bibfnamefont {C.}~\bibnamefont {Caprini}}, \ and\ \bibinfo {author} {\bibfnamefont {D.}~\bibnamefont {Semikoz}},\ }\href {\doibase 10.1103/PhysRevD.103.L041302} {\bibfield  {journal} {\bibinfo  {journal} {Phys. Rev. D}\ }\textbf {\bibinfo {volume} {103}},\ \bibinfo {pages} {041302} (\bibinfo {year} {2021})},\ \Eprint {http://arxiv.org/abs/2009.14174} {arXiv:2009.14174 [astro-ph.CO]} \BibitemShut {NoStop}%
\bibitem [{\citenamefont {Li}\ \emph {et~al.}(2021)\citenamefont {Li}, \citenamefont {Shao}, \citenamefont {Wu},\ and\ \citenamefont {Yu}}]{Li:2021qer}%
  \BibitemOpen
  \bibfield  {author} {\bibinfo {author} {\bibfnamefont {S.~L.}\ \bibnamefont {Li}}, \bibinfo {author} {\bibfnamefont {L.~J.}\ \bibnamefont {Shao}}, \bibinfo {author} {\bibfnamefont {P.~X.}\ \bibnamefont {Wu}}, \ and\ \bibinfo {author} {\bibfnamefont {H.}~\bibnamefont {Yu}},\ }\href {\doibase 10.1103/PhysRevD.104.043510} {\bibfield  {journal} {\bibinfo  {journal} {Phys. Rev. D}\ }\textbf {\bibinfo {volume} {104}},\ \bibinfo {pages} {043510} (\bibinfo {year} {2021})},\ \Eprint {http://arxiv.org/abs/2101.08012} {arXiv:2101.08012 [astro-ph.CO]} \BibitemShut {NoStop}%
\bibitem [{\citenamefont {Nakai}\ \emph {et~al.}(2021)\citenamefont {Nakai}, \citenamefont {Suzuki}, \citenamefont {Takahashi},\ and\ \citenamefont {Yamada}}]{Nakai:2020oit}%
  \BibitemOpen
  \bibfield  {author} {\bibinfo {author} {\bibfnamefont {Y.}~\bibnamefont {Nakai}}, \bibinfo {author} {\bibfnamefont {M.}~\bibnamefont {Suzuki}}, \bibinfo {author} {\bibfnamefont {F.}~\bibnamefont {Takahashi}}, \ and\ \bibinfo {author} {\bibfnamefont {M.}~\bibnamefont {Yamada}},\ }\href {\doibase 10.1016/j.physletb.2021.136238} {\bibfield  {journal} {\bibinfo  {journal} {Phys. Lett. B}\ }\textbf {\bibinfo {volume} {816}},\ \bibinfo {pages} {136238} (\bibinfo {year} {2021})},\ \Eprint {http://arxiv.org/abs/2009.09754} {arXiv:2009.09754 [astro-ph.CO]} \BibitemShut {NoStop}%
\bibitem [{\citenamefont {Ratzinger}\ and\ \citenamefont {Schwaller}(2021)}]{Ratzinger:2020koh}%
  \BibitemOpen
  \bibfield  {author} {\bibinfo {author} {\bibfnamefont {W.}~\bibnamefont {Ratzinger}}\ and\ \bibinfo {author} {\bibfnamefont {P.}~\bibnamefont {Schwaller}},\ }\href {\doibase 10.21468/SciPostPhys.10.2.047} {\bibfield  {journal} {\bibinfo  {journal} {SciPost Phys.}\ }\textbf {\bibinfo {volume} {10}},\ \bibinfo {pages} {047} (\bibinfo {year} {2021})},\ \Eprint {http://arxiv.org/abs/2009.11875} {arXiv:2009.11875 [astro-ph.CO]} \BibitemShut {NoStop}%
\bibitem [{\citenamefont {Kosowsky}\ \emph {et~al.}(1992)\citenamefont {Kosowsky}, \citenamefont {Turner},\ and\ \citenamefont {Watkins}}]{Kosowsky:1992rz}%
  \BibitemOpen
  \bibfield  {author} {\bibinfo {author} {\bibfnamefont {A.}~\bibnamefont {Kosowsky}}, \bibinfo {author} {\bibfnamefont {M.~S.}\ \bibnamefont {Turner}}, \ and\ \bibinfo {author} {\bibfnamefont {R.}~\bibnamefont {Watkins}},\ }\href {\doibase 10.1103/PhysRevLett.69.2026} {\bibfield  {journal} {\bibinfo  {journal} {Phys. Rev. Lett.}\ }\textbf {\bibinfo {volume} {69}},\ \bibinfo {pages} {2026} (\bibinfo {year} {1992})}\BibitemShut {NoStop}%
\bibitem [{\citenamefont {Caprini}\ \emph {et~al.}(2010)\citenamefont {Caprini}, \citenamefont {Durrer},\ and\ \citenamefont {Siemens}}]{Caprini:2010xv}%
  \BibitemOpen
  \bibfield  {author} {\bibinfo {author} {\bibfnamefont {C.}~\bibnamefont {Caprini}}, \bibinfo {author} {\bibfnamefont {R.}~\bibnamefont {Durrer}}, \ and\ \bibinfo {author} {\bibfnamefont {X.}~\bibnamefont {Siemens}},\ }\href {\doibase 10.1103/PhysRevD.82.063511} {\bibfield  {journal} {\bibinfo  {journal} {Phys. Rev. D}\ }\textbf {\bibinfo {volume} {82}},\ \bibinfo {pages} {063511} (\bibinfo {year} {2010})},\ \Eprint {http://arxiv.org/abs/1007.1218} {arXiv:1007.1218 [astro-ph.CO]} \BibitemShut {NoStop}%
\bibitem [{\citenamefont {Xue}\ \emph {et~al.}(2021)\citenamefont {Xue} \emph {et~al.}}]{Xue:2021gyq}%
  \BibitemOpen
  \bibfield  {author} {\bibinfo {author} {\bibfnamefont {X.}~\bibnamefont {Xue}} \emph {et~al.},\ }\href {\doibase 10.1103/PhysRevLett.127.251303} {\bibfield  {journal} {\bibinfo  {journal} {Phys. Rev. Lett.}\ }\textbf {\bibinfo {volume} {127}},\ \bibinfo {pages} {251303} (\bibinfo {year} {2021})},\ \Eprint {http://arxiv.org/abs/2110.03096} {arXiv:2110.03096 [astro-ph.CO]} \BibitemShut {NoStop}%
\bibitem [{\citenamefont {Ellis}\ \emph {et~al.}(2023)\citenamefont {Ellis}, \citenamefont {Lewicki}, \citenamefont {Lin},\ and\ \citenamefont {Vaskonen}}]{Ellis:2023tsl}%
  \BibitemOpen
  \bibfield  {author} {\bibinfo {author} {\bibfnamefont {J.}~\bibnamefont {Ellis}}, \bibinfo {author} {\bibfnamefont {M.}~\bibnamefont {Lewicki}}, \bibinfo {author} {\bibfnamefont {C.}~\bibnamefont {Lin}}, \ and\ \bibinfo {author} {\bibfnamefont {V.}~\bibnamefont {Vaskonen}},\ }\href {\doibase 10.1103/PhysRevD.108.103511} {\bibfield  {journal} {\bibinfo  {journal} {Phys. Rev. D}\ }\textbf {\bibinfo {volume} {108}},\ \bibinfo {pages} {103511} (\bibinfo {year} {2023})},\ \Eprint {http://arxiv.org/abs/2306.17147} {arXiv:2306.17147 [astro-ph.CO]} \BibitemShut {NoStop}%
\bibitem [{\citenamefont {Addazi}\ \emph {et~al.}(2024)\citenamefont {Addazi}, \citenamefont {Cai}, \citenamefont {Marciano},\ and\ \citenamefont {Visinelli}}]{Addazi:2023jvg}%
  \BibitemOpen
  \bibfield  {author} {\bibinfo {author} {\bibfnamefont {A.}~\bibnamefont {Addazi}}, \bibinfo {author} {\bibfnamefont {Y.-F.}\ \bibnamefont {Cai}}, \bibinfo {author} {\bibfnamefont {A.}~\bibnamefont {Marciano}}, \ and\ \bibinfo {author} {\bibfnamefont {L.}~\bibnamefont {Visinelli}},\ }\href {\doibase 10.1103/PhysRevD.109.015028} {\bibfield  {journal} {\bibinfo  {journal} {Phys. Rev. D}\ }\textbf {\bibinfo {volume} {109}},\ \bibinfo {pages} {015028} (\bibinfo {year} {2024})},\ \Eprint {http://arxiv.org/abs/2306.17205} {arXiv:2306.17205 [astro-ph.CO]} \BibitemShut {NoStop}%
\bibitem [{\citenamefont {Ghosh}\ \emph {et~al.}(2024)\citenamefont {Ghosh}, \citenamefont {Ghoshal}, \citenamefont {Guo}, \citenamefont {Hajkarim}, \citenamefont {King}, \citenamefont {Sinha}, \citenamefont {Wang},\ and\ \citenamefont {White}}]{Ghosh:2023aum}%
  \BibitemOpen
  \bibfield  {author} {\bibinfo {author} {\bibfnamefont {T.}~\bibnamefont {Ghosh}}, \bibinfo {author} {\bibfnamefont {A.}~\bibnamefont {Ghoshal}}, \bibinfo {author} {\bibfnamefont {H.-K.}\ \bibnamefont {Guo}}, \bibinfo {author} {\bibfnamefont {F.}~\bibnamefont {Hajkarim}}, \bibinfo {author} {\bibfnamefont {S.~F.}\ \bibnamefont {King}}, \bibinfo {author} {\bibfnamefont {K.}~\bibnamefont {Sinha}}, \bibinfo {author} {\bibfnamefont {X.}~\bibnamefont {Wang}}, \ and\ \bibinfo {author} {\bibfnamefont {G.}~\bibnamefont {White}},\ }\href {\doibase 10.1088/1475-7516/2024/05/100} {\bibfield  {journal} {\bibinfo  {journal} {JCAP}\ }\textbf {\bibinfo {volume} {05}},\ \bibinfo {pages} {100} (\bibinfo {year} {2024})},\ \Eprint {http://arxiv.org/abs/2307.02259} {arXiv:2307.02259 [astro-ph.HE]} \BibitemShut {NoStop}%
\bibitem [{\citenamefont {Rezapour}\ \emph {et~al.}(2022)\citenamefont {Rezapour}, \citenamefont {Bitaghsir~Fadafan},\ and\ \citenamefont {Ahmadvand}}]{Rezapour:2022iqq}%
  \BibitemOpen
  \bibfield  {author} {\bibinfo {author} {\bibfnamefont {S.}~\bibnamefont {Rezapour}}, \bibinfo {author} {\bibfnamefont {K.}~\bibnamefont {Bitaghsir~Fadafan}}, \ and\ \bibinfo {author} {\bibfnamefont {M.}~\bibnamefont {Ahmadvand}},\ }\href {\doibase 10.1088/1402-4896/ac4f03} {\bibfield  {journal} {\bibinfo  {journal} {Phys. Scripta}\ }\textbf {\bibinfo {volume} {97}},\ \bibinfo {pages} {035301} (\bibinfo {year} {2022})}\BibitemShut {NoStop}%
\bibitem [{\citenamefont {Ahmadvand}\ and\ \citenamefont {Bitaghsir~Fadafan}(2017)}]{Ahmadvand:2017xrw}%
  \BibitemOpen
  \bibfield  {author} {\bibinfo {author} {\bibfnamefont {M.}~\bibnamefont {Ahmadvand}}\ and\ \bibinfo {author} {\bibfnamefont {K.}~\bibnamefont {Bitaghsir~Fadafan}},\ }\href {\doibase 10.1016/j.physletb.2017.07.039} {\bibfield  {journal} {\bibinfo  {journal} {Phys. Lett. B}\ }\textbf {\bibinfo {volume} {772}},\ \bibinfo {pages} {747} (\bibinfo {year} {2017})},\ \Eprint {http://arxiv.org/abs/1703.02801} {arXiv:1703.02801 [hep-th]} \BibitemShut {NoStop}%
\bibitem [{\citenamefont {Athron}\ \emph {et~al.}(2024)\citenamefont {Athron}, \citenamefont {Fowlie}, \citenamefont {Lu}, \citenamefont {Morris}, \citenamefont {Wu}, \citenamefont {Wu},\ and\ \citenamefont {Xu}}]{Athron:2023mer}%
  \BibitemOpen
  \bibfield  {author} {\bibinfo {author} {\bibfnamefont {P.}~\bibnamefont {Athron}}, \bibinfo {author} {\bibfnamefont {A.}~\bibnamefont {Fowlie}}, \bibinfo {author} {\bibfnamefont {C.-T.}\ \bibnamefont {Lu}}, \bibinfo {author} {\bibfnamefont {L.}~\bibnamefont {Morris}}, \bibinfo {author} {\bibfnamefont {L.}~\bibnamefont {Wu}}, \bibinfo {author} {\bibfnamefont {Y.}~\bibnamefont {Wu}}, \ and\ \bibinfo {author} {\bibfnamefont {Z.}~\bibnamefont {Xu}},\ }\href {\doibase 10.1103/PhysRevLett.132.221001} {\bibfield  {journal} {\bibinfo  {journal} {Phys. Rev. Lett.}\ }\textbf {\bibinfo {volume} {132}},\ \bibinfo {pages} {221001} (\bibinfo {year} {2024})},\ \Eprint {http://arxiv.org/abs/2306.17239} {arXiv:2306.17239 [hep-ph]} \BibitemShut {NoStop}%
\bibitem [{\citenamefont {Han}\ \emph {et~al.}(2023)\citenamefont {Han}, \citenamefont {Xie}, \citenamefont {Yang},\ and\ \citenamefont {Zhang}}]{Han:2023olf}%
  \BibitemOpen
  \bibfield  {author} {\bibinfo {author} {\bibfnamefont {C.}~\bibnamefont {Han}}, \bibinfo {author} {\bibfnamefont {K.-P.}\ \bibnamefont {Xie}}, \bibinfo {author} {\bibfnamefont {J.~M.}\ \bibnamefont {Yang}}, \ and\ \bibinfo {author} {\bibfnamefont {M.}~\bibnamefont {Zhang}},\ }\href@noop {} {\  (\bibinfo {year} {2023})},\ \Eprint {http://arxiv.org/abs/2306.16966} {arXiv:2306.16966 [hep-ph]} \BibitemShut {NoStop}%
\bibitem [{\citenamefont {Salvio}(2024)}]{Salvio:2023blb}%
  \BibitemOpen
  \bibfield  {author} {\bibinfo {author} {\bibfnamefont {A.}~\bibnamefont {Salvio}},\ }\href {\doibase 10.1016/j.physletb.2024.138639} {\bibfield  {journal} {\bibinfo  {journal} {Phys. Lett. B}\ }\textbf {\bibinfo {volume} {852}},\ \bibinfo {pages} {138639} (\bibinfo {year} {2024})},\ \Eprint {http://arxiv.org/abs/2312.04628} {arXiv:2312.04628 [hep-ph]} \BibitemShut {NoStop}%
\bibitem [{\citenamefont {Qin}\ and\ \citenamefont {Bian}(2024)}]{Qin:2024idc}%
  \BibitemOpen
  \bibfield  {author} {\bibinfo {author} {\bibfnamefont {R.}~\bibnamefont {Qin}}\ and\ \bibinfo {author} {\bibfnamefont {L.}~\bibnamefont {Bian}},\ }\href@noop {} {\  (\bibinfo {year} {2024})},\ \Eprint {http://arxiv.org/abs/2407.01981} {arXiv:2407.01981 [hep-ph]} \BibitemShut {NoStop}%
\bibitem [{\citenamefont {Gouttenoire}(2023)}]{Gouttenoire:2023bqy}%
  \BibitemOpen
  \bibfield  {author} {\bibinfo {author} {\bibfnamefont {Y.}~\bibnamefont {Gouttenoire}},\ }\href {\doibase 10.1103/PhysRevLett.131.171404} {\bibfield  {journal} {\bibinfo  {journal} {Phys. Rev. Lett.}\ }\textbf {\bibinfo {volume} {131}},\ \bibinfo {pages} {171404} (\bibinfo {year} {2023})},\ \Eprint {http://arxiv.org/abs/2307.04239} {arXiv:2307.04239 [hep-ph]} \BibitemShut {NoStop}%
\bibitem [{\citenamefont {Siemens}\ \emph {et~al.}(2007)\citenamefont {Siemens}, \citenamefont {Mandic},\ and\ \citenamefont {Creighton}}]{Siemens:2006yp}%
  \BibitemOpen
  \bibfield  {author} {\bibinfo {author} {\bibfnamefont {X.}~\bibnamefont {Siemens}}, \bibinfo {author} {\bibfnamefont {V.}~\bibnamefont {Mandic}}, \ and\ \bibinfo {author} {\bibfnamefont {J.}~\bibnamefont {Creighton}},\ }\href {\doibase 10.1103/PhysRevLett.98.111101} {\bibfield  {journal} {\bibinfo  {journal} {Phys. Rev. Lett.}\ }\textbf {\bibinfo {volume} {98}},\ \bibinfo {pages} {111101} (\bibinfo {year} {2007})},\ \Eprint {http://arxiv.org/abs/astro-ph/0610920} {arXiv:astro-ph/0610920} \BibitemShut {NoStop}%
\bibitem [{\citenamefont {Blasi}\ \emph {et~al.}(2021)\citenamefont {Blasi}, \citenamefont {Brdar},\ and\ \citenamefont {Schmitz}}]{Blasi:2020mfx}%
  \BibitemOpen
  \bibfield  {author} {\bibinfo {author} {\bibfnamefont {S.}~\bibnamefont {Blasi}}, \bibinfo {author} {\bibfnamefont {V.}~\bibnamefont {Brdar}}, \ and\ \bibinfo {author} {\bibfnamefont {K.}~\bibnamefont {Schmitz}},\ }\href {\doibase 10.1103/PhysRevLett.126.041305} {\bibfield  {journal} {\bibinfo  {journal} {Phys. Rev. Lett.}\ }\textbf {\bibinfo {volume} {126}},\ \bibinfo {pages} {041305} (\bibinfo {year} {2021})},\ \Eprint {http://arxiv.org/abs/2009.06607} {arXiv:2009.06607 [astro-ph.CO]} \BibitemShut {NoStop}%
\bibitem [{\citenamefont {Buchmuller}\ \emph {et~al.}(2020)\citenamefont {Buchmuller}, \citenamefont {Domcke},\ and\ \citenamefont {Schmitz}}]{Buchmuller:2020lbh}%
  \BibitemOpen
  \bibfield  {author} {\bibinfo {author} {\bibfnamefont {W.}~\bibnamefont {Buchmuller}}, \bibinfo {author} {\bibfnamefont {V.}~\bibnamefont {Domcke}}, \ and\ \bibinfo {author} {\bibfnamefont {K.}~\bibnamefont {Schmitz}},\ }\href {\doibase 10.1016/j.physletb.2020.135914} {\bibfield  {journal} {\bibinfo  {journal} {Phys. Lett. B}\ }\textbf {\bibinfo {volume} {811}},\ \bibinfo {pages} {135914} (\bibinfo {year} {2020})},\ \Eprint {http://arxiv.org/abs/2009.10649} {arXiv:2009.10649 [astro-ph.CO]} \BibitemShut {NoStop}%
\bibitem [{\citenamefont {Lazarides}\ \emph {et~al.}(2023)\citenamefont {Lazarides}, \citenamefont {Maji},\ and\ \citenamefont {Shafi}}]{Lazarides:2023ksx}%
  \BibitemOpen
  \bibfield  {author} {\bibinfo {author} {\bibfnamefont {G.}~\bibnamefont {Lazarides}}, \bibinfo {author} {\bibfnamefont {R.}~\bibnamefont {Maji}}, \ and\ \bibinfo {author} {\bibfnamefont {Q.}~\bibnamefont {Shafi}},\ }\href {\doibase 10.1103/PhysRevD.108.095041} {\bibfield  {journal} {\bibinfo  {journal} {Phys. Rev. D}\ }\textbf {\bibinfo {volume} {108}},\ \bibinfo {pages} {095041} (\bibinfo {year} {2023})},\ \Eprint {http://arxiv.org/abs/2306.17788} {arXiv:2306.17788 [hep-ph]} \BibitemShut {NoStop}%
\bibitem [{\citenamefont {Wang}\ \emph {et~al.}(2023)\citenamefont {Wang}, \citenamefont {Lei}, \citenamefont {Jiao}, \citenamefont {Feng},\ and\ \citenamefont {Fan}}]{Wang:2023len}%
  \BibitemOpen
  \bibfield  {author} {\bibinfo {author} {\bibfnamefont {Z.}~\bibnamefont {Wang}}, \bibinfo {author} {\bibfnamefont {L.}~\bibnamefont {Lei}}, \bibinfo {author} {\bibfnamefont {H.}~\bibnamefont {Jiao}}, \bibinfo {author} {\bibfnamefont {L.}~\bibnamefont {Feng}}, \ and\ \bibinfo {author} {\bibfnamefont {Y.-Z.}\ \bibnamefont {Fan}},\ }\href {\doibase 10.1007/s11433-023-2262-0} {\bibfield  {journal} {\bibinfo  {journal} {Sci. China Phys. Mech. Astron.}\ }\textbf {\bibinfo {volume} {66}},\ \bibinfo {pages} {120403} (\bibinfo {year} {2023})},\ \Eprint {http://arxiv.org/abs/2306.17150} {arXiv:2306.17150 [astro-ph.HE]} \BibitemShut {NoStop}%
\bibitem [{\citenamefont {Samanta}\ and\ \citenamefont {Datta}(2021)}]{Samanta:2020cdk}%
  \BibitemOpen
  \bibfield  {author} {\bibinfo {author} {\bibfnamefont {R.}~\bibnamefont {Samanta}}\ and\ \bibinfo {author} {\bibfnamefont {S.}~\bibnamefont {Datta}},\ }\href {\doibase 10.1007/JHEP05(2021)211} {\bibfield  {journal} {\bibinfo  {journal} {JHEP}\ }\textbf {\bibinfo {volume} {05}},\ \bibinfo {pages} {211} (\bibinfo {year} {2021})},\ \Eprint {http://arxiv.org/abs/2009.13452} {arXiv:2009.13452 [hep-ph]} \BibitemShut {NoStop}%
\bibitem [{\citenamefont {Bian}\ \emph {et~al.}(2022{\natexlab{a}})\citenamefont {Bian}, \citenamefont {Shu}, \citenamefont {Wang}, \citenamefont {Yuan},\ and\ \citenamefont {Zong}}]{Bian:2022tju}%
  \BibitemOpen
  \bibfield  {author} {\bibinfo {author} {\bibfnamefont {L.}~\bibnamefont {Bian}}, \bibinfo {author} {\bibfnamefont {J.}~\bibnamefont {Shu}}, \bibinfo {author} {\bibfnamefont {B.}~\bibnamefont {Wang}}, \bibinfo {author} {\bibfnamefont {Q.}~\bibnamefont {Yuan}}, \ and\ \bibinfo {author} {\bibfnamefont {J.}~\bibnamefont {Zong}},\ }\href {\doibase 10.1103/PhysRevD.106.L101301} {\bibfield  {journal} {\bibinfo  {journal} {Phys. Rev. D}\ }\textbf {\bibinfo {volume} {106}},\ \bibinfo {pages} {L101301} (\bibinfo {year} {2022}{\natexlab{a}})},\ \Eprint {http://arxiv.org/abs/2205.07293} {arXiv:2205.07293 [hep-ph]} \BibitemShut {NoStop}%
\bibitem [{\citenamefont {Hiramatsu}\ \emph {et~al.}(2014)\citenamefont {Hiramatsu}, \citenamefont {Kawasaki},\ and\ \citenamefont {Saikawa}}]{Hiramatsu:2013qaa}%
  \BibitemOpen
  \bibfield  {author} {\bibinfo {author} {\bibfnamefont {T.}~\bibnamefont {Hiramatsu}}, \bibinfo {author} {\bibfnamefont {M.}~\bibnamefont {Kawasaki}}, \ and\ \bibinfo {author} {\bibfnamefont {K.}~\bibnamefont {Saikawa}},\ }\href {\doibase 10.1088/1475-7516/2014/02/031} {\bibfield  {journal} {\bibinfo  {journal} {JCAP}\ }\textbf {\bibinfo {volume} {02}},\ \bibinfo {pages} {031} (\bibinfo {year} {2014})},\ \Eprint {http://arxiv.org/abs/1309.5001} {arXiv:1309.5001 [astro-ph.CO]} \BibitemShut {NoStop}%
\bibitem [{\citenamefont {Ferreira}\ \emph {et~al.}(2023)\citenamefont {Ferreira}, \citenamefont {Notari}, \citenamefont {Pujolas},\ and\ \citenamefont {Rompineve}}]{Ferreira:2022zzo}%
  \BibitemOpen
  \bibfield  {author} {\bibinfo {author} {\bibfnamefont {R.~Z.}\ \bibnamefont {Ferreira}}, \bibinfo {author} {\bibfnamefont {A.}~\bibnamefont {Notari}}, \bibinfo {author} {\bibfnamefont {O.}~\bibnamefont {Pujolas}}, \ and\ \bibinfo {author} {\bibfnamefont {F.}~\bibnamefont {Rompineve}},\ }\href {\doibase 10.1088/1475-7516/2023/02/001} {\bibfield  {journal} {\bibinfo  {journal} {JCAP}\ }\textbf {\bibinfo {volume} {02}},\ \bibinfo {pages} {001} (\bibinfo {year} {2023})},\ \Eprint {http://arxiv.org/abs/2204.04228} {arXiv:2204.04228 [astro-ph.CO]} \BibitemShut {NoStop}%
\bibitem [{\citenamefont {Kitajima}\ \emph {et~al.}(2024)\citenamefont {Kitajima}, \citenamefont {Lee}, \citenamefont {Murai}, \citenamefont {Takahashi},\ and\ \citenamefont {Yin}}]{Kitajima:2023cek}%
  \BibitemOpen
  \bibfield  {author} {\bibinfo {author} {\bibfnamefont {N.}~\bibnamefont {Kitajima}}, \bibinfo {author} {\bibfnamefont {J.}~\bibnamefont {Lee}}, \bibinfo {author} {\bibfnamefont {K.}~\bibnamefont {Murai}}, \bibinfo {author} {\bibfnamefont {F.}~\bibnamefont {Takahashi}}, \ and\ \bibinfo {author} {\bibfnamefont {W.}~\bibnamefont {Yin}},\ }\href {\doibase 10.1016/j.physletb.2024.138586} {\bibfield  {journal} {\bibinfo  {journal} {Phys. Lett. B}\ }\textbf {\bibinfo {volume} {851}},\ \bibinfo {pages} {138586} (\bibinfo {year} {2024})},\ \Eprint {http://arxiv.org/abs/2306.17146} {arXiv:2306.17146 [hep-ph]} \BibitemShut {NoStop}%
\bibitem [{\citenamefont {Blasi}\ \emph {et~al.}(2023)\citenamefont {Blasi}, \citenamefont {Mariotti}, \citenamefont {Rase},\ and\ \citenamefont {Sevrin}}]{Blasi:2023sej}%
  \BibitemOpen
  \bibfield  {author} {\bibinfo {author} {\bibfnamefont {S.}~\bibnamefont {Blasi}}, \bibinfo {author} {\bibfnamefont {A.}~\bibnamefont {Mariotti}}, \bibinfo {author} {\bibfnamefont {A.}~\bibnamefont {Rase}}, \ and\ \bibinfo {author} {\bibfnamefont {A.}~\bibnamefont {Sevrin}},\ }\href {\doibase 10.1007/JHEP11(2023)169} {\bibfield  {journal} {\bibinfo  {journal} {JHEP}\ }\textbf {\bibinfo {volume} {11}},\ \bibinfo {pages} {169} (\bibinfo {year} {2023})},\ \Eprint {http://arxiv.org/abs/2306.17830} {arXiv:2306.17830 [hep-ph]} \BibitemShut {NoStop}%
\bibitem [{\citenamefont {Li}\ \emph {et~al.}(2023)\citenamefont {Li}, \citenamefont {Bian},\ and\ \citenamefont {Jia}}]{Li:2023yzq}%
  \BibitemOpen
  \bibfield  {author} {\bibinfo {author} {\bibfnamefont {Y.}~\bibnamefont {Li}}, \bibinfo {author} {\bibfnamefont {L.}~\bibnamefont {Bian}}, \ and\ \bibinfo {author} {\bibfnamefont {Y.}~\bibnamefont {Jia}},\ }\href@noop {} {\  (\bibinfo {year} {2023})},\ \Eprint {http://arxiv.org/abs/2304.05220} {arXiv:2304.05220 [hep-ph]} \BibitemShut {NoStop}%
\bibitem [{\citenamefont {Bian}\ \emph {et~al.}(2022{\natexlab{b}})\citenamefont {Bian}, \citenamefont {Ge}, \citenamefont {Li}, \citenamefont {Shu},\ and\ \citenamefont {Zong}}]{Bian:2022qbh}%
  \BibitemOpen
  \bibfield  {author} {\bibinfo {author} {\bibfnamefont {L.}~\bibnamefont {Bian}}, \bibinfo {author} {\bibfnamefont {S.}~\bibnamefont {Ge}}, \bibinfo {author} {\bibfnamefont {C.}~\bibnamefont {Li}}, \bibinfo {author} {\bibfnamefont {J.}~\bibnamefont {Shu}}, \ and\ \bibinfo {author} {\bibfnamefont {J.}~\bibnamefont {Zong}},\ }\href@noop {} {\  (\bibinfo {year} {2022}{\natexlab{b}})},\ \Eprint {http://arxiv.org/abs/2212.07871} {arXiv:2212.07871 [hep-ph]} \BibitemShut {NoStop}%
\bibitem [{\citenamefont {Babichev}\ \emph {et~al.}(2023)\citenamefont {Babichev}, \citenamefont {Gorbunov}, \citenamefont {Ramazanov}, \citenamefont {Samanta},\ and\ \citenamefont {Vikman}}]{Babichev:2023pbf}%
  \BibitemOpen
  \bibfield  {author} {\bibinfo {author} {\bibfnamefont {E.}~\bibnamefont {Babichev}}, \bibinfo {author} {\bibfnamefont {D.}~\bibnamefont {Gorbunov}}, \bibinfo {author} {\bibfnamefont {S.}~\bibnamefont {Ramazanov}}, \bibinfo {author} {\bibfnamefont {R.}~\bibnamefont {Samanta}}, \ and\ \bibinfo {author} {\bibfnamefont {A.}~\bibnamefont {Vikman}},\ }\href {\doibase 10.1103/PhysRevD.108.123529} {\bibfield  {journal} {\bibinfo  {journal} {Phys. Rev. D}\ }\textbf {\bibinfo {volume} {108}},\ \bibinfo {pages} {123529} (\bibinfo {year} {2023})},\ \Eprint {http://arxiv.org/abs/2307.04582} {arXiv:2307.04582 [hep-ph]} \BibitemShut {NoStop}%
\bibitem [{\citenamefont {Sakharov}\ \emph {et~al.}(2021)\citenamefont {Sakharov}, \citenamefont {Eroshenko},\ and\ \citenamefont {Rubin}}]{Sakharov:2021dim}%
  \BibitemOpen
  \bibfield  {author} {\bibinfo {author} {\bibfnamefont {A.~S.}\ \bibnamefont {Sakharov}}, \bibinfo {author} {\bibfnamefont {Y.~N.}\ \bibnamefont {Eroshenko}}, \ and\ \bibinfo {author} {\bibfnamefont {S.~G.}\ \bibnamefont {Rubin}},\ }\href {\doibase 10.1103/PhysRevD.104.043005} {\bibfield  {journal} {\bibinfo  {journal} {Phys. Rev. D}\ }\textbf {\bibinfo {volume} {104}},\ \bibinfo {pages} {043005} (\bibinfo {year} {2021})},\ \Eprint {http://arxiv.org/abs/2104.08750} {arXiv:2104.08750 [hep-ph]} \BibitemShut {NoStop}%
\bibitem [{\citenamefont {Zhang}\ \emph {et~al.}(2023)\citenamefont {Zhang}, \citenamefont {Cai}, \citenamefont {Su}, \citenamefont {Wang}, \citenamefont {Yu},\ and\ \citenamefont {Zhang}}]{Zhang:2023nrs}%
  \BibitemOpen
  \bibfield  {author} {\bibinfo {author} {\bibfnamefont {Z.}~\bibnamefont {Zhang}}, \bibinfo {author} {\bibfnamefont {C.}~\bibnamefont {Cai}}, \bibinfo {author} {\bibfnamefont {Y.-H.}\ \bibnamefont {Su}}, \bibinfo {author} {\bibfnamefont {S.}~\bibnamefont {Wang}}, \bibinfo {author} {\bibfnamefont {Z.-H.}\ \bibnamefont {Yu}}, \ and\ \bibinfo {author} {\bibfnamefont {H.-H.}\ \bibnamefont {Zhang}},\ }\href {\doibase 10.1103/PhysRevD.108.095037} {\bibfield  {journal} {\bibinfo  {journal} {Phys. Rev. D}\ }\textbf {\bibinfo {volume} {108}},\ \bibinfo {pages} {095037} (\bibinfo {year} {2023})},\ \Eprint {http://arxiv.org/abs/2307.11495} {arXiv:2307.11495 [hep-ph]} \BibitemShut {NoStop}%
\bibitem [{\citenamefont {Gouttenoire}\ and\ \citenamefont {Vitagliano}(2024)}]{Gouttenoire:2023ftk}%
  \BibitemOpen
  \bibfield  {author} {\bibinfo {author} {\bibfnamefont {Y.}~\bibnamefont {Gouttenoire}}\ and\ \bibinfo {author} {\bibfnamefont {E.}~\bibnamefont {Vitagliano}},\ }\href {\doibase 10.1103/PhysRevD.110.L061306} {\bibfield  {journal} {\bibinfo  {journal} {Phys. Rev. D}\ }\textbf {\bibinfo {volume} {110}},\ \bibinfo {pages} {L061306} (\bibinfo {year} {2024})},\ \Eprint {http://arxiv.org/abs/2306.17841} {arXiv:2306.17841 [gr-qc]} \BibitemShut {NoStop}%
\bibitem [{\citenamefont {Afzal}\ \emph {et~al.}(2023)\citenamefont {Afzal} \emph {et~al.}}]{NANOGrav:2023hvm}%
  \BibitemOpen
  \bibfield  {author} {\bibinfo {author} {\bibfnamefont {A.}~\bibnamefont {Afzal}} \emph {et~al.} (\bibinfo {collaboration} {NANOGrav}),\ }\href {\doibase 10.3847/2041-8213/acdc91} {\bibfield  {journal} {\bibinfo  {journal} {Astrophys. J. Lett.}\ }\textbf {\bibinfo {volume} {951}},\ \bibinfo {pages} {L11} (\bibinfo {year} {2023})},\ \Eprint {http://arxiv.org/abs/2306.16219} {arXiv:2306.16219 [astro-ph.HE]} \BibitemShut {NoStop}%
\bibitem [{\citenamefont {Bian}\ \emph {et~al.}(2024)\citenamefont {Bian}, \citenamefont {Ge}, \citenamefont {Shu}, \citenamefont {Wang}, \citenamefont {Yang},\ and\ \citenamefont {Zong}}]{Bian:2023dnv}%
  \BibitemOpen
  \bibfield  {author} {\bibinfo {author} {\bibfnamefont {L.}~\bibnamefont {Bian}}, \bibinfo {author} {\bibfnamefont {S.}~\bibnamefont {Ge}}, \bibinfo {author} {\bibfnamefont {J.}~\bibnamefont {Shu}}, \bibinfo {author} {\bibfnamefont {B.}~\bibnamefont {Wang}}, \bibinfo {author} {\bibfnamefont {X.-Y.}\ \bibnamefont {Yang}}, \ and\ \bibinfo {author} {\bibfnamefont {J.}~\bibnamefont {Zong}},\ }\href {\doibase 10.1103/PhysRevD.109.L101301} {\bibfield  {journal} {\bibinfo  {journal} {Phys. Rev. D}\ }\textbf {\bibinfo {volume} {109}},\ \bibinfo {pages} {L101301} (\bibinfo {year} {2024})},\ \Eprint {http://arxiv.org/abs/2307.02376} {arXiv:2307.02376 [astro-ph.HE]} \BibitemShut {NoStop}%
\bibitem [{\citenamefont {Bai}\ and\ \citenamefont {Korwar}(2022)}]{Bai:2021ibt}%
  \BibitemOpen
  \bibfield  {author} {\bibinfo {author} {\bibfnamefont {Y.}~\bibnamefont {Bai}}\ and\ \bibinfo {author} {\bibfnamefont {M.}~\bibnamefont {Korwar}},\ }\href {\doibase 10.1103/PhysRevD.105.095015} {\bibfield  {journal} {\bibinfo  {journal} {Phys. Rev. D}\ }\textbf {\bibinfo {volume} {105}},\ \bibinfo {pages} {095015} (\bibinfo {year} {2022})},\ \Eprint {http://arxiv.org/abs/2109.14765} {arXiv:2109.14765 [hep-ph]} \BibitemShut {NoStop}%
\bibitem [{\citenamefont {Gao}\ and\ \citenamefont {Oldengott}(2022)}]{Gao:2021nwz}%
  \BibitemOpen
  \bibfield  {author} {\bibinfo {author} {\bibfnamefont {F.}~\bibnamefont {Gao}}\ and\ \bibinfo {author} {\bibfnamefont {I.~M.}\ \bibnamefont {Oldengott}},\ }\href {\doibase 10.1103/PhysRevLett.128.131301} {\bibfield  {journal} {\bibinfo  {journal} {Phys. Rev. Lett.}\ }\textbf {\bibinfo {volume} {128}},\ \bibinfo {pages} {131301} (\bibinfo {year} {2022})},\ \Eprint {http://arxiv.org/abs/2106.11991} {arXiv:2106.11991 [hep-ph]} \BibitemShut {NoStop}%
\bibitem [{\citenamefont {Gao}\ \emph {et~al.}(2023)\citenamefont {Gao}, \citenamefont {Harz}, \citenamefont {Hati}, \citenamefont {Lu}, \citenamefont {Oldengott},\ and\ \citenamefont {White}}]{Gao:2023djs}%
  \BibitemOpen
  \bibfield  {author} {\bibinfo {author} {\bibfnamefont {F.}~\bibnamefont {Gao}}, \bibinfo {author} {\bibfnamefont {J.}~\bibnamefont {Harz}}, \bibinfo {author} {\bibfnamefont {C.}~\bibnamefont {Hati}}, \bibinfo {author} {\bibfnamefont {Y.}~\bibnamefont {Lu}}, \bibinfo {author} {\bibfnamefont {I.~M.}\ \bibnamefont {Oldengott}}, \ and\ \bibinfo {author} {\bibfnamefont {G.}~\bibnamefont {White}},\ }\href@noop {} {\  (\bibinfo {year} {2023})},\ \Eprint {http://arxiv.org/abs/2309.00672} {arXiv:2309.00672 [hep-ph]} \BibitemShut {NoStop}%
\bibitem [{\citenamefont {Lu}\ \emph {et~al.}(2024)\citenamefont {Lu}, \citenamefont {Gao}, \citenamefont {Liu},\ and\ \citenamefont {Pawlowski}}]{Lu:2023mkn}%
  \BibitemOpen
  \bibfield  {author} {\bibinfo {author} {\bibfnamefont {Y.}~\bibnamefont {Lu}}, \bibinfo {author} {\bibfnamefont {F.}~\bibnamefont {Gao}}, \bibinfo {author} {\bibfnamefont {Y.-X.}\ \bibnamefont {Liu}}, \ and\ \bibinfo {author} {\bibfnamefont {J.~M.}\ \bibnamefont {Pawlowski}},\ }\href {\doibase 10.1103/PhysRevD.110.014036} {\bibfield  {journal} {\bibinfo  {journal} {Phys. Rev. D}\ }\textbf {\bibinfo {volume} {110}},\ \bibinfo {pages} {014036} (\bibinfo {year} {2024})},\ \Eprint {http://arxiv.org/abs/2310.18383} {arXiv:2310.18383 [hep-ph]} \BibitemShut {NoStop}%
\bibitem [{\citenamefont {Zheng}\ \emph {et~al.}(2024)\citenamefont {Zheng}, \citenamefont {Lu}, \citenamefont {Gao}, \citenamefont {Qin},\ and\ \citenamefont {Liu}}]{Zheng:2023tbv}%
  \BibitemOpen
  \bibfield  {author} {\bibinfo {author} {\bibfnamefont {H.~W.}\ \bibnamefont {Zheng}}, \bibinfo {author} {\bibfnamefont {Y.}~\bibnamefont {Lu}}, \bibinfo {author} {\bibfnamefont {F.}~\bibnamefont {Gao}}, \bibinfo {author} {\bibfnamefont {S.~X.}\ \bibnamefont {Qin}}, \ and\ \bibinfo {author} {\bibfnamefont {Y.~X.}\ \bibnamefont {Liu}},\ }\href {\doibase 10.1103/PhysRevD.109.114013} {\bibfield  {journal} {\bibinfo  {journal} {Phys. Rev. D}\ }\textbf {\bibinfo {volume} {109}},\ \bibinfo {pages} {114013} (\bibinfo {year} {2024})},\ \Eprint {http://arxiv.org/abs/2312.00382} {arXiv:2312.00382 [hep-ph]} \BibitemShut {NoStop}%
\bibitem [{\citenamefont {Roberts}\ and\ \citenamefont {Williams}(1994)}]{Roberts:1994dr}%
  \BibitemOpen
  \bibfield  {author} {\bibinfo {author} {\bibfnamefont {C.~D.}\ \bibnamefont {Roberts}}\ and\ \bibinfo {author} {\bibfnamefont {A.~G.}\ \bibnamefont {Williams}},\ }\href {\doibase 10.1016/0146-6410(94)90049-3} {\bibfield  {journal} {\bibinfo  {journal} {Prog. Part. Nucl. Phys.}\ }\textbf {\bibinfo {volume} {33}},\ \bibinfo {pages} {477} (\bibinfo {year} {1994})},\ \Eprint {http://arxiv.org/abs/hep-ph/9403224} {arXiv:hep-ph/9403224} \BibitemShut {NoStop}%
\bibitem [{\citenamefont {Alkofer}\ and\ \citenamefont {von Smekal}(2001)}]{Alkofer:2000wg}%
  \BibitemOpen
  \bibfield  {author} {\bibinfo {author} {\bibfnamefont {R.}~\bibnamefont {Alkofer}}\ and\ \bibinfo {author} {\bibfnamefont {L.}~\bibnamefont {von Smekal}},\ }\href {\doibase 10.1016/S0370-1573(01)00010-2} {\bibfield  {journal} {\bibinfo  {journal} {Phys. Rept.}\ }\textbf {\bibinfo {volume} {353}},\ \bibinfo {pages} {281} (\bibinfo {year} {2001})},\ \Eprint {http://arxiv.org/abs/hep-ph/0007355} {arXiv:hep-ph/0007355} \BibitemShut {NoStop}%
\bibitem [{\citenamefont {Fischer}(2006)}]{Fischer:2006ub}%
  \BibitemOpen
  \bibfield  {author} {\bibinfo {author} {\bibfnamefont {C.~S.}\ \bibnamefont {Fischer}},\ }\href {\doibase 10.1088/0954-3899/32/8/R02} {\bibfield  {journal} {\bibinfo  {journal} {J. Phys. G}\ }\textbf {\bibinfo {volume} {32}},\ \bibinfo {pages} {R253} (\bibinfo {year} {2006})},\ \Eprint {http://arxiv.org/abs/hep-ph/0605173} {arXiv:hep-ph/0605173} \BibitemShut {NoStop}%
\bibitem [{\citenamefont {Gao}\ \emph {et~al.}(2021)\citenamefont {Gao}, \citenamefont {Papavassiliou},\ and\ \citenamefont {Pawlowski}}]{Gao:2021wun}%
  \BibitemOpen
  \bibfield  {author} {\bibinfo {author} {\bibfnamefont {F.}~\bibnamefont {Gao}}, \bibinfo {author} {\bibfnamefont {J.}~\bibnamefont {Papavassiliou}}, \ and\ \bibinfo {author} {\bibfnamefont {J.~M.}\ \bibnamefont {Pawlowski}},\ }\href {\doibase 10.1103/PhysRevD.103.094013} {\bibfield  {journal} {\bibinfo  {journal} {Phys. Rev. D}\ }\textbf {\bibinfo {volume} {103}},\ \bibinfo {pages} {094013} (\bibinfo {year} {2021})},\ \Eprint {http://arxiv.org/abs/2102.13053} {arXiv:2102.13053 [hep-ph]} \BibitemShut {NoStop}%
\bibitem [{\citenamefont {Schwarz}\ and\ \citenamefont {Stuke}(2009)}]{Schwarz:2009ii}%
  \BibitemOpen
  \bibfield  {author} {\bibinfo {author} {\bibfnamefont {D.~J.}\ \bibnamefont {Schwarz}}\ and\ \bibinfo {author} {\bibfnamefont {M.}~\bibnamefont {Stuke}},\ }\href {\doibase 10.1088/1475-7516/2009/11/025} {\bibfield  {journal} {\bibinfo  {journal} {JCAP}\ }\textbf {\bibinfo {volume} {11}},\ \bibinfo {pages} {025} (\bibinfo {year} {2009})},\ \bibinfo {note} {[Erratum: JCAP 10, E01 (2010)]},\ \Eprint {http://arxiv.org/abs/0906.3434} {arXiv:0906.3434 [hep-ph]} \BibitemShut {NoStop}%
\bibitem [{\citenamefont {Aghanim}\ \emph {et~al.}(2020)\citenamefont {Aghanim} \emph {et~al.}}]{Planck:2018vyg}%
  \BibitemOpen
  \bibfield  {author} {\bibinfo {author} {\bibfnamefont {N.}~\bibnamefont {Aghanim}} \emph {et~al.} (\bibinfo {collaboration} {Planck}),\ }\href {\doibase 10.1051/0004-6361/201833910} {\bibfield  {journal} {\bibinfo  {journal} {Astron. Astrophys.}\ }\textbf {\bibinfo {volume} {641}},\ \bibinfo {pages} {A6} (\bibinfo {year} {2020})},\ \bibinfo {note} {[Erratum: Astron.Astrophys. 652, C4 (2021)]},\ \Eprint {http://arxiv.org/abs/1807.06209} {arXiv:1807.06209 [astro-ph.CO]} \BibitemShut {NoStop}%
\bibitem [{\citenamefont {Pitrou}\ \emph {et~al.}(2018)\citenamefont {Pitrou}, \citenamefont {Coc}, \citenamefont {Uzan},\ and\ \citenamefont {Vangioni}}]{Pitrou:2018cgg}%
  \BibitemOpen
  \bibfield  {author} {\bibinfo {author} {\bibfnamefont {C.}~\bibnamefont {Pitrou}}, \bibinfo {author} {\bibfnamefont {A.}~\bibnamefont {Coc}}, \bibinfo {author} {\bibfnamefont {J.~P.}\ \bibnamefont {Uzan}}, \ and\ \bibinfo {author} {\bibfnamefont {E.}~\bibnamefont {Vangioni}},\ }\href {\doibase 10.1016/j.physrep.2018.04.005} {\bibfield  {journal} {\bibinfo  {journal} {Phys. Rept.}\ }\textbf {\bibinfo {volume} {754}},\ \bibinfo {pages} {1} (\bibinfo {year} {2018})},\ \Eprint {http://arxiv.org/abs/1801.08023} {arXiv:1801.08023 [astro-ph.CO]} \BibitemShut {NoStop}%
\bibitem [{\citenamefont {Oldengott}\ and\ \citenamefont {Schwarz}(2017)}]{Oldengott:2017tzj}%
  \BibitemOpen
  \bibfield  {author} {\bibinfo {author} {\bibfnamefont {I.~M.}\ \bibnamefont {Oldengott}}\ and\ \bibinfo {author} {\bibfnamefont {D.~J.}\ \bibnamefont {Schwarz}},\ }\href {\doibase 10.1209/0295-5075/119/29001} {\bibfield  {journal} {\bibinfo  {journal} {EPL}\ }\textbf {\bibinfo {volume} {119}},\ \bibinfo {pages} {29001} (\bibinfo {year} {2017})},\ \Eprint {http://arxiv.org/abs/1706.01705} {arXiv:1706.01705 [astro-ph.CO]} \BibitemShut {NoStop}%
\bibitem [{\citenamefont {Popa}\ and\ \citenamefont {Vasile}(2008)}]{Popa:2008tb}%
  \BibitemOpen
  \bibfield  {author} {\bibinfo {author} {\bibfnamefont {L.~A.}\ \bibnamefont {Popa}}\ and\ \bibinfo {author} {\bibfnamefont {A.}~\bibnamefont {Vasile}},\ }\href {\doibase 10.1088/1475-7516/2008/06/028} {\bibfield  {journal} {\bibinfo  {journal} {JCAP}\ }\textbf {\bibinfo {volume} {06}},\ \bibinfo {pages} {028} (\bibinfo {year} {2008})},\ \Eprint {http://arxiv.org/abs/0804.2971} {arXiv:0804.2971 [astro-ph]} \BibitemShut {NoStop}%
\bibitem [{\citenamefont {Simha}\ and\ \citenamefont {Steigman}(2008)}]{Simha:2008mt}%
  \BibitemOpen
  \bibfield  {author} {\bibinfo {author} {\bibfnamefont {V.}~\bibnamefont {Simha}}\ and\ \bibinfo {author} {\bibfnamefont {G.}~\bibnamefont {Steigman}},\ }\href {\doibase 10.1088/1475-7516/2008/08/011} {\bibfield  {journal} {\bibinfo  {journal} {JCAP}\ }\textbf {\bibinfo {volume} {08}},\ \bibinfo {pages} {011} (\bibinfo {year} {2008})},\ \Eprint {http://arxiv.org/abs/0806.0179} {arXiv:0806.0179 [hep-ph]} \BibitemShut {NoStop}%
\bibitem [{\citenamefont {Serpico}\ and\ \citenamefont {Raffelt}(2005)}]{Serpico:2005bc}%
  \BibitemOpen
  \bibfield  {author} {\bibinfo {author} {\bibfnamefont {P.~D.}\ \bibnamefont {Serpico}}\ and\ \bibinfo {author} {\bibfnamefont {G.~G.}\ \bibnamefont {Raffelt}},\ }\href {\doibase 10.1103/PhysRevD.71.127301} {\bibfield  {journal} {\bibinfo  {journal} {Phys. Rev. D}\ }\textbf {\bibinfo {volume} {71}},\ \bibinfo {pages} {127301} (\bibinfo {year} {2005})},\ \Eprint {http://arxiv.org/abs/astro-ph/0506162} {arXiv:astro-ph/0506162} \BibitemShut {NoStop}%
\bibitem [{\citenamefont {Gelmini}\ \emph {et~al.}(2020)\citenamefont {Gelmini}, \citenamefont {Kawasaki}, \citenamefont {Kusenko}, \citenamefont {Murai},\ and\ \citenamefont {Takhistov}}]{Gelmini:2020ekg}%
  \BibitemOpen
  \bibfield  {author} {\bibinfo {author} {\bibfnamefont {G.~B.}\ \bibnamefont {Gelmini}}, \bibinfo {author} {\bibfnamefont {M.}~\bibnamefont {Kawasaki}}, \bibinfo {author} {\bibfnamefont {A.}~\bibnamefont {Kusenko}}, \bibinfo {author} {\bibfnamefont {K.}~\bibnamefont {Murai}}, \ and\ \bibinfo {author} {\bibfnamefont {V.}~\bibnamefont {Takhistov}},\ }\href {\doibase 10.1088/1475-7516/2020/09/051} {\bibfield  {journal} {\bibinfo  {journal} {JCAP}\ }\textbf {\bibinfo {volume} {09}},\ \bibinfo {pages} {051} (\bibinfo {year} {2020})},\ \Eprint {http://arxiv.org/abs/2005.06721} {arXiv:2005.06721 [hep-ph]} \BibitemShut {NoStop}%
\bibitem [{\citenamefont {Pastor}\ \emph {et~al.}(2009)\citenamefont {Pastor}, \citenamefont {Pinto},\ and\ \citenamefont {Raffelt}}]{Pastor:2008ti}%
  \BibitemOpen
  \bibfield  {author} {\bibinfo {author} {\bibfnamefont {S.}~\bibnamefont {Pastor}}, \bibinfo {author} {\bibfnamefont {T.}~\bibnamefont {Pinto}}, \ and\ \bibinfo {author} {\bibfnamefont {G.~G.}\ \bibnamefont {Raffelt}},\ }\href {\doibase 10.1103/PhysRevLett.102.241302} {\bibfield  {journal} {\bibinfo  {journal} {Phys. Rev. Lett.}\ }\textbf {\bibinfo {volume} {102}},\ \bibinfo {pages} {241302} (\bibinfo {year} {2009})},\ \Eprint {http://arxiv.org/abs/0808.3137} {arXiv:0808.3137 [astro-ph]} \BibitemShut {NoStop}%
\bibitem [{\citenamefont {Mangano}\ \emph {et~al.}(2011)\citenamefont {Mangano}, \citenamefont {Miele}, \citenamefont {Pastor}, \citenamefont {Pisanti},\ and\ \citenamefont {Sarikas}}]{Mangano:2010ei}%
  \BibitemOpen
  \bibfield  {author} {\bibinfo {author} {\bibfnamefont {G.}~\bibnamefont {Mangano}}, \bibinfo {author} {\bibfnamefont {G.}~\bibnamefont {Miele}}, \bibinfo {author} {\bibfnamefont {S.}~\bibnamefont {Pastor}}, \bibinfo {author} {\bibfnamefont {O.}~\bibnamefont {Pisanti}}, \ and\ \bibinfo {author} {\bibfnamefont {S.}~\bibnamefont {Sarikas}},\ }\href {\doibase 10.1088/1475-7516/2011/03/035} {\bibfield  {journal} {\bibinfo  {journal} {JCAP}\ }\textbf {\bibinfo {volume} {03}},\ \bibinfo {pages} {035} (\bibinfo {year} {2011})},\ \Eprint {http://arxiv.org/abs/1011.0916} {arXiv:1011.0916 [astro-ph.CO]} \BibitemShut {NoStop}%
\bibitem [{\citenamefont {Mangano}\ \emph {et~al.}(2012)\citenamefont {Mangano}, \citenamefont {Miele}, \citenamefont {Pastor}, \citenamefont {Pisanti},\ and\ \citenamefont {Sarikas}}]{Mangano:2011ip}%
  \BibitemOpen
  \bibfield  {author} {\bibinfo {author} {\bibfnamefont {G.}~\bibnamefont {Mangano}}, \bibinfo {author} {\bibfnamefont {G.}~\bibnamefont {Miele}}, \bibinfo {author} {\bibfnamefont {S.}~\bibnamefont {Pastor}}, \bibinfo {author} {\bibfnamefont {O.}~\bibnamefont {Pisanti}}, \ and\ \bibinfo {author} {\bibfnamefont {S.}~\bibnamefont {Sarikas}},\ }\href {\doibase 10.1016/j.physletb.2012.01.015} {\bibfield  {journal} {\bibinfo  {journal} {Phys. Lett. B}\ }\textbf {\bibinfo {volume} {708}},\ \bibinfo {pages} {1} (\bibinfo {year} {2012})},\ \Eprint {http://arxiv.org/abs/1110.4335} {arXiv:1110.4335 [hep-ph]} \BibitemShut {NoStop}%
\bibitem [{\citenamefont {Middeldorf-Wygas}\ \emph {et~al.}(2022)\citenamefont {Middeldorf-Wygas}, \citenamefont {Oldengott}, \citenamefont {B\"odeker},\ and\ \citenamefont {Schwarz}}]{Middeldorf-Wygas:2020glx}%
  \BibitemOpen
  \bibfield  {author} {\bibinfo {author} {\bibfnamefont {M.~M.}\ \bibnamefont {Middeldorf-Wygas}}, \bibinfo {author} {\bibfnamefont {I.~M.}\ \bibnamefont {Oldengott}}, \bibinfo {author} {\bibfnamefont {D.}~\bibnamefont {B\"odeker}}, \ and\ \bibinfo {author} {\bibfnamefont {D.~J.}\ \bibnamefont {Schwarz}},\ }\href {\doibase 10.1103/PhysRevD.105.123533} {\bibfield  {journal} {\bibinfo  {journal} {Phys. Rev. D}\ }\textbf {\bibinfo {volume} {105}},\ \bibinfo {pages} {123533} (\bibinfo {year} {2022})},\ \Eprint {http://arxiv.org/abs/2009.00036} {arXiv:2009.00036 [hep-ph]} \BibitemShut {NoStop}%
\bibitem [{\citenamefont {Franciolini}\ \emph {et~al.}(2024)\citenamefont {Franciolini}, \citenamefont {Racco},\ and\ \citenamefont {Rompineve}}]{Franciolini:2023wjm}%
  \BibitemOpen
  \bibfield  {author} {\bibinfo {author} {\bibfnamefont {G.}~\bibnamefont {Franciolini}}, \bibinfo {author} {\bibfnamefont {D.}~\bibnamefont {Racco}}, \ and\ \bibinfo {author} {\bibfnamefont {F.}~\bibnamefont {Rompineve}},\ }\href {\doibase 10.1103/PhysRevLett.132.081001} {\bibfield  {journal} {\bibinfo  {journal} {Phys. Rev. Lett.}\ }\textbf {\bibinfo {volume} {132}},\ \bibinfo {pages} {081001} (\bibinfo {year} {2024})},\ \Eprint {http://arxiv.org/abs/2306.17136} {arXiv:2306.17136 [astro-ph.CO]} \BibitemShut {NoStop}%
\bibitem [{\citenamefont {Jinno}\ and\ \citenamefont {Takimoto}(2017)}]{Jinno:2016vai}%
  \BibitemOpen
  \bibfield  {author} {\bibinfo {author} {\bibfnamefont {R.}~\bibnamefont {Jinno}}\ and\ \bibinfo {author} {\bibfnamefont {M.}~\bibnamefont {Takimoto}},\ }\href {\doibase 10.1103/PhysRevD.95.024009} {\bibfield  {journal} {\bibinfo  {journal} {Phys. Rev. D}\ }\textbf {\bibinfo {volume} {95}},\ \bibinfo {pages} {024009} (\bibinfo {year} {2017})},\ \Eprint {http://arxiv.org/abs/1605.01403} {arXiv:1605.01403 [astro-ph.CO]} \BibitemShut {NoStop}%
\bibitem [{\citenamefont {Hindmarsh}\ \emph {et~al.}(2015)\citenamefont {Hindmarsh}, \citenamefont {Huber}, \citenamefont {Rummukainen},\ and\ \citenamefont {Weir}}]{Hindmarsh:2015qta}%
  \BibitemOpen
  \bibfield  {author} {\bibinfo {author} {\bibfnamefont {M.}~\bibnamefont {Hindmarsh}}, \bibinfo {author} {\bibfnamefont {S.~J.}\ \bibnamefont {Huber}}, \bibinfo {author} {\bibfnamefont {K.}~\bibnamefont {Rummukainen}}, \ and\ \bibinfo {author} {\bibfnamefont {D.~J.}\ \bibnamefont {Weir}},\ }\href {\doibase 10.1103/PhysRevD.92.123009} {\bibfield  {journal} {\bibinfo  {journal} {Phys. Rev. D}\ }\textbf {\bibinfo {volume} {92}},\ \bibinfo {pages} {123009} (\bibinfo {year} {2015})},\ \Eprint {http://arxiv.org/abs/1504.03291} {arXiv:1504.03291 [astro-ph.CO]} \BibitemShut {NoStop}%
\bibitem [{\citenamefont {Espinosa}\ \emph {et~al.}(2010)\citenamefont {Espinosa}, \citenamefont {Konstandin}, \citenamefont {No},\ and\ \citenamefont {Servant}}]{Espinosa:2010hh}%
  \BibitemOpen
  \bibfield  {author} {\bibinfo {author} {\bibfnamefont {J.~R.}\ \bibnamefont {Espinosa}}, \bibinfo {author} {\bibfnamefont {T.}~\bibnamefont {Konstandin}}, \bibinfo {author} {\bibfnamefont {J.~M.}\ \bibnamefont {No}}, \ and\ \bibinfo {author} {\bibfnamefont {G.}~\bibnamefont {Servant}},\ }\href {\doibase 10.1088/1475-7516/2010/06/028} {\bibfield  {journal} {\bibinfo  {journal} {JCAP}\ }\textbf {\bibinfo {volume} {06}},\ \bibinfo {pages} {028} (\bibinfo {year} {2010})},\ \Eprint {http://arxiv.org/abs/1004.4187} {arXiv:1004.4187 [hep-ph]} \BibitemShut {NoStop}%
\bibitem [{\citenamefont {Ellis}\ \emph {et~al.}(2020)\citenamefont {Ellis}, \citenamefont {Lewicki},\ and\ \citenamefont {No}}]{Ellis:2020awk}%
  \BibitemOpen
  \bibfield  {author} {\bibinfo {author} {\bibfnamefont {J.}~\bibnamefont {Ellis}}, \bibinfo {author} {\bibfnamefont {M.}~\bibnamefont {Lewicki}}, \ and\ \bibinfo {author} {\bibfnamefont {J.~M.}\ \bibnamefont {No}},\ }\href {\doibase 10.1088/1475-7516/2020/07/050} {\bibfield  {journal} {\bibinfo  {journal} {JCAP}\ }\textbf {\bibinfo {volume} {07}},\ \bibinfo {pages} {050} (\bibinfo {year} {2020})},\ \Eprint {http://arxiv.org/abs/2003.07360} {arXiv:2003.07360 [hep-ph]} \BibitemShut {NoStop}%
\bibitem [{\citenamefont {Ellis}\ \emph {et~al.}(2019{\natexlab{a}})\citenamefont {Ellis}, \citenamefont {Lewicki}, \citenamefont {No},\ and\ \citenamefont {Vaskonen}}]{Ellis:2019oqb}%
  \BibitemOpen
  \bibfield  {author} {\bibinfo {author} {\bibfnamefont {J.}~\bibnamefont {Ellis}}, \bibinfo {author} {\bibfnamefont {M.}~\bibnamefont {Lewicki}}, \bibinfo {author} {\bibfnamefont {J.~M.}\ \bibnamefont {No}}, \ and\ \bibinfo {author} {\bibfnamefont {V.}~\bibnamefont {Vaskonen}},\ }\href {\doibase 10.1088/1475-7516/2019/06/024} {\bibfield  {journal} {\bibinfo  {journal} {JCAP}\ }\textbf {\bibinfo {volume} {06}},\ \bibinfo {pages} {024} (\bibinfo {year} {2019}{\natexlab{a}})},\ \Eprint {http://arxiv.org/abs/1903.09642} {arXiv:1903.09642 [hep-ph]} \BibitemShut {NoStop}%
\bibitem [{\citenamefont {Guo}\ \emph {et~al.}(2021)\citenamefont {Guo}, \citenamefont {Sinha}, \citenamefont {Vagie},\ and\ \citenamefont {White}}]{Guo:2020grp}%
  \BibitemOpen
  \bibfield  {author} {\bibinfo {author} {\bibfnamefont {H.~K.}\ \bibnamefont {Guo}}, \bibinfo {author} {\bibfnamefont {K.}~\bibnamefont {Sinha}}, \bibinfo {author} {\bibfnamefont {D.}~\bibnamefont {Vagie}}, \ and\ \bibinfo {author} {\bibfnamefont {G.}~\bibnamefont {White}},\ }\href {\doibase 10.1088/1475-7516/2021/01/001} {\bibfield  {journal} {\bibinfo  {journal} {JCAP}\ }\textbf {\bibinfo {volume} {01}},\ \bibinfo {pages} {001} (\bibinfo {year} {2021})},\ \Eprint {http://arxiv.org/abs/2007.08537} {arXiv:2007.08537 [hep-ph]} \BibitemShut {NoStop}%
\bibitem [{\citenamefont {Weir}(2018)}]{Weir:2017wfa}%
  \BibitemOpen
  \bibfield  {author} {\bibinfo {author} {\bibfnamefont {D.~J.}\ \bibnamefont {Weir}},\ }\href {\doibase 10.1098/rsta.2017.0126} {\bibfield  {journal} {\bibinfo  {journal} {Phil. Trans. Roy. Soc. Lond. A}\ }\textbf {\bibinfo {volume} {376}},\ \bibinfo {pages} {20170126} (\bibinfo {year} {2018})},\ \bibinfo {note} {[Erratum: Phil.Trans.Roy.Soc.Lond.A 381, 20230212 (2023)]},\ \Eprint {http://arxiv.org/abs/1705.01783} {arXiv:1705.01783 [hep-ph]} \BibitemShut {NoStop}%
\bibitem [{\citenamefont {Haymaker}\ \emph {et~al.}(1987)\citenamefont {Haymaker}, \citenamefont {Matsuki},\ and\ \citenamefont {Cooper}}]{Haymaker:1986tt}%
  \BibitemOpen
  \bibfield  {author} {\bibinfo {author} {\bibfnamefont {R.~W.}\ \bibnamefont {Haymaker}}, \bibinfo {author} {\bibfnamefont {T.}~\bibnamefont {Matsuki}}, \ and\ \bibinfo {author} {\bibfnamefont {F.}~\bibnamefont {Cooper}},\ }\href {\doibase 10.1103/PhysRevD.35.2567} {\bibfield  {journal} {\bibinfo  {journal} {Phys. Rev. D}\ }\textbf {\bibinfo {volume} {35}},\ \bibinfo {pages} {2567} (\bibinfo {year} {1987})}\BibitemShut {NoStop}%
\bibitem [{\citenamefont {Ekstedt}\ \emph {et~al.}(2023)\citenamefont {Ekstedt}, \citenamefont {Gould},\ and\ \citenamefont {Hirvonen}}]{Ekstedt:2023sqc}%
  \BibitemOpen
  \bibfield  {author} {\bibinfo {author} {\bibfnamefont {A.}~\bibnamefont {Ekstedt}}, \bibinfo {author} {\bibfnamefont {O.}~\bibnamefont {Gould}}, \ and\ \bibinfo {author} {\bibfnamefont {J.}~\bibnamefont {Hirvonen}},\ }\href {\doibase 10.1007/JHEP12(2023)056} {\bibfield  {journal} {\bibinfo  {journal} {JHEP}\ }\textbf {\bibinfo {volume} {12}},\ \bibinfo {pages} {056} (\bibinfo {year} {2023})},\ \Eprint {http://arxiv.org/abs/2308.15652} {arXiv:2308.15652 [hep-ph]} \BibitemShut {NoStop}%
\bibitem [{\citenamefont {Ai}\ \emph {et~al.}(2023)\citenamefont {Ai}, \citenamefont {Laurent},\ and\ \citenamefont {van~de Vis}}]{Ai:2023see}%
  \BibitemOpen
  \bibfield  {author} {\bibinfo {author} {\bibfnamefont {W.-Y.}\ \bibnamefont {Ai}}, \bibinfo {author} {\bibfnamefont {B.}~\bibnamefont {Laurent}}, \ and\ \bibinfo {author} {\bibfnamefont {J.}~\bibnamefont {van~de Vis}},\ }\href {\doibase 10.1088/1475-7516/2023/07/002} {\bibfield  {journal} {\bibinfo  {journal} {JCAP}\ }\textbf {\bibinfo {volume} {07}},\ \bibinfo {pages} {002} (\bibinfo {year} {2023})},\ \Eprint {http://arxiv.org/abs/2303.10171} {arXiv:2303.10171 [astro-ph.CO]} \BibitemShut {NoStop}%
\bibitem [{\citenamefont {Krajewski}\ \emph {et~al.}(2024)\citenamefont {Krajewski}, \citenamefont {Lewicki},\ and\ \citenamefont {Zych}}]{Krajewski:2024gma}%
  \BibitemOpen
  \bibfield  {author} {\bibinfo {author} {\bibfnamefont {T.}~\bibnamefont {Krajewski}}, \bibinfo {author} {\bibfnamefont {M.}~\bibnamefont {Lewicki}}, \ and\ \bibinfo {author} {\bibfnamefont {M.}~\bibnamefont {Zych}},\ }\href {\doibase 10.1007/JHEP05(2024)011} {\bibfield  {journal} {\bibinfo  {journal} {JHEP}\ }\textbf {\bibinfo {volume} {05}},\ \bibinfo {pages} {011} (\bibinfo {year} {2024})},\ \Eprint {http://arxiv.org/abs/2402.15408} {arXiv:2402.15408 [astro-ph.CO]} \BibitemShut {NoStop}%
\bibitem [{\citenamefont {Giese}\ \emph {et~al.}(2020)\citenamefont {Giese}, \citenamefont {Konstandin},\ and\ \citenamefont {van~de Vis}}]{Giese:2020rtr}%
  \BibitemOpen
  \bibfield  {author} {\bibinfo {author} {\bibfnamefont {F.}~\bibnamefont {Giese}}, \bibinfo {author} {\bibfnamefont {T.}~\bibnamefont {Konstandin}}, \ and\ \bibinfo {author} {\bibfnamefont {J.}~\bibnamefont {van~de Vis}},\ }\href {\doibase 10.1088/1475-7516/2020/07/057} {\bibfield  {journal} {\bibinfo  {journal} {JCAP}\ }\textbf {\bibinfo {volume} {07}},\ \bibinfo {pages} {057} (\bibinfo {year} {2020})},\ \Eprint {http://arxiv.org/abs/2004.06995} {arXiv:2004.06995 [astro-ph.CO]} \BibitemShut {NoStop}%
\bibitem [{\citenamefont {Bodeker}\ and\ \citenamefont {Moore}(2009)}]{Bodeker:2009qy}%
  \BibitemOpen
  \bibfield  {author} {\bibinfo {author} {\bibfnamefont {D.}~\bibnamefont {Bodeker}}\ and\ \bibinfo {author} {\bibfnamefont {G.~D.}\ \bibnamefont {Moore}},\ }\href {\doibase 10.1088/1475-7516/2009/05/009} {\bibfield  {journal} {\bibinfo  {journal} {JCAP}\ }\textbf {\bibinfo {volume} {05}},\ \bibinfo {pages} {009} (\bibinfo {year} {2009})},\ \Eprint {http://arxiv.org/abs/0903.4099} {arXiv:0903.4099 [hep-ph]} \BibitemShut {NoStop}%
\bibitem [{\citenamefont {Reardon}\ \emph {et~al.}(2023{\natexlab{b}})\citenamefont {Reardon} \emph {et~al.}}]{Reardon:2023zen}%
  \BibitemOpen
  \bibfield  {author} {\bibinfo {author} {\bibfnamefont {D.~J.}\ \bibnamefont {Reardon}} \emph {et~al.},\ }\href {\doibase 10.3847/2041-8213/acdd03} {\bibfield  {journal} {\bibinfo  {journal} {Astrophys. J. Lett.}\ }\textbf {\bibinfo {volume} {951}},\ \bibinfo {pages} {L7} (\bibinfo {year} {2023}{\natexlab{b}})},\ \Eprint {http://arxiv.org/abs/2306.16229} {arXiv:2306.16229 [astro-ph.HE]} \BibitemShut {NoStop}%
\bibitem [{\citenamefont {Zic}\ \emph {et~al.}(2023)\citenamefont {Zic} \emph {et~al.}}]{Zic:2023gta}%
  \BibitemOpen
  \bibfield  {author} {\bibinfo {author} {\bibfnamefont {A.}~\bibnamefont {Zic}} \emph {et~al.},\ }\href {\doibase 10.1017/pasa.2023.36} {\bibfield  {journal} {\bibinfo  {journal} {Publ. Astron. Soc. Austral.}\ }\textbf {\bibinfo {volume} {40}},\ \bibinfo {pages} {e049} (\bibinfo {year} {2023})},\ \Eprint {http://arxiv.org/abs/2306.16230} {arXiv:2306.16230 [astro-ph.HE]} \BibitemShut {NoStop}%
\bibitem [{\citenamefont {Agazie}\ \emph {et~al.}(2023{\natexlab{b}})\citenamefont {Agazie} \emph {et~al.}}]{NANOGrav:2023hde}%
  \BibitemOpen
  \bibfield  {author} {\bibinfo {author} {\bibfnamefont {G.}~\bibnamefont {Agazie}} \emph {et~al.} (\bibinfo {collaboration} {NANOGrav}),\ }\href {\doibase 10.3847/2041-8213/acda9a} {\bibfield  {journal} {\bibinfo  {journal} {Astrophys. J. Lett.}\ }\textbf {\bibinfo {volume} {951}},\ \bibinfo {pages} {L9} (\bibinfo {year} {2023}{\natexlab{b}})},\ \Eprint {http://arxiv.org/abs/2306.16217} {arXiv:2306.16217 [astro-ph.HE]} \BibitemShut {NoStop}%
\bibitem [{\citenamefont {B\"odeker}(2021)}]{Bodeker:2021mcj}%
  \BibitemOpen
  \bibfield  {author} {\bibinfo {author} {\bibfnamefont {D.}~\bibnamefont {B\"odeker}},\ }\href {\doibase 10.1103/PhysRevD.104.L111501} {\bibfield  {journal} {\bibinfo  {journal} {Phys. Rev. D}\ }\textbf {\bibinfo {volume} {104}},\ \bibinfo {pages} {L111501} (\bibinfo {year} {2021})},\ \Eprint {http://arxiv.org/abs/2108.11966} {arXiv:2108.11966 [hep-ph]} \BibitemShut {NoStop}%
\bibitem [{\citenamefont {Sagunski}\ \emph {et~al.}(2023)\citenamefont {Sagunski}, \citenamefont {Schicho},\ and\ \citenamefont {Schmitt}}]{Sagunski:2023ynd}%
  \BibitemOpen
  \bibfield  {author} {\bibinfo {author} {\bibfnamefont {L.}~\bibnamefont {Sagunski}}, \bibinfo {author} {\bibfnamefont {P.}~\bibnamefont {Schicho}}, \ and\ \bibinfo {author} {\bibfnamefont {D.}~\bibnamefont {Schmitt}},\ }\href {\doibase 10.1103/PhysRevD.107.123512} {\bibfield  {journal} {\bibinfo  {journal} {Phys. Rev. D}\ }\textbf {\bibinfo {volume} {107}},\ \bibinfo {pages} {123512} (\bibinfo {year} {2023})},\ \Eprint {http://arxiv.org/abs/2303.02450} {arXiv:2303.02450 [hep-ph]} \BibitemShut {NoStop}%
\bibitem [{\citenamefont {Giese}\ \emph {et~al.}(2021)\citenamefont {Giese}, \citenamefont {Konstandin}, \citenamefont {Schmitz},\ and\ \citenamefont {van~de Vis}}]{Giese:2020znk}%
  \BibitemOpen
  \bibfield  {author} {\bibinfo {author} {\bibfnamefont {F.}~\bibnamefont {Giese}}, \bibinfo {author} {\bibfnamefont {T.}~\bibnamefont {Konstandin}}, \bibinfo {author} {\bibfnamefont {K.}~\bibnamefont {Schmitz}}, \ and\ \bibinfo {author} {\bibfnamefont {J.}~\bibnamefont {van~de Vis}},\ }\href {\doibase 10.1088/1475-7516/2021/01/072} {\bibfield  {journal} {\bibinfo  {journal} {JCAP}\ }\textbf {\bibinfo {volume} {01}},\ \bibinfo {pages} {072} (\bibinfo {year} {2021})},\ \Eprint {http://arxiv.org/abs/2010.09744} {arXiv:2010.09744 [astro-ph.CO]} \BibitemShut {NoStop}%
\bibitem [{\citenamefont {von Harling}\ and\ \citenamefont {Servant}(2018)}]{vonHarling:2017yew}%
  \BibitemOpen
  \bibfield  {author} {\bibinfo {author} {\bibfnamefont {B.}~\bibnamefont {von Harling}}\ and\ \bibinfo {author} {\bibfnamefont {G.}~\bibnamefont {Servant}},\ }\href {\doibase 10.1007/JHEP01(2018)159} {\bibfield  {journal} {\bibinfo  {journal} {JHEP}\ }\textbf {\bibinfo {volume} {01}},\ \bibinfo {pages} {159} (\bibinfo {year} {2018})},\ \Eprint {http://arxiv.org/abs/1711.11554} {arXiv:1711.11554 [hep-ph]} \BibitemShut {NoStop}%
\bibitem [{\citenamefont {Iso}\ \emph {et~al.}(2017)\citenamefont {Iso}, \citenamefont {Serpico},\ and\ \citenamefont {Shimada}}]{Iso:2017uuu}%
  \BibitemOpen
  \bibfield  {author} {\bibinfo {author} {\bibfnamefont {S.}~\bibnamefont {Iso}}, \bibinfo {author} {\bibfnamefont {P.~D.}\ \bibnamefont {Serpico}}, \ and\ \bibinfo {author} {\bibfnamefont {K.}~\bibnamefont {Shimada}},\ }\href {\doibase 10.1103/PhysRevLett.119.141301} {\bibfield  {journal} {\bibinfo  {journal} {Phys. Rev. Lett.}\ }\textbf {\bibinfo {volume} {119}},\ \bibinfo {pages} {141301} (\bibinfo {year} {2017})},\ \Eprint {http://arxiv.org/abs/1704.04955} {arXiv:1704.04955 [hep-ph]} \BibitemShut {NoStop}%
\bibitem [{\citenamefont {Gao}\ and\ \citenamefont {Pawlowski}(2020)}]{Gao:2020qsj}%
  \BibitemOpen
  \bibfield  {author} {\bibinfo {author} {\bibfnamefont {F.}~\bibnamefont {Gao}}\ and\ \bibinfo {author} {\bibfnamefont {J.~M.}\ \bibnamefont {Pawlowski}},\ }\href {\doibase 10.1103/PhysRevD.102.034027} {\bibfield  {journal} {\bibinfo  {journal} {Phys. Rev. D}\ }\textbf {\bibinfo {volume} {102}},\ \bibinfo {pages} {034027} (\bibinfo {year} {2020})},\ \Eprint {http://arxiv.org/abs/2002.07500} {arXiv:2002.07500 [hep-ph]} \BibitemShut {NoStop}%
\bibitem [{\citenamefont {Philipsen}(2013)}]{Philipsen:2012nu}%
  \BibitemOpen
  \bibfield  {author} {\bibinfo {author} {\bibfnamefont {O.}~\bibnamefont {Philipsen}},\ }\href {\doibase 10.1016/j.ppnp.2012.09.003} {\bibfield  {journal} {\bibinfo  {journal} {Prog. Part. Nucl. Phys.}\ }\textbf {\bibinfo {volume} {70}},\ \bibinfo {pages} {55} (\bibinfo {year} {2013})},\ \Eprint {http://arxiv.org/abs/1207.5999} {arXiv:1207.5999 [hep-lat]} \BibitemShut {NoStop}%
\bibitem [{\citenamefont {Guenther}\ \emph {et~al.}(2017)\citenamefont {Guenther}, \citenamefont {Bellwied}, \citenamefont {Borsanyi}, \citenamefont {Fodor}, \citenamefont {Katz}, \citenamefont {Pasztor}, \citenamefont {Ratti},\ and\ \citenamefont {Szab\'o}}]{Guenther:2017hnx}%
  \BibitemOpen
  \bibfield  {author} {\bibinfo {author} {\bibfnamefont {J.~N.}\ \bibnamefont {Guenther}}, \bibinfo {author} {\bibfnamefont {R.}~\bibnamefont {Bellwied}}, \bibinfo {author} {\bibfnamefont {S.}~\bibnamefont {Borsanyi}}, \bibinfo {author} {\bibfnamefont {Z.}~\bibnamefont {Fodor}}, \bibinfo {author} {\bibfnamefont {S.~D.}\ \bibnamefont {Katz}}, \bibinfo {author} {\bibfnamefont {A.}~\bibnamefont {Pasztor}}, \bibinfo {author} {\bibfnamefont {C.}~\bibnamefont {Ratti}}, \ and\ \bibinfo {author} {\bibfnamefont {K.~K.}\ \bibnamefont {Szab\'o}},\ }\href {\doibase 10.1016/j.nuclphysa.2017.05.044} {\bibfield  {journal} {\bibinfo  {journal} {Nucl. Phys. A}\ }\textbf {\bibinfo {volume} {967}},\ \bibinfo {pages} {720} (\bibinfo {year} {2017})},\ \Eprint {http://arxiv.org/abs/1607.02493} {arXiv:1607.02493 [hep-lat]} \BibitemShut {NoStop}%
\bibitem [{\citenamefont {Gao}\ \emph {et~al.}(2016)\citenamefont {Gao}, \citenamefont {Chen}, \citenamefont {Liu}, \citenamefont {Qin}, \citenamefont {Roberts},\ and\ \citenamefont {Schmidt}}]{Gao:2015kea}%
  \BibitemOpen
  \bibfield  {author} {\bibinfo {author} {\bibfnamefont {F.}~\bibnamefont {Gao}}, \bibinfo {author} {\bibfnamefont {J.}~\bibnamefont {Chen}}, \bibinfo {author} {\bibfnamefont {Y.~X.}\ \bibnamefont {Liu}}, \bibinfo {author} {\bibfnamefont {S.~X.}\ \bibnamefont {Qin}}, \bibinfo {author} {\bibfnamefont {C.~D.}\ \bibnamefont {Roberts}}, \ and\ \bibinfo {author} {\bibfnamefont {S.~M.}\ \bibnamefont {Schmidt}},\ }\href {\doibase 10.1103/PhysRevD.93.094019} {\bibfield  {journal} {\bibinfo  {journal} {Phys. Rev. D}\ }\textbf {\bibinfo {volume} {93}},\ \bibinfo {pages} {094019} (\bibinfo {year} {2016})},\ \Eprint {http://arxiv.org/abs/1507.00875} {arXiv:1507.00875 [nucl-th]} \BibitemShut {NoStop}%
\bibitem [{\citenamefont {Isserstedt}\ \emph {et~al.}(2019)\citenamefont {Isserstedt}, \citenamefont {Buballa}, \citenamefont {Fischer},\ and\ \citenamefont {Gunkel}}]{Isserstedt:2019pgx}%
  \BibitemOpen
  \bibfield  {author} {\bibinfo {author} {\bibfnamefont {P.}~\bibnamefont {Isserstedt}}, \bibinfo {author} {\bibfnamefont {M.}~\bibnamefont {Buballa}}, \bibinfo {author} {\bibfnamefont {C.~S.}\ \bibnamefont {Fischer}}, \ and\ \bibinfo {author} {\bibfnamefont {P.~J.}\ \bibnamefont {Gunkel}},\ }\href {\doibase 10.1103/PhysRevD.100.074011} {\bibfield  {journal} {\bibinfo  {journal} {Phys. Rev. D}\ }\textbf {\bibinfo {volume} {100}},\ \bibinfo {pages} {074011} (\bibinfo {year} {2019})},\ \Eprint {http://arxiv.org/abs/1906.11644} {arXiv:1906.11644 [hep-ph]} \BibitemShut {NoStop}%
\bibitem [{\citenamefont {Fu}\ and\ \citenamefont {Pawlowski}(2015)}]{Fu:2015naa}%
  \BibitemOpen
  \bibfield  {author} {\bibinfo {author} {\bibfnamefont {W.~J.}\ \bibnamefont {Fu}}\ and\ \bibinfo {author} {\bibfnamefont {J.~M.}\ \bibnamefont {Pawlowski}},\ }\href {\doibase 10.1103/PhysRevD.92.116006} {\bibfield  {journal} {\bibinfo  {journal} {Phys. Rev. D}\ }\textbf {\bibinfo {volume} {92}},\ \bibinfo {pages} {116006} (\bibinfo {year} {2015})},\ \Eprint {http://arxiv.org/abs/1508.06504} {arXiv:1508.06504 [hep-ph]} \BibitemShut {NoStop}%
\bibitem [{\citenamefont {Lu}\ \emph {et~al.}(2023)\citenamefont {Lu}, \citenamefont {Gao}, \citenamefont {Fu}, \citenamefont {Song},\ and\ \citenamefont {Liu}}]{Lu:2023msn}%
  \BibitemOpen
  \bibfield  {author} {\bibinfo {author} {\bibfnamefont {Y.}~\bibnamefont {Lu}}, \bibinfo {author} {\bibfnamefont {F.}~\bibnamefont {Gao}}, \bibinfo {author} {\bibfnamefont {B.~C.}\ \bibnamefont {Fu}}, \bibinfo {author} {\bibfnamefont {H.~C.}\ \bibnamefont {Song}}, \ and\ \bibinfo {author} {\bibfnamefont {Y.~X.}\ \bibnamefont {Liu}},\ }\href@noop {} {\  (\bibinfo {year} {2023})},\ \Eprint {http://arxiv.org/abs/2310.16345} {arXiv:2310.16345 [hep-ph]} \BibitemShut {NoStop}%
\bibitem [{\citenamefont {Fischer}\ \emph {et~al.}(2014)\citenamefont {Fischer}, \citenamefont {Luecker},\ and\ \citenamefont {Welzbacher}}]{Fischer:2014ata}%
  \BibitemOpen
  \bibfield  {author} {\bibinfo {author} {\bibfnamefont {C.~S.}\ \bibnamefont {Fischer}}, \bibinfo {author} {\bibfnamefont {J.}~\bibnamefont {Luecker}}, \ and\ \bibinfo {author} {\bibfnamefont {C.~A.}\ \bibnamefont {Welzbacher}},\ }\href {\doibase 10.1103/PhysRevD.90.034022} {\bibfield  {journal} {\bibinfo  {journal} {Phys. Rev. D}\ }\textbf {\bibinfo {volume} {90}},\ \bibinfo {pages} {034022} (\bibinfo {year} {2014})},\ \Eprint {http://arxiv.org/abs/1405.4762} {arXiv:1405.4762 [hep-ph]} \BibitemShut {NoStop}%
\bibitem [{\citenamefont {Eichhorn}\ \emph {et~al.}(2021)\citenamefont {Eichhorn}, \citenamefont {Lumma}, \citenamefont {Pawlowski}, \citenamefont {Reichert},\ and\ \citenamefont {Yamada}}]{Eichhorn:2020upj}%
  \BibitemOpen
  \bibfield  {author} {\bibinfo {author} {\bibfnamefont {A.}~\bibnamefont {Eichhorn}}, \bibinfo {author} {\bibfnamefont {J.}~\bibnamefont {Lumma}}, \bibinfo {author} {\bibfnamefont {J.~M.}\ \bibnamefont {Pawlowski}}, \bibinfo {author} {\bibfnamefont {M.}~\bibnamefont {Reichert}}, \ and\ \bibinfo {author} {\bibfnamefont {M.}~\bibnamefont {Yamada}},\ }\href {\doibase 10.1088/1475-7516/2021/05/006} {\bibfield  {journal} {\bibinfo  {journal} {JCAP}\ }\textbf {\bibinfo {volume} {05}},\ \bibinfo {pages} {006} (\bibinfo {year} {2021})},\ \Eprint {http://arxiv.org/abs/2010.00017} {arXiv:2010.00017 [hep-ph]} \BibitemShut {NoStop}%
\bibitem [{\citenamefont {Lewicki}\ \emph {et~al.}(2022)\citenamefont {Lewicki}, \citenamefont {Merchand},\ and\ \citenamefont {Zych}}]{Lewicki:2021pgr}%
  \BibitemOpen
  \bibfield  {author} {\bibinfo {author} {\bibfnamefont {M.}~\bibnamefont {Lewicki}}, \bibinfo {author} {\bibfnamefont {M.}~\bibnamefont {Merchand}}, \ and\ \bibinfo {author} {\bibfnamefont {M.}~\bibnamefont {Zych}},\ }\href {\doibase 10.1007/JHEP02(2022)017} {\bibfield  {journal} {\bibinfo  {journal} {JHEP}\ }\textbf {\bibinfo {volume} {02}},\ \bibinfo {pages} {017} (\bibinfo {year} {2022})},\ \Eprint {http://arxiv.org/abs/2111.02393} {arXiv:2111.02393 [astro-ph.CO]} \BibitemShut {NoStop}%
\bibitem [{\citenamefont {Kamionkowski}\ \emph {et~al.}(1994)\citenamefont {Kamionkowski}, \citenamefont {Kosowsky},\ and\ \citenamefont {Turner}}]{Kamionkowski:1993fg}%
  \BibitemOpen
  \bibfield  {author} {\bibinfo {author} {\bibfnamefont {M.}~\bibnamefont {Kamionkowski}}, \bibinfo {author} {\bibfnamefont {A.}~\bibnamefont {Kosowsky}}, \ and\ \bibinfo {author} {\bibfnamefont {M.~S.}\ \bibnamefont {Turner}},\ }\href {\doibase 10.1103/PhysRevD.49.2837} {\bibfield  {journal} {\bibinfo  {journal} {Phys. Rev. D}\ }\textbf {\bibinfo {volume} {49}},\ \bibinfo {pages} {2837} (\bibinfo {year} {1994})},\ \Eprint {http://arxiv.org/abs/astro-ph/9310044} {arXiv:astro-ph/9310044} \BibitemShut {NoStop}%
\bibitem [{\citenamefont {Steinhardt}(1982)}]{Steinhardt:1981ct}%
  \BibitemOpen
  \bibfield  {author} {\bibinfo {author} {\bibfnamefont {P.~J.}\ \bibnamefont {Steinhardt}},\ }\href {\doibase 10.1103/PhysRevD.25.2074} {\bibfield  {journal} {\bibinfo  {journal} {Phys. Rev. D}\ }\textbf {\bibinfo {volume} {25}},\ \bibinfo {pages} {2074} (\bibinfo {year} {1982})}\BibitemShut {NoStop}%
\bibitem [{\citenamefont {Ellis}\ \emph {et~al.}(2019{\natexlab{b}})\citenamefont {Ellis}, \citenamefont {Lewicki},\ and\ \citenamefont {No}}]{Ellis:2018mja}%
  \BibitemOpen
  \bibfield  {author} {\bibinfo {author} {\bibfnamefont {J.}~\bibnamefont {Ellis}}, \bibinfo {author} {\bibfnamefont {M.}~\bibnamefont {Lewicki}}, \ and\ \bibinfo {author} {\bibfnamefont {J.~M.}\ \bibnamefont {No}},\ }\href {\doibase 10.1088/1475-7516/2019/04/003} {\bibfield  {journal} {\bibinfo  {journal} {JCAP}\ }\textbf {\bibinfo {volume} {04}},\ \bibinfo {pages} {003} (\bibinfo {year} {2019}{\natexlab{b}})},\ \Eprint {http://arxiv.org/abs/1809.08242} {arXiv:1809.08242 [hep-ph]} \BibitemShut {NoStop}%
\end{thebibliography}%
\clearpage

\onecolumngrid
\begin{center}
  \textbf{\large Supplemental Material}\\[.2cm]
\end{center}

    \section{Thermal quantities and Cosmic trajectory during the QCD transition}
    {
    At finite temperature and chemical potential, the general form of the solution of the quark DSE can be expressed as:
    \begin{equation}
    {{S}^{-1}}({{{\tilde{\omega }}}_{n}},\vec{p})= i\vec{\gamma }\cdot \vec{p}A({{{\tilde{\omega }}}_{n}},\vec{p})+i{{\gamma }_{4}}{{{\tilde{\omega }}}_{n}}C({{{\tilde{\omega }}}_{n}},\vec{p})+B({{{\tilde{\omega }}}_{n}},\vec{p}).
    \end{equation}
    with $\vec{p}$ the momentum, ${{{\tilde{\omega }}}_{n}}={{\omega }_{n}}+i{{\mu }_{q}}$, ${{\omega }_{n}}=(2n+1)\pi T$ the Matsubara frequency and $\mu_q$ the quark chemical potential. Then, the dynamic quark mass is defined as:
    \begin{equation}
        {{M}_{q}}=\operatorname{Re}\left[ \frac{B({{{\tilde{\omega }}}_{0}},\vec{p}=0)}{C({{{\tilde{\omega }}}_{0}},\vec{p}=0)} \right].
    \end{equation}
    At finite temperature and chemical potential, the quark gap equation can be written as:
    \begin{align}    \label{eq:gap}
        {{S}^{-1}}({{{\tilde{\omega }}}_{n}},\vec{p})=(i{{\gamma }_{4}}{{{\tilde{\omega }}}_{n}}+i\vec{\gamma }\cdot \vec{p})+{{m}_{\zeta }}+\Sigma ({{{\tilde{\omega }}}_{n}},\vec{p})
        \end{align}
    with self energy~\cite{Roberts:1994dr}:
    \begin{align}
    \Sigma ({{{\tilde{\omega }}}_{n}},\vec{\,p})=& \, \frac{4}{3}{{g}^{2}}T\sum\limits_{m=-\infty }^{\infty }{\int_{q}^{\Lambda }{{{D}_{\mu \nu }}}({{\Omega }_{nm}},\vec{k};T,{{\mu }_{q}})} {{\gamma }_{\mu }}S({{{\tilde{\omega }}}_{m}},\vec{q}){{\Gamma }_{\nu }}({{{\tilde{\omega }}}_{n}},\vec{\,p},{{{\tilde{\omega }}}_{m}},\vec{\,q};T,{{\mu }_{q}} ) \, ,
    \end{align}
    which is defined in the Euclidean space,
    with $\mu_{q}$ the quark chemical potential,
    ${{\omega }_{n}}=(2n+1)\pi T$ the Matsubara frequency and ${{{\tilde{\omega }}}_{n}}={{\omega }_{n}}+i{{\mu }_{q}}$ for the quarks.
    ${{\Omega }_{nm}}={{\omega }_{n}}-{{\omega }_{m}}$ is the Matsubara frequency of the gluon.
    $m_{\zeta}$ is the current quark mass at the renormalization point $\zeta$,
    $\int_{\vec{q}}^{\Lambda}\equiv \int _0^{\Lambda} d^3q/(2\pi)^3$ with $\Lambda$ the energy scale of the loop momentum regularisation.
    Finally, $\Gamma_{\nu}$ is the dressed quark-gluon interaction vertex and $D_{\mu\nu}^{ab}$ is the dressed gluon propagator.
    }

    The dressed gluon propagator is fitted from the lattice results, and its general form in finite temperature and density is:
    \begin{align}
    {{D}_{\mu \nu }}({{\Omega }_{nm}},\vec{k})=P_{\mu \nu }^{T}{{D}_{T}}({{\Omega }_{nm}},\vec{k})+P_{\mu \nu }^{L}{{D}_{L}}({{\Omega }_{nm}},\vec{k}),
    \end{align}
    where $P_{\mu \nu }^{L,T}$ are the longitudinal and transverse projection operators:
    \begin{equation}\label{projct}
    \begin{split}
     P_{\mu \nu }^{T}=(1-{{\delta }_{\mu 4}})(1-{{\delta }_{\nu 4}})({{\delta }_{\mu \nu }}-\frac{{{k}_{\mu }}{{k}_{\nu }}}{{{k}^{2}}}), \quad P_{\mu \nu }^{L}=({{\delta }_{\mu \nu }}-\frac{{{k}_{\mu }}{{k}_{\nu }}}{{{k}^{2}}})-P_{\mu \nu }^{T},
    \end{split}
    \end{equation}
    where $k=({{\Omega }_{nm}},\vec{k})$.

    Taking the gauge symmetry in consideration, including the longitudinal and transverse Ward-Green-Takahashi identity, we take the minimal \textit{Ans\"atze} of the dressed quark-gluon interaction vertex as~\cite{Gao:2020qsj}:
    \begin{equation}          \label{vertex: qg}
    \begin{split}
    {{\Gamma }_{\mu }}({{{\tilde{\omega }}}_{n}},\vec{p},{{{\tilde{\omega }}}_{m}},\vec{q})=  F({{k}^{2}})\frac{A({{{\tilde{\omega }}}_{n}},\vec{p})+A({{{\tilde{\omega }}}_{m}},\vec{q})}{2}{{\gamma }_{\mu }} +Z_{A,L}^{-1/2}({{k}^{2}}){{\Delta }_{B}}P_{\mu \nu }^{L}{{\sigma }_{\mu \nu }}{{k}_{\nu }}+Z_{A,T}^{-1/2}({{k}^{2}}){{\Delta }_{B}}P_{\mu \nu }^{T}{{\sigma }_{\mu \nu }}{{k}_{\nu }}
    \end{split}
    \end{equation}
    with
    %
    \begin{align}
    {{\Delta }_{B}}=\frac{\tilde{B}({{{\tilde{\omega }}}_{n}},\vec{p})-\tilde{B}({{{\tilde{\omega }}}_{m}},\vec{q})}{{{{\tilde{\omega }}}_{n}}^{2}+{{{\vec{p}}}^{2}}-{{{\tilde{\omega }}}_{m}}^{2}-{{{\vec{q}}}^{2}}}, \quad \tilde{B}({{{\tilde{\omega }}}_{n}},\vec{p})=B({{\omega }_{0}}~\text{sgn}({{\omega }_{\text{n}}})+i{{\mu }_{q}},{{l}_{p}}),
    \end{align}
    {where $k=({{\Omega }_{nm}},\vec{q}-\vec{p})$, ${{l}_{p}}=
    {{({{{\vec{p}}}^{2}}+{{\omega }_{n}}^{2}-{{\omega }_{0}}^{2})}^{1/2}}$}, ${{Z}_{A,L}}={{D}_{L}}({{k}^{2}}){{k}^{2}}$ and ${{Z}_{A,T}}={{D}_{T}}({{k}^{2}}){{k}^{2}}$ are the gluon dressing functions, and $F(k^2)$ is the ghost dressing function that is fitted from the functional renormalization group (fRG).  To include the gluon pressure, we use the lattice QCD results to obtain the entropy density at vanishing chemical potential as parametrized in the lattice QCD~\cite{Philipsen:2012nu}.
    The full entropy density can then be  expressed as:

    \begin{equation}\label{eq:s_QCD}
        s_{QCD}(T,\mu )={{s}_{latt}}(T,\mu=0 )+\delta s(T,\mu ).
    \end{equation}
    \begin{figure}
        \centering
        \includegraphics[width=8.5 cm]{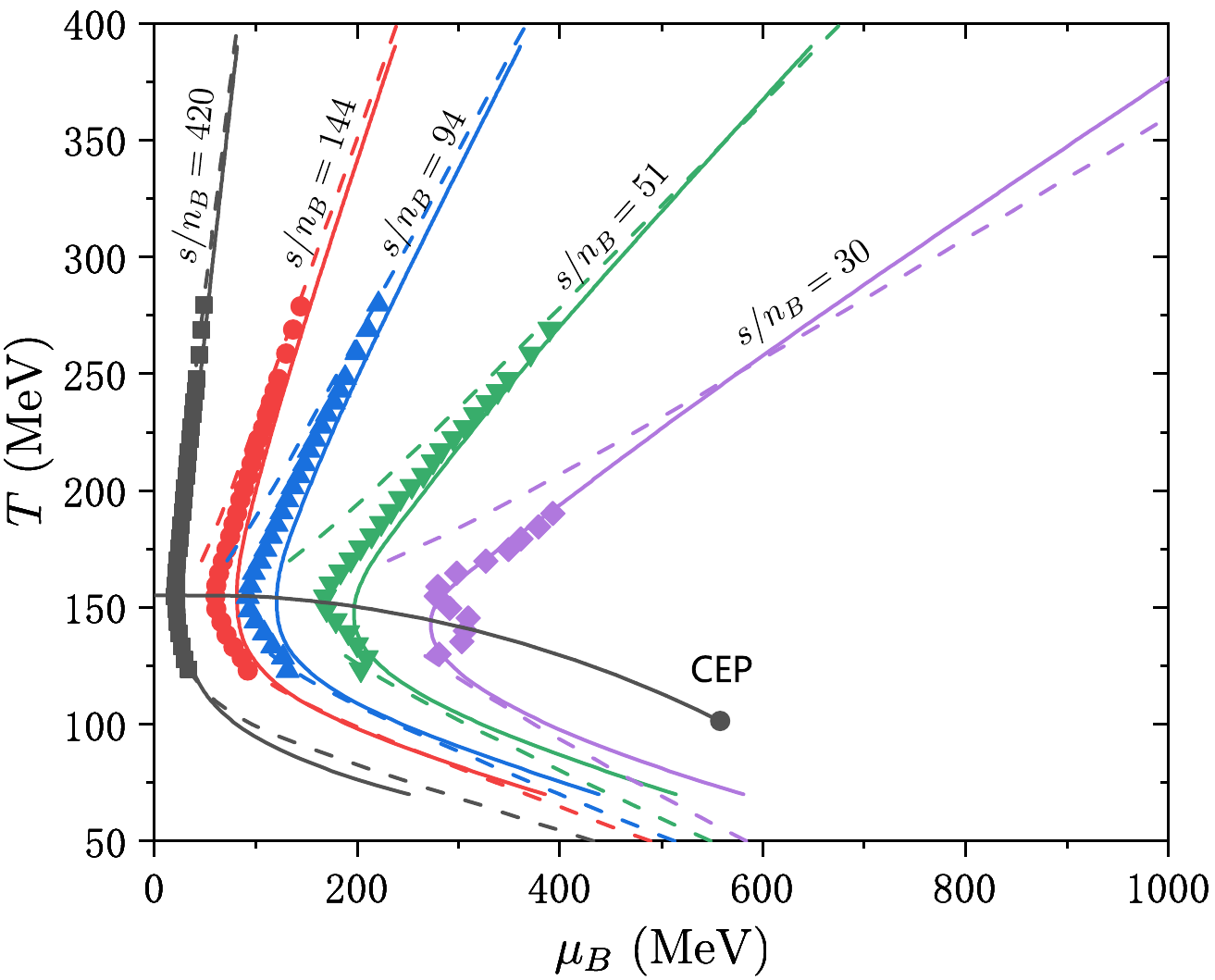}
        \includegraphics[width=8.5 cm]{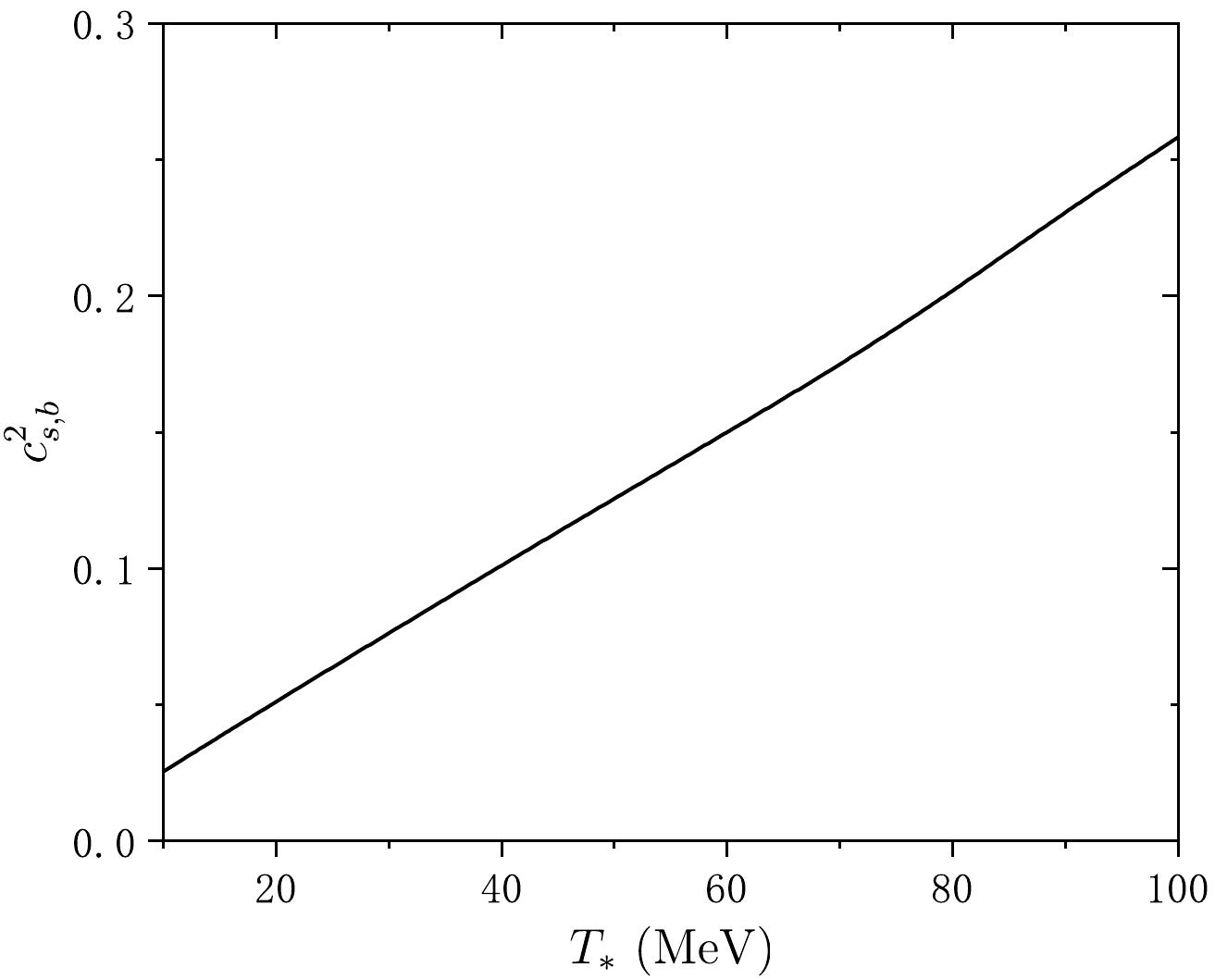}
        \caption{\emph{left panel:} Isentropic trajectories at $s/n_B=420,\,144,\,94,\,51$ and $30$. Solid line: DSE from this work; scatter points: the lattice
        ~\cite{Guenther:2017hnx}
        ; dashed line (above): the improved ideal quark gas; dashed line (below): HRG. The solid line with a point represents the crossover of QCD phase transition, with the point corresponding to the critical end point (CEP).\emph{right panel:}The speed of sound $c_{s,b}^2$ of the broken phase along the percolation line.}
        \label{fig:isentraj}
    \end{figure}
    
    The thermal quantities of QCD, including the number density and the entropy density, can be obtained by solving DSE~\cite{Gao:2015kea,Isserstedt:2019pgx,Lu:2023mkn}.
    The quark number density $n_q$ and the entropy $s$ can be written as:
    \begin{align}
         {{n}_{q}} =\frac{\partial P}{\partial {{\mu }_{q}}}, \quad 
         s =\frac{\partial P}{\partial T}, \label{eq:s_P}
    \end{align}
    with $P$ the pressure and $\mu_q$ the quark chemical potential.
    The number density of $u/\,d$ and $s$ quark can be expressed in terms of the dynamical quark mass $M_q$ and the traced Polyakov loop $\Phi$ in phase diagram ($T,\mu_q$) plane~\cite{Fu:2015naa,Lu:2023msn}:
    \begin{subequations}
        \begin{align}
          {{n}_{q}}(T,{{\mu }_{q}})&=2{{N}_{c}}\int{\frac{{{d}^{3}}k}{{{(2\pi )}^{3}}}[f_{q}^{+}(k;T,{{\mu }_{q}})-f_{q}^{-}(k;T,{{\mu }_{q}})]}, \\ 
          f_{q}^{\pm }(k;T,{{\mu }_{q}})&=\frac{\Phi (T,{{\mu }_{q}})x_{\pm }^{2}+2\Phi (T,{{\mu }_{q}}){{x}_{\pm }}+1}{x_{\pm }^{3}+3\Phi (T,{{\mu }_{q}})x_{\pm }^{2}+3\Phi (T,{{\mu }_{q}}){{x}_{\pm }}+1}, \\
\mathrm{with}\quad{{x}_{\pm }}(k;T,{{\mu }_{q}})&=\exp [({{E}_{q}}(k;T,{{\mu }_{q}})\mp {{\mu }_{q}})/T], \quad {{E}_{q}}(k;T,{{\mu }_{q}})=\sqrt{{{k}^{2}}+M_{q}^{2}},\notag
        \end{align}
    \end{subequations}
   and  with $N_c$ the color number. The traced Polyakov loop $\Phi$ at vanishing quark chemical potential is parameterized as~\cite{Lu:2023msn,Fischer:2014ata}:
    \begin{subequations}
        \begin{align}
          \Phi (T,{{\mu }_{q}})&=L({{t}_{{{\mu }_{q}}}}), \\ 
        \quad {{t}_{{{\mu }_{q}}}}=\frac{T}{{{T}_{c}(0)}}+\kappa {{\left( \frac{3{{\mu }_{q}}}{{{T}_{c}(0)}} \right)}^{2}}, &\quad 
         L\left( t \right)=\frac{2}{1+\exp \left( \frac{{{a}_{1}}+{{b}_{1}}{{t}^{3}}}{t+{{c}_{1}}{{t}^{3}}+{{d}_{1}}{{t}^{6}}} \right)},
        \end{align}
    \end{subequations}
    where $T_c(0)=155.2\,{\rm MeV}$ the PT temperature at vanishing chemical potential and $\kappa=0.0184$ the curvature of the transition line. The fit parameters are $[a_1,\,b_1,\,c_1,\,d_1]=[2.5996,\,-0.2046,\,-1.2346,\,2.2308]$.
    From the definition of  the Eqs.~(\ref{eq:s_P}), the chemical potential dependence of the entropy density can be expressed as the integral along $\mu_q$~\cite{Lu:2023mkn}:
    \begin{equation}
        \delta s(T,\mu )=s(T,\mu )-s(T,0)=\int_{0}^{\mu_q }{d{\mu_q }'\frac{\partial {{n}_{q}}(T,{\mu }')}{\partial T}.}
    \end{equation}

    Besides, the pseudotrace anomaly takes into account the speed of sound, \(c_{s,b}^2\) of the broken phase, which is written as:
    \begin{equation}
        c_{s,b}^2 = {{\left. \frac{dp_b}{de_b} \right|}_{s}},
    \end{equation}    
    with $p_b$ the pressure density of the broken phase, $e_b$ the energy density of the broken phase and $s$ the fixed entropy density. As shown in Fig.~\ref{fig:isentraj}, \(c_{s,b}^2\) is smaller at lower temperatures, leading to a pseudotrace anomaly that is larger than the traditional trace anomaly.

    \begin{figure}
            \centering
            \includegraphics[width=0.5\linewidth]{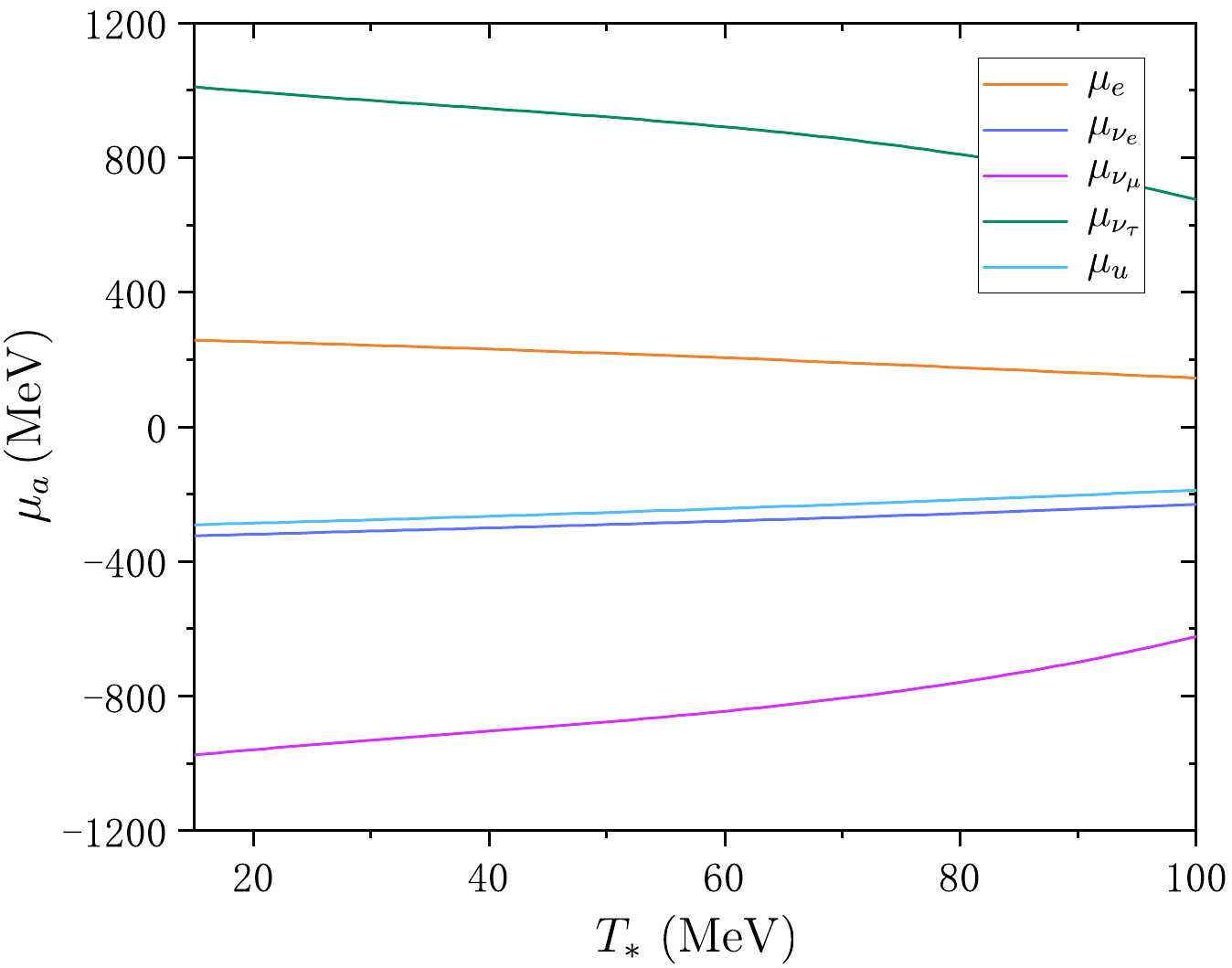}
            \caption{Chemical potentials of $e,\,\mu_e,\,\mu_\mu,\,\mu_\tau,u$ when the cosmic trajectories of different $Y_{L_\mu}$  reach percolation.}
            \label{fig:cos_lepton}
    \end{figure}

    We compare the isentropic trajectories $s/n_B=const$ from DSE with the result of the improved ideal quark gas, HRG and the lattice in Fig.~(\ref{fig:isentraj}), showing an well consistency. The improved ideal quark gas includes the lattice QCD entropy ${{s}_{latt}}(T,\mu=0 )$ in Eq.~(\ref{eq:s_QCD}), and apply  the Fermi-Dirac distribution for the quark number densities. The application of  ${{s}_{latt}}(T,\mu=0 )$ is due to the gluons do not behave like the idea gas in the QCD phase transition region. The consistency shown in Fig.~(\ref{fig:isentraj}) makes  the calculation of the five conservation equations of the cosmic trajectories more reliable, because these equations also satisfies a form similar to $s/n_B=const$.

    During the cosmic QCD transition, the kinetic and the chemical equilibrium are excellent approximation~\cite{Eichhorn:2020upj}.
    %
    The $2\leftrightarrow 2$ process is the primary mechanism, responsible for establishing both kinetic and chemical equilibrium among all particles, which constrains the chemical potentials of leptons as ${{\mu }_{\mu }}={{\mu }_{e}}-{{\mu }_{{{\nu }_{e}}}}+{{\mu }_{{{\nu }_{\mu }}}}$, and the leptons and quarks as ${{\mu }_{d}}={{\mu }_{e}}-{{\mu }_{{{\nu }_{e}}}}+{{\mu }_{u}}$.
    Chemical equilibrium constrains the chemical potentials of different particle species to only five independent chemical potentials for the universe. The chemical potentials of photons and gluons are zero because the numbers of photons and gluons are not conserved. The chemical potentials of particles and antiparticles are equal in magnitude, but opposite in sign. For instance, if a particle has a chemical potential of $\mu_i$, its antiparticle has a chemical potential of $-\mu_i$. 
    Due to flavor mixing, we only need to distinguish the chemical potentials of up and down quarks: ${{\mu }_{u}}={{\mu }_{c}}={{\mu }_{t}}$, and ${{\mu }_{d}}={{\mu }_{s}}={{\mu }_{b}}$~\cite{Schwarz:2009ii}. 
    Therefore we are left with ${{\mu }_{u}},\,{{\mu }_{e}},\,{{\mu }_{{{\nu }_{e}}}},\,{{\mu }_{{{\nu }_{\mu }}}},\,{{\mu }_{{{\nu }_{\tau }}}}$ as the independent variables for the cosmic trajectory. Then the chemical potentials of conserved charges (${{Y}_{{{L}_f}}},\,{{Y}_{B}},\,Q$) can be expressed in terms of particle chemical potential as:
    \begin{subequations}
        \begin{align}
          {{\mu }_{{{L}_f}}}&={{\mu }_{{{\nu }_f}}}, \\ 
         {{\mu }_{Q}}&={{\mu }_{u}}-{{\mu }_{d}}, \\ 
         {{\mu }_{B}}&={{\mu }_{u}}+2{{\mu }_{d}}.
        \end{align}
    \end{subequations}
    Then the temperature dependence of the five chemical potentials (${{\mu }_{{{L}_f}}},\,{{\mu }_{Q}},\,{{\mu }_{B}}$) are the cosmic trajectory. We will focus on the ($T,\mu_q$) plane in QCD phase diagram, especially in the first-order phase transition region, where the CEP locates at $(T^E,\,{{\mu }_{q}^E})=(101.3,\ 186.0)\ \rm{MeV}$.

 When the cosmic trajectory starts percolation, the conservation conditions   with a large lepton asymmetry $Y_{L_\mu}$ result in large lepton chemical potential as shown in Fig.~\ref{fig:cos_lepton}.
    The chemical potentials of $e$ and $\nu_e$ are of the same order as that of $u$ quark, while the chemical potentials of $\nu_\mu$ and $\nu_\tau$ are more than three times that of $u$ quark.
    Such large lepton chemical potentials are due to the conservation conditions. This, in turn, results in a large total energy density of the Universe, and then significantly influences the GWs spectrum.

\section{PT parameters of gravitational waves in finite temperature and chemical potential}
   
    {First one may determine the  the bubble wall velocity $v_w$,   as in Ref.~\cite{Lewicki:2021pgr}:
        \begin{equation}
            {{v}_{w}}=\left\{ \begin{matrix}
               \sqrt{\frac{\Delta {{V}_{eff}}}{{{\alpha }_{*}}{{\rho }_{r}}}}\,\,\, {\rm for}\ \sqrt{\frac{\Delta {{V}_{eff}}}{{{\alpha }_{*}}{{\rho }_{r}}}}<{{v}_{J}}  \\
             \quad \quad 1\quad \ {\rm for}\ \sqrt{\frac{\Delta {{V}_{eff}}}{{{\alpha }_{*}}{{\rho }_{r}}}}\ge {{v}_{J}}  \\
            \end{matrix} \right.,
        \end{equation}
        with $v_J$  the Jouguet velocity as~\cite{Kamionkowski:1993fg,Steinhardt:1981ct,Espinosa:2010hh}
        \begin{equation}
            {{v}_{J}}=\frac{1}{\sqrt{3}}\frac{1+\sqrt{3\alpha _{*}^{2}+2{{\alpha }_{*}}}}{1+{{\alpha }_{*}}}.
        \end{equation}
         $c_s$ is the speed of sound in the radiation. The results are shown in   in Fig.~\ref{fig:vw}.

        \begin{figure}[t]
            \centering
            \includegraphics[width=8.5 cm]{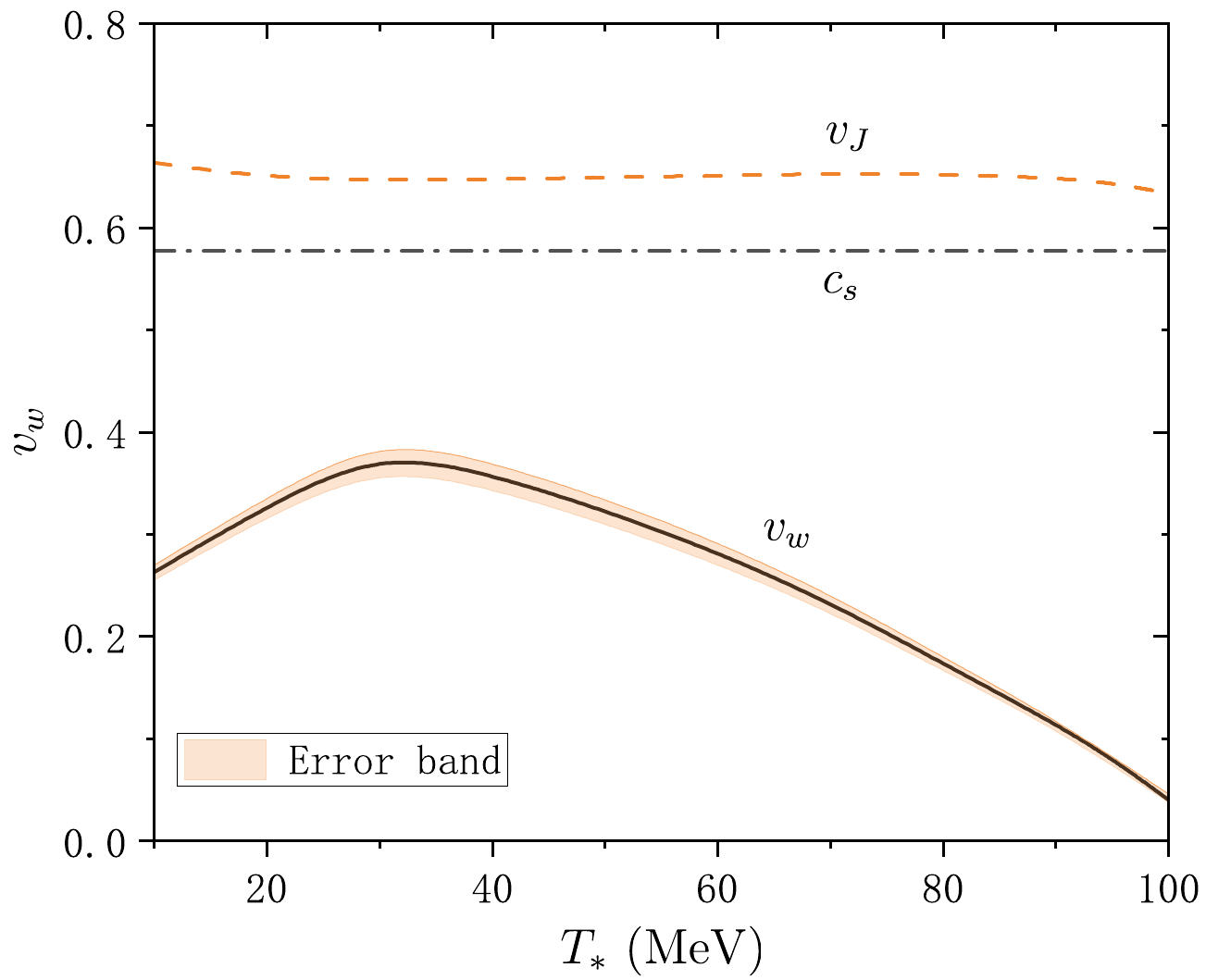}
            \caption{The bubble wall velocity on the percolation line. Solid line: the bubble wall velocity; dashed line: the Jouguet velocity; dot-dashed line: the speed of sound in the background radiation.}
            \label{fig:vw}
        \end{figure}
 }
    
     For the subsonic deflagrations, the efficiency coefficient $\kappa_s$ that measures how much of the vacuum energy goes into bulk kinetic energy is written as~\cite{Espinosa:2010hh}:
        \begin{equation}
            {{\kappa }_{s}}(v_w\leq c_s)=\frac{c_{s}^{11/5}{{\kappa }_{A}}{{\kappa }_{B}}}{(c_{s}^{11/5}-v_{w}^{11/5}){{\kappa }_{B}}+{{v}_{w}}c_{s}^{6/5}{{\kappa }_{A}}},
        \end{equation}
        with
        \begin{align}
              {{\kappa }_{A}} =v_{w}^{6/5}\frac{6.9{{\alpha }_{*}}}{1.36-0.037\sqrt{{{\alpha }_{*}}}+{{\alpha }_{*}}}, \quad
             {{\kappa }_{B}} =\frac{\alpha _{*}^{2/5}}{0.017+{{(0.997+{{\alpha }_{*}})}^{2/5}}},  
        \end{align}
        where $\alpha_*$ is the PT strength, $v_w$ is the bubble wall velocity and $c_s$ is the sound speed in the radiation.

    \begin{figure}[htbp]
         \includegraphics[width=8.5 cm]
            {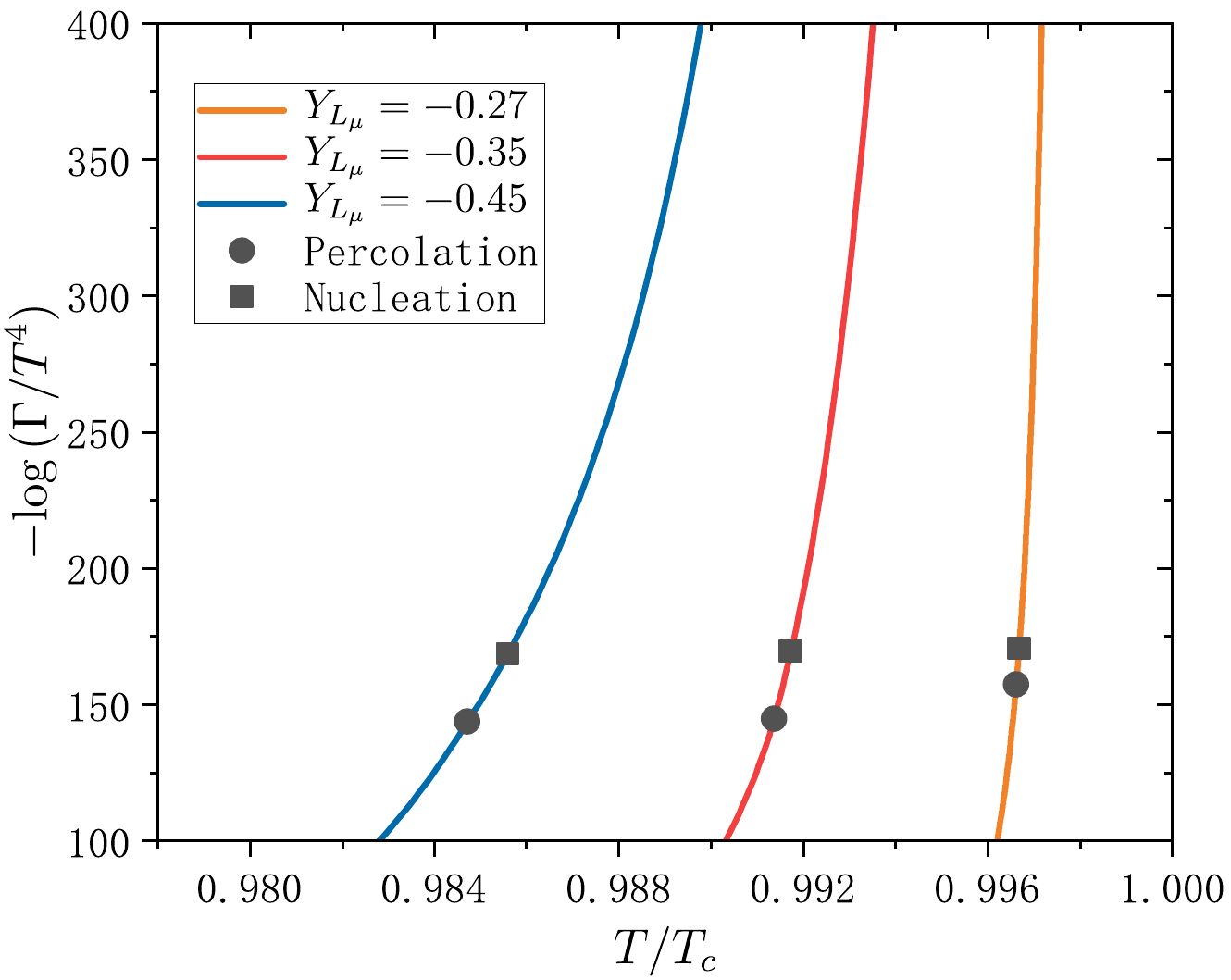}
            \includegraphics[width=8.5 cm]{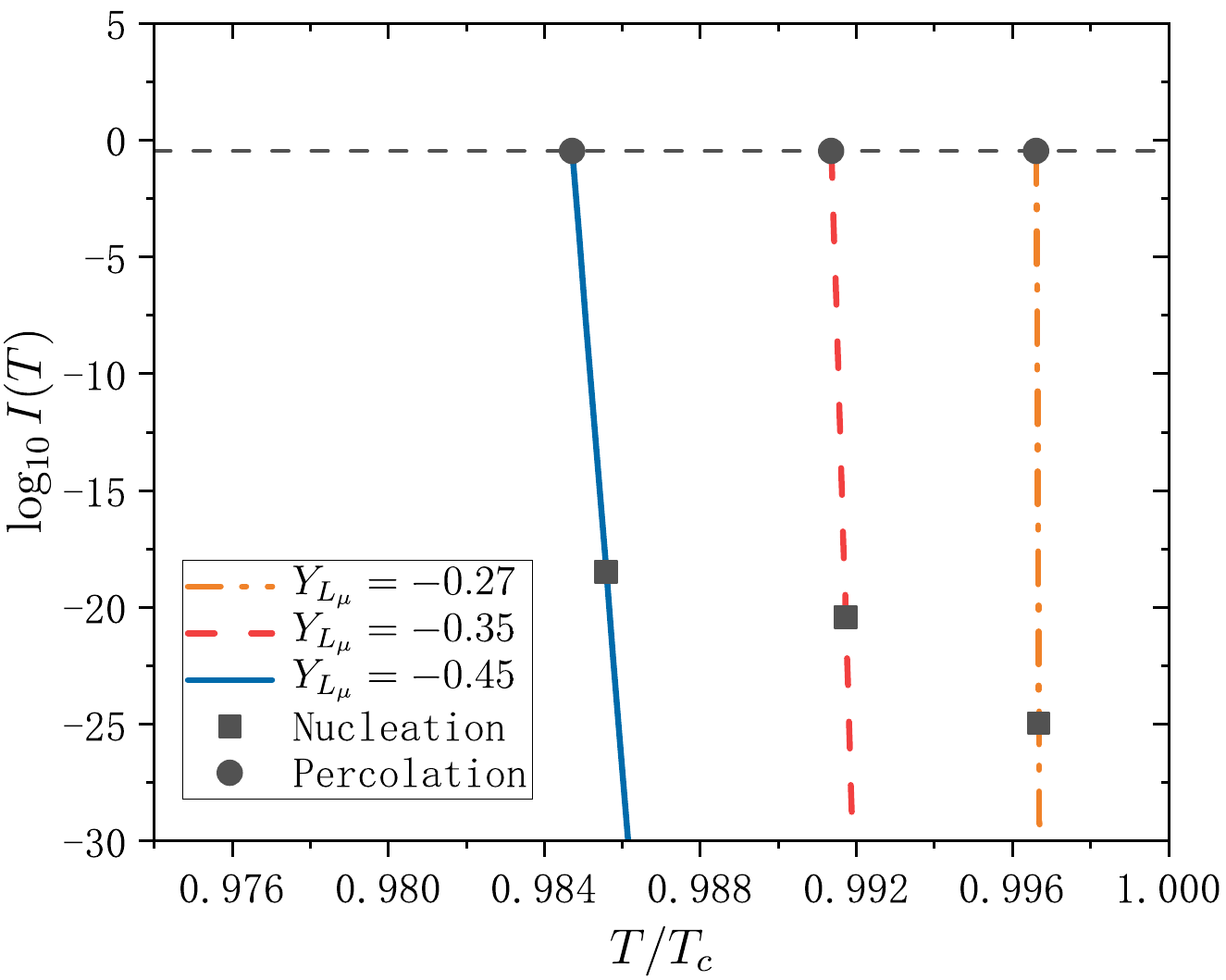}
            \caption{ The evolution of  thermally-induced decay rate along the cosmic trajectory \emph{(left panel)} and evolution of the false vacuum $I(T)$    along the cosmic trajectory \emph{(right panel)}.Black dots: percolation; Black squares: nucleation}
            \label{fig:Gamma_evo}
    \end{figure}
    
    {A critical value of $\alpha$, which can determine whether the bubble walls can runaway, is defined as~\cite{Ellis:2019oqb}:
    \begin{equation}
        {{\alpha }_{\infty }}=\frac{\Delta {{P}_{LO}}}{{{\rho }_{r}}},
    \end{equation}
    where~\cite{Bodeker:2009qy}
    \begin{equation}
        \Delta {{P}_{LO}}=\sum\limits_{{a}'}{\int_{{{m}_{f}}}^{{{m}_{t}}}{d{{m}^{2}}}\int{\frac{{{d}^{3}}p}{{{(2\pi )}^{3}}}\frac{{{f}_{F/B}}(E,T,{{\mu }_{{a}'}})}{2E}}},
    \end{equation}
    with $\sum\limits_{{a}'}$ summing over all particles taking part in the phase transition, $m_f$ ($m_t$) the particle mass in false (true) vacuum and $f_{F/B}$ the Fermi–Dirac (Bose–Einstein) distribution function.}

    For runaway case, i.e., $\alpha>\alpha_\infty$, both bubble collisions and sound wave effects should be taken into account for the GW spectrum. The corresponding efficiency factors are:
    \begin{align}
          {{\kappa }_{b}}= 1-\frac{{{\alpha }_{\infty }}}{\alpha }, \quad 
         {{\kappa }_{s}}= \frac{c_{s}^{11/5}{{\kappa }_{A}}{{\kappa }_{B}}}{(c_{s}^{11/5}-v_{w}^{11/5}){{\kappa }_{B}}+{{v}_{w}}c_{s}^{6/5}{{\kappa }_{A}}}. \label{kappa_s}
    \end{align}

    As for the non-runaway case, i.e., $\alpha<\alpha_\infty$, only the sound wave contribution should be considered. Therefore, we apply Eq.~(\ref{kappa_s}) and focus on the sound wave GW spectrum.


    To obtain the PT parameters that take the derivative along the cosmic trajectory at finite temperature and chemical potential, we start with the fluid equation:
    \begin{equation}\label{eq:fluid_eq}
        \dot{\rho_r }+3\frac{{\dot{a}}}{a}(\rho_r +p)=0,
    \end{equation}
    with
    \begin{equation}\label{eq:eos_rad}
        p=\frac{1}{3}\rho_r,
    \end{equation}
    with $a$ the scale factor, $\rho_r$ the energy density of radiation and $p$ the pressure. Eq.~(\ref{eq:eos_rad}) is valid even for radiation in finite chemical potential. The term $\rho_r$ considers only the radiation (except for muon due to its mass is close to the PT temperature.) because the matter energy density is much smaller than that of the radiation. The $u/d$ quarks in the false vacuum, instead of the true vacuum, are considered in $\rho_r$  because the decay rate $\Gamma(T,\mu)$ is that of the false vacuum. Combining Eq.~(\ref{eq:fluid_eq}) and Eq.~(\ref{eq:eos_rad}), we obtain:
    \begin{align}
       dt =-\frac{d\rho_r }{4H\rho_r },  \quad \rho_r  ={{\rho }_{r0}}{{a}^{-4}},\label{eq:dt_drho}
    \end{align}
    with $t$ the time along the cosmic trajectory, $H=\frac{{\dot{a}}}{a}$ the Hubble parameter and $\rho_{r0}$ the energy density at a given time.  The evolution of the decay rate is calculated using the Python package \texttt{BubbleDet}~\cite{Ekstedt:2023sqc}. In Fig.~\ref{fig:Gamma_evo}, the value of $-\log (\Gamma /{{T}^{4}})$ is significantly influenced by the Hubble parameter that is enlarged from the energy density $\rho_r$.

    The probability that a point still remains in the false vacuum is written as~\cite{Ellis:2018mja}:
    \begin{align}
      P(t) ={{e}^{-I(t)}},\quad I(t) =\frac{4\pi }{3}\int_{{{t}_{c}}}^{t}{d{t}'\Gamma \left( {{t}'} \right)a{{({t}')}^{3}}{{\left( \int_{{{t}'}}^{t}{\frac{{{v}_{w}}d{t}''}{a({t}'')}} \right)}^{3}}}.\label{eq:I(t)}
    \end{align}

{  After that, the formula to determine the percolation temperature and chemical potential $(T_*,\mu_*)$ can be derived by applying Eq.~(\ref{eq:dt_drho})  to Eq.~(\ref{eq:I(t)}):
    \begin{align}
         I\left( T,\mu \right) = \frac{4\pi }{3}\int_{\rho_r }^{{{\rho }^c_{r}}}{d{\rho_r }'\frac{1}{4{{{{\rho_r }'}}^{7/4}}}\frac{\Gamma \left( {T}',{\mu }' \right)}     {H\left( {T}',{\mu }' \right)}}  \times {{\left( \int_{\rho_r }^{{{\rho_r }'}}{d{\rho_r }''\frac{{{v}_{w}}}{4H\left( {T}'',{\mu }'' \right){{{{\rho_r }''}}^{3/4}}}} \right)}^{3}},
    \end{align}    
    with 
$H=\sqrt{\frac{{{\rho }_{r}}+\Delta \Gamma_\Sigma}{3M_{p}^{2}}},$
     and  $(T,\mu)$ varying along the cosmic trajectory, $H$ the Hubble parameter, $\rho_r=\rho_r(T, \mu)$ the energy density of the radiation, $\rho^c_{r}$ the energy density when the false and the true vacuum are degenerate, and $\Delta \Gamma_\Sigma=-\Delta P$ the energy difference between false and true vacuum.  The percolation temperature $T_*$ is obtained  by typically choosing $I(T_*,\mu_*)\simeq0.34$~\cite{Ellis:2018mja}, as shown in Fig.~\ref{fig:Gamma_evo}. Unlike the case of vanishing chemical potential~\cite{Sagunski:2023ynd}, the evolution of \( I(T) \) is approximately linear, because its evolution primarily depends on the chemical potential.}

    Similarly, the inverse PT duration $\beta/H_*$ can be derived with Eq.~(\ref{eq:dt_drho}):
    \begin{equation}
        \beta =-\frac{d}{dt}{{\left. \left( \frac{{{S}_{3}}}{T} \right) \right|}_{{{t}_{*}}}}=4H_*\rho_r \frac{d}{d\rho_r }{{\left. \left( \frac{{{S}_{3}}}{T} \right) \right|}_{({{T}_{*}},{{\mu }_{*}})}},
    \end{equation}
    where the derivative lies along the cosmic trajectory.

\end{document}